    \documentclass[a4paper,fleqn,usenatbib]{mnras}
    \usepackage{newtxtext,newtxmath}
    \usepackage[usenames,dvipsnames]{color}
    \usepackage[T1]{fontenc}
    \usepackage{ae,aecompl}

    \usepackage[flushleft]{threeparttable}
    \usepackage{graphicx}   
    \usepackage{soul}
    \DeclareGraphicsExtensions{.ps,.pdf,.png}
    \usepackage{amsmath}    
    \usepackage{amssymb}    
    \def \mstar{$M_{*}$}

    \def \MBH{$M_{\rm BH}$}

    \def \LAGN{$L_{\rm AGN}$}

    \def\Hubble{\ifmmode {\rm km\,s}^{-1}\,{\rm Mpc}^{-1}\else km\,s$^{-1}$\,Mpc$^{-1}$\fi}
\def\Ho{\ifmmode H_{0} \else $H_{0}$\fi}
    \def\omm{\ifmmode \Omega_{{\rm M}} \else $\Omega_{{\rm M}}$\fi}
\def\omlam{\ifmmode \Omega_{\Lambda} \else $\Omega_{\Lambda}$\fi}
\usepackage{calc}
\newsavebox\CBox
\newcommand\hcancel[2][0.5pt]{%
  \ifmmode\sbox\CBox{$#2$}\else\sbox\CBox{#2}\fi%
  \makebox[0pt][l]{\usebox\CBox}%
  \rule[0.5\ht\CBox-#1/2]{\wd\CBox}{#1}}
\makeatletter
\newcommand*{\rom}[1]{\expandafter\@slowromancap\romannumeral #1@}
\usepackage{ulem}
\makeatother

\def \mic {$\mu \rm m$}
\def \LAGN {L$_{\rm AGN}$}
    \def\Msun{\ifmmode M_{\odot} \else $M_{\odot}$\fi}
    \def\Lsun{\ifmmode L_{\odot} \else $L_{\odot}$\fi}
    \def\msun{\ifmmode M_{\odot} \else $M_{\odot}$\fi}
    \def\lsun{\ifmmode L_{\odot} \else $L_{\odot}$\fi}
    
        \usepackage{psfrag}
    \usepackage{longtable}
\usepackage{rotating}

\usepackage{caption}
    \title{Intrinsic AGN SED \& black hole growth in the Palomar$-$Green quasars}
    \author[Lani et al.]{
    Caterina Lani$^{1}$\thanks{E-mail: caterina@wise.tau.ac.il},
    Hagai Netzer$^{1}$
     and Dieter Lutz$^{2}$
    \\
    $^{1}$School of Physics and Astronomy, Tel Aviv University, Tel Aviv 69978, Israel\\
    $^{2}$Max-Planck-Institut f\"ur extraterrestrische Physik,
             Giessenbachstra\ss\/e, 85748 Garching, Germany\\
    }
    \date{Accepted 2017 June 1. Received 2017 May 28; in original form 2017 February 26}
    \pubyear{2015}
    
\begin{document}
    \label{firstpage}
    \pagerange{\pageref{firstpage}--\pageref{lastpage}}
    \maketitle
    \begin{abstract}
We present a new analysis of the PG quasar sample based on $\it Spitzer$ and $\it Herschel$ observations. (\rom{1}) Assuming PAH-based star formation luminosities ($L_{\rm SF}$) similar to \citet[][S16]{Symeonidis2016}, we find mean and median intrinsic AGN spectral energy distributions (SEDs). These, in the FIR, appear hotter and significantly less luminous than the S16 mean intrinsic AGN SED. The differences are mostly due to our normalization of the individual SEDs, that properly accounts for a small number of very FIR-luminous quasars. Our median, PAH-based SED represents $\sim6$\% increase on the $1-243$\mic\ luminosity of the extended \citet[][EM12]{Mor2012} torus SED, while S16 find a significantly larger difference. It requires large-scale dust with $T\sim 20-30\,\rm K$ which, if optically thin and heated by the AGN, would be outside the host galaxy. (\rom{2}) We also explore the black hole and stellar mass growths, using $L_{\rm SF}$ estimates from fitting $\it Herschel$/PACS observations after subtracting the EM12 torus contribution. We use rough estimates of stellar mass, based on scaling relations, to divide our sample into groups: on, below and above the star formation main sequence (SFMS). Objects on the SFMS show a strong correlation between star formation luminosity and AGN bolometric luminosity, with a logarithmic slope of $\sim0.7$. Finally we derive the relative duty cycles of this and another sample of very luminous AGN at $z=2-3.5$. Large differences in this quantity indicate different evolutionary pathways for these two populations characterised by significantly different black hole masses.

    \end{abstract}
    \begin{keywords}
    galaxies: active -- quasars: supermassive black holes -- galaxies: star formation
    \end{keywords}
    \section{Introduction}
    \label{sec:intro}
    A long-standing quest in the field of extra-galactic astronomy is the
    understanding of how Active Galactic Nuclei (AGNs) and their host galaxies
    impact on the respective evolution.
    
 The well-known correlation
    between the mass of the central Super Massive Black Hole (SMBH) and the
    velocity dispersion of the bulge of its host ($M-\sigma$ relation;
    \citealt{Ferrarese2000}, \citealt{Gebhardt2000}) led to the postulation of a
    scenario where feedback, driven by the central SMBH, impacts on the galactic
    environment and ultimately regulates the bulge's star formation activity and
    mass growth. Although the exact relationship is still highly debated 
    (e.g. \citealt{Kormendy2013}, 
    \citealt{Lasker2014a}, \citealt{Lasker2014b}, \citealt{ Reines2015}, \citealt{Lasker2016}, \citealt{Shankar2016}), this is an indication that the evolution of the host galaxy and
    that of the central SMBH may be closely related.
    \newline\indent Both star formation activity and black hole (BH)
    accretion feed from the same materials.
    Although these processes take place on very different spatial (typically
    sub-parsec for the BH accretion and kpc for star formation activity) and temporal (typically
    a few million years for the BH accretion and a few hundreds million years for star formation activity) scales, they may compete for their fuel. In favour of an evolution ``in tandem" for
    these two processes, is the observation that the cosmic star formation rate
    density, and the cosmic accretion activity of SMBHs, rise and fall at a
    similar pace, with both peaking at around redshift $z\sim2$
    \citep[e.g.][]{Madau1996,Ueda2003,Heckman2004,Hopkins2006, Madau2014}. We caution, however, that this
    picture only holds when the overall galaxy population is considered. In
    contrast, when only the most luminous quasars in the Universe are taken into
    consideration, star formation activity and BH accretion appear to peak at
    different redshift ($z\sim4$ for the former and $z\sim2.5$ for the latter,
    e.g. \citealt{Netzer2016}). This may be an indication that BHs with significantly different masses evolve along different pathways with respect to their hosts.
    
 One avenue to observationally explore a potential co-evolution is
    to consider the black hole accretion rate and star formation activity in
    AGNs and their hosts, over a range in redshift and AGN bolometric
    luminosity. This is a difficult task because it requires the determination
    of the relative contribution from star formation and black hole accretion to
    the NIR--MIR BH+host SED. A variety of methods has been used to tackle this
    issue. Some consider polycyclic aromatic hydrocarbons (PAHs) spectral
    features ($\lambda=6-13\,\mu m$), as these originate in the proximity of
    young massive stars and therefore constitute good tracers of pure star
    formation activity \citep[e.g.][]{Gillet1975,Willner1977,Genzel1998, Laurent2000,Sturm2000,HernanCaballero2015,GarciaBernete2015}.
    Given
    that the conversion from PAH flux to SF is non-trivial, and given that
    the use of PAH features requires spectroscopic data, many studies prefer SED
    fitting as a mean to estimate the star formation activity in AGN hosts. This approach requires far infrared (FIR)  and millimeter observations. It also necessitates prior-knowledge (or
    assumption) of both the shape and contribution of the intrinsic AGN emission to
    the overall AGN+host SED.
    
As explained, the PAH-based method relies on MIR
    spectroscopy to identify the most appropriate SF template (from a wide range
    of libraries) which, in turn, is used to subtract the flux attributable to
    SF in the host galaxy from the AGN+host SED. The result is therefore the
    intrinsic AGN SED. Unsurprisingly, this procedure does not yield an
    uncertainty-free intrinsic AGN SED. For example, accurate estimates for the
    equivalent width of the $7.7\rm\,\mu m$ PAH feature is difficult due to its
    blending with the neighbouring $8.6\rm\,\mu m$ feature. Most works converge on a
    typical intrinsic AGN SED shape, characterised by two broad bumps at
    $\sim\,10$ and $\sim\,20-40\rm\,\mu m$ due to silicate emission, and a fast
    decay long-ward of the second silicate feature. The slope of the decay and
    the exact location of the turn-off can however strongly vary from one work
    to the next. Some find the turn-off is located at about $\lambda>30\,\rm\mu m$ \citep[e.g.][]{Netzer2007,Mor2012,Xu2015,Lyu2017}, others find it at moderately
    longer wavelengths depending on AGN luminosity \citep[e.g.][]{Mullaney2011}
    and at much longer wavelengths (e.g. \citealt{Symeonidis2016} at $\sim100\,\rm\mu m$). The intrinsic
    uncertainty and difficulties involved in the accurate determination of
    star-formation with PAHs in AGNs are the most likely culprits for the
    observed discrepancies. In order to test the turn-off and slope of the
    intrinsic AGN SED in the FIR, \cite{Netzer2016} considered a large number of
    FIR upper limits for the most luminous quasars at redshift $z=2-3.5$
    ($\log\,L_{\rm AGN}\geq46.5$), and found that these are fully consistent with
    the extended \cite{Mor2012} intrinsic AGN SED (see \S\ref{sec: intro to SFR and AGN SED} and \citealt{Netzer2016}), mostly consistent with the \cite{Mullaney2011}
    intrinsic AGN SED (that based on the more luminous sources in their sample,
    $\log\left( L_{\rm 2-10\,KeV}\right) \geq42.9$) and inconsistent with the \cite{Polletta2006,Polletta2007} SED as used by \cite{Tsai2015}. 
     
     Very recently \cite{Lyu2016} considered the entire sample of Palomar$-$Green quasars \citep{Schmidt1983} and identified a secure sub-sample 
      ($\sim10$ per cent) of sources that show evidence of warm/hot dust
       deficiency in their NIR-MIR SEDs. Regardless of their spectral shape
        at shorter wavelengths these quasars show a FIR decay that is broadly
         consistent with earlier works \citep[e.g.][]{Xu2015,Netzer2016}, 
         but starts at shorter wavelengths ($2-5\mu \rm m$) and is steeper. We 
       note that these IR-weak sources had already been identified, although 
       not as systematically characterised, in \cite{Mor2011} and \cite{Mor2012}. 
       Despite the consideration of nearly an identical sample in the current work
        and in \cite{Lyu2016}, we focus on the longer wavelengths (FIR)
         of the intrinsic AGN SED. We will, therefore, not discuss the results
          from \cite{Lyu2016} on this specific topic any further in the remainder of this work.\\
\\
\indent  Irrespective of the differences in the adopted intrinsic AGN SEDs,
    the sample selection methods and the treatment of the data in hand (e.g.
    survival analysis vs. stacking), a plethora of works has identified some general trends of AGN luminosity
    ($\equiv \,L_{\rm AGN}$ hereafter) with SF luminosity (obtained by
     integrating the relevant SF templates between $8\,\mu\rm m$ and $1000\,\mu\rm m$, $L_{\rm SF}$
    hereafter). On the one hand, in the regime where AGN activity dominates,
    i.e. $L_{\rm AGN}>L_{\rm SF}$,  there are hints for a correlation:
     $L_{\rm SF}\propto L_{\rm AGN}^{0.6-0.7}$ 
     \citep[e.g.][]{Netzer2009, Rosario2012, Netzer2016, Ichikawa2016}, 
     at least in the local Universe. Some studies confirm the persistence of this correlation at high redshift (e.g. $z=2-3.5$, \citealt{Mullaney2012,Rovilos2012,Netzer2016}), while others 
     point towards its disappearance for $z>1$ (e.g. \citealt{Rosario2012, Stanley2015}).
      On the other hand, there is on average a close to linear relation, $L_{\rm SF}\sim L_{\rm AGN}$, for FIR selected samples \citep[e.g.][]{Mullaney2012,Chen2013,Delvecchio2015}, and no
    significant correlation for X-ray selected samples
    \citep[e.g.][]{Rosario2012,Stanley2015,Shimizu2016}. 

 The same matter can be analysed in terms of the ratio between star formation rate (SFR) and black hole accretion rate
    (BHAR), e.g \cite{Mullaney2012}. This represents the instantaneous growth of stellar mass to that of the
    black hole. For example \cite{Netzer2016} found that the most luminous AGNs
    in the redshift range $z\sim\,2-3.5$, that are detected in $Herschel$/SPIRE,
    show a ratio of SFR/BHAR similar to the ratio of stellar mass to black hole
    mass observed in the local Universe for the most massive, spheroidal
    galaxies.
    This is consistent with a scenario where the most luminous sources in the
    Universe have built up a significant fraction of their stellar mass during
    periods when the black hole was extremely active.\\

    In this paper we focus on the well-studied sample of Palomar--Green
    quasars \citep{Schmidt1983} and their recent $Herschel$\footnote{Herschel is an
 ESA space observatory with science instruments provided by European-led
 Principal Investigator consortia and with important participation from
 NASA.} observations. This
    allows us to extend the study by \cite{Netzer2016} to a sample of black
    holes with lower mass and at lower redshift, with the aim of determining how
    their behaviour differs from that of the most massive BHs. Furthermore, we
    make use of their extensive spectroscopic and photometric data to revisit
    the intrinsic AGN SED shape. \footnote{A new paper by \cite{Lyu2017} appeared on the arXiv two days before we received our referee report. The paper addresses several issues we also discuss in this work and a complete reference to all these would completely alter the structure of our manuscript. Because of this, we only refer to the main points that are relevant to the present work, mostly at the end of \S\ref{sec:results}.}
 
 The sample, the data and basic quantities (i.e. BH mass) are described in \S\ref{sec:sample & data}. 
   The two sets of SF luminosities (PAH-based and SED-fitting-based) considered in our work, a new median 
   PAH-based intrinsic AGN SED, and our SED fitting decomposition method are detailed in \S\ref{sec: intro to SFR and AGN SED}. 
   In \S\ref{sec:results} we discuss our findings. In \S\ref{sec: L_SFus vs L_SFshi} we contrast the two sets of SF
    luminosities dealt with in this work, while in \S\ref{sec: revisiting AGN SED} we characterise the new median 
    PAH-based intrinsic AGN SED. In \S\ref{sec:torus covering factor} we investigate the typical covering factor
     that we found for the PG sample, and in \S\ref{sec:non-torus covering factor etc.} that of the potential 
     non-torus-related, non-SF-related dust. We conclude by focusing on the relative instantaneous growth 
     rates of BH and stellar mass in \S\ref{sec:SFR BHAR}, where we also consider results from a previous 
     study in order to gain a more complete understanding of these quantities over a range in redshift and AGN luminosity/BH mass. Our results are summarised in \S\ref{sec:conc}. Throughout this paper we assume $\Ho=70$ \Hubble, $\omm = 0.3$ and $\omlam = 0.7$

    \section{Sample and Data}
    \label{sec:sample & data}
    \subsection{Sample}
    \label{sec:sample}
    We consider a sample of Palomar--Green Quasars (PG QSOs), which
    was in turn selected from the Palomar Bright Quasar Survey Catalogue from
    \cite{Schmidt1983}. While below we summarise its characteristic, we refer
    the reader to \cite{Schmidt1983}, \cite{Petric2015} and references therein
    for more details regarding this sample.
    
\begin{figure*}
    \begin{minipage}[c]{\textwidth}
    \begin{center}
    \begin{tabular}{c c}
    {\psfrag{y2}[][][1][0]{normalised fraction}
    \psfrag{x2}[][][1][0]{ $\rm\log \left(L_{\rm 5100}/erg\,s^{-1}\right) $}
    \psfrag{blah1}[][][0.9][0]{} \hspace{-2.7cm}\includegraphics[trim=11cm 0cm
    16cm 1cm, clip=true,
    width=0.425\textwidth]{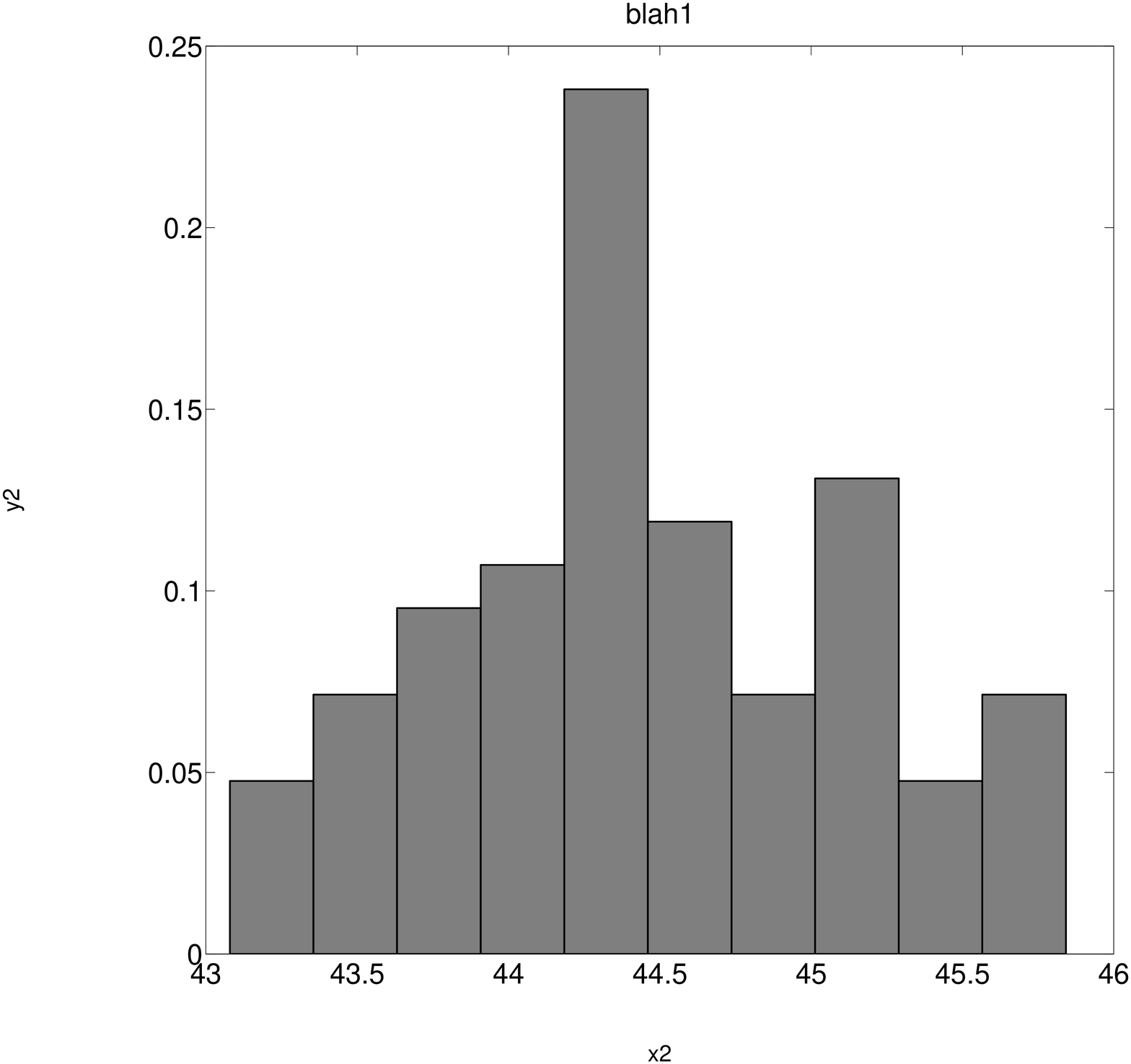}} &
    {\psfrag{y}[][][1][0]{normalised fraction}
    \psfrag{x}[][][1][0]{z}
    \psfrag{blah4}[][][0.9][0]{} \hspace{.8cm}\includegraphics[trim=11cm 0cm
    16cm 1cm, clip=true,
    width=0.425\textwidth]{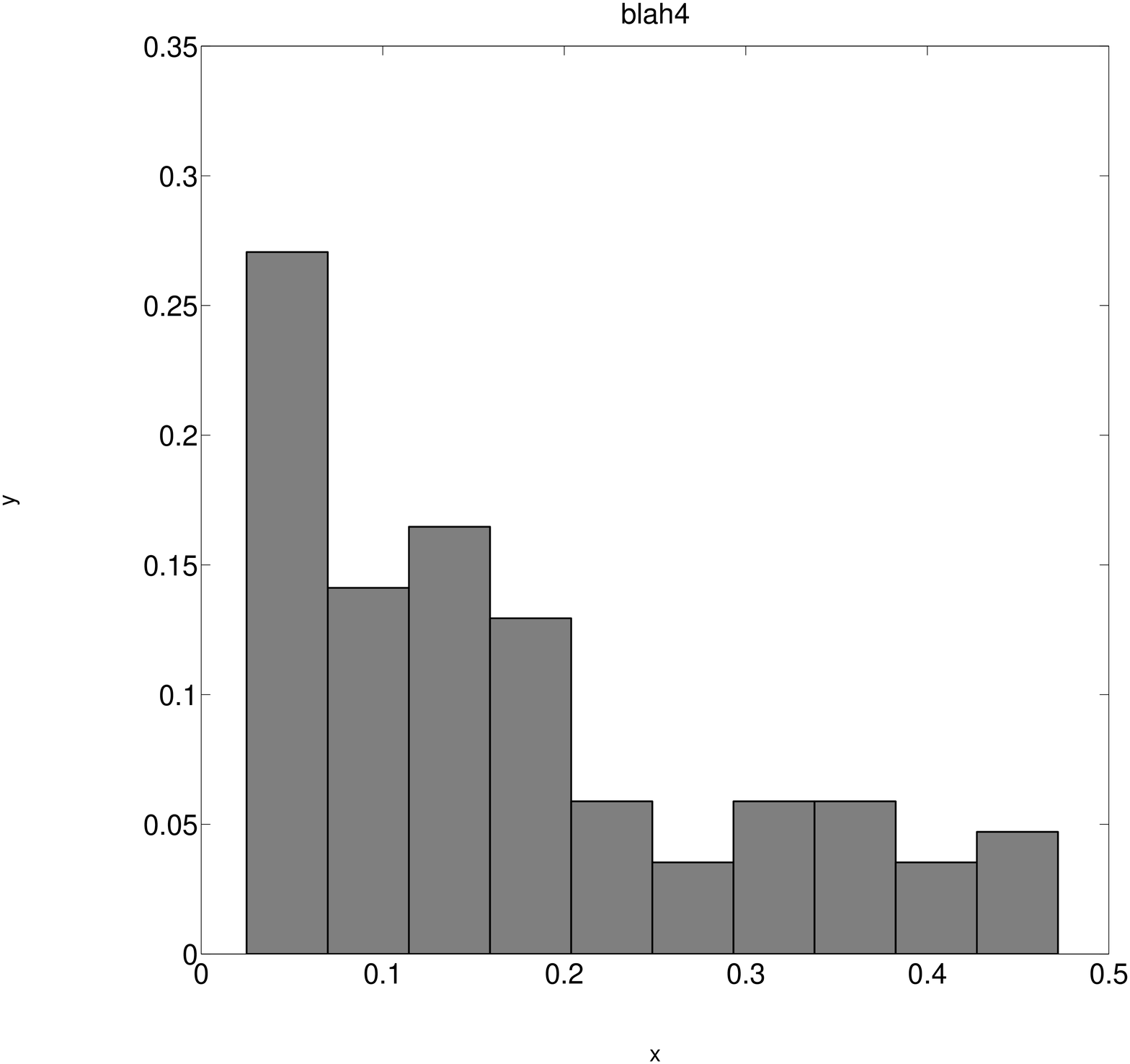}} \\
    \vspace{1cm}
    {\psfrag{x}[][][1][0]{$\rm\log \left(M_{\rm BH}/M_{\odot}\right)$}
    \psfrag{y}[][][1][0]{normalised fraction }
    \psfrag{blah3}[][][0.9][0]{} \hspace{-2.7cm}\includegraphics[trim=11cm 0cm
    16cm 0cm, clip=true,
    width=0.425\textwidth]{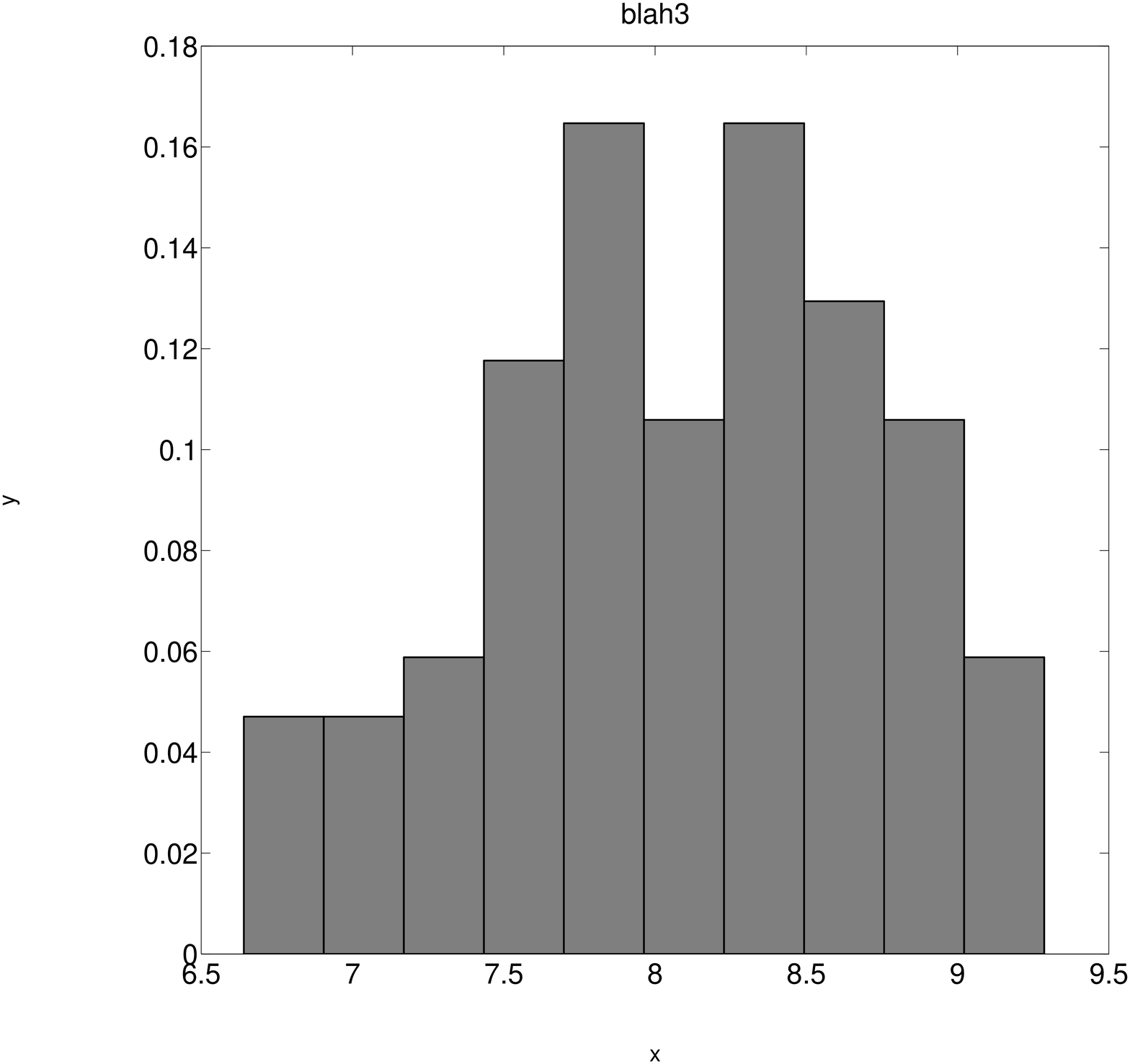}} &
    {\psfrag{blah5}[][][1][0]{} \psfrag{y}[][][1][0]{normalised fraction}
    \psfrag{x}[][][1][0]{$\rm L/L_{Edd}$}
    \hspace{.8cm}\includegraphics[trim=11cm 0cm 16cm 0cm, clip=true,
    width=0.425\textwidth]{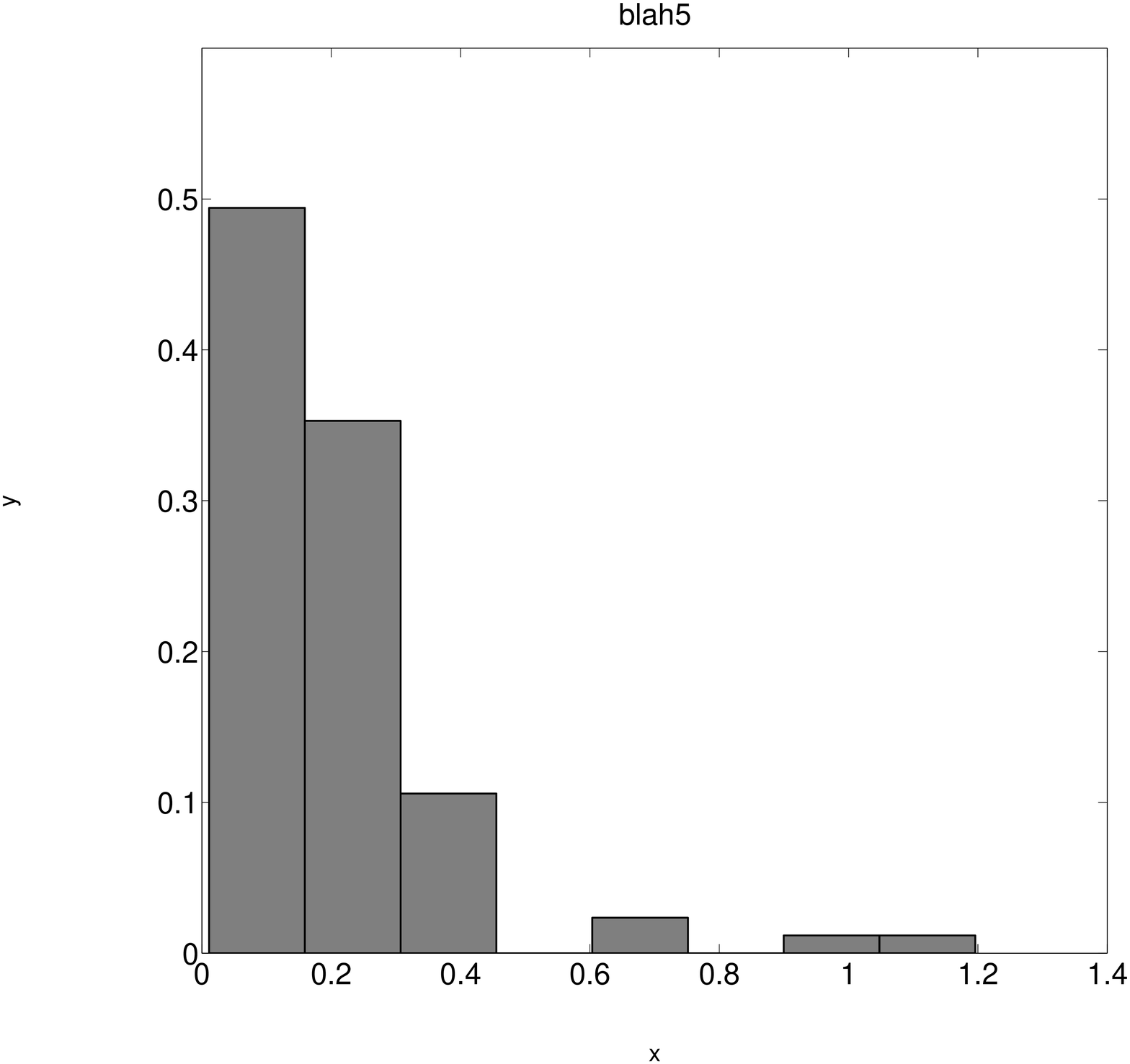}} \\
     \end{tabular}
       \end{center}
    \end{minipage}
    \vspace{0.2cm}
      \caption{The four histograms show the properties of the PG quasars sample considered 
      in the current work. The top left panel shows the normalised fraction of objects as a
       function of optical (at 5100$\AA$) luminosity (not corrected for host galaxy contamination,
        see \S\ref{sec:sample}), while the top right panel shows that as a function of redshift. 
        The bottom panels show the normalised fraction of PG quasars as a function of black hole mass (left)
        and Eddington ratio (right). The description of how the quantities along  the x-axes were measured is provided in \S\ref{sec:Data}.}
    \label{fig:sample_charct}
    \end{figure*}

 PG QSOs are optically
    luminous ($B-$band magnitude $M_{\rm B}<16.16$), blue type-I quasars. Their
    optical colour is $U-B<-0.44$ \citep{Schmidt1983}, their redshifts range
    from $z\sim0$ to $z\sim0.5$ \citep{Boroson1992}, and their black hole masses
    and Eddington ratios cover the ranges $M_{\rm BH}=4\times10^{6}-2\times10^{9}\,M_{\odot}$ and $L_{\rm AGN}/L_{\rm Edd}=0.01-1.2$
     respectively (for details on how we measured these quantities
    see \S\ref{sec:Data}). All these properties are illustrated in Figure
    \ref{fig:sample_charct}. While the $L_{\rm 5100}$ values employed in our work were not corrected for host
     galaxy contamination, the observed $EW\left( \rm H \rm \beta\right)$ distribution suggests a typical host galaxy
      contamination to $L_{5100}$ of $\sim 20$ per cent. Finally, \cite{Jester2005} show that, although incomplete in
    $U-B$, PG QSOs are representative of the general population of
    optically-bright, optically-selected quasars.
    
    \subsection{Data}
    \label{sec:Data}
    In our work we make use of both photometric and spectroscopic IR data. The
    former cover the following wavelength range: NIR from the 2MASS survey
    (\citealt{Skrutskie2006}); MIR from $WISE$ \citep{Wright2010} and FIR from
    the $Herschel\,\,Space\,\, Observatory$ \citep{Pilbratt2010}. In particular, the
    $J$, $H$, $K$ and $WISE$ data were taken from the NASA/IPAC IR science archive, from the  
    2MASS All-Sky Point Source Catalog and the AllWISE Source Catalog respectively. Far-infrared 70, 100, and 160$\mu$m $Herschel$/PACS (Poglitsch et al. 2010) 
    data were retrieved from the $Herschel$ archive and reduced as described in and used for \cite{Lutz2016}. In summary, we used a standard ``masked highpass filtering"
     reduction with a circular mask centered on the source position. Fluxes are measured by aperture photometry, and noise is estimated by placing additional apertures in regions off the source. 
     Most sources were observed twice to obtain both 70 and 100$\mu$m photometry, and the parallel 160$\mu$m results for these were averaged. In two cases we do not quote a 160$\mu$m flux due 
     to blending with a nearby source. Furthermore, no PACS observations are in the Herschel archive for PG1226+023 and PG1444+407. Most observations of PG QSOs
      have been obtained in program OT1\_lho\_1 
     (see \citealt{Petric2015} for first results), with a few others from OT1\_hnetzer\_1 and OT1\_rmushotz\_1 (see \citealt{Melendez2014}). 
     The 250 350, and 500$\mu$m $Herschel$/SPIRE data were taken from \cite[][their Table 1]{Petric2015}.
    
 The spectroscopic
    data were taken from \cite{Shi2014} and are a compilation of spectra
    observed as part of a programme (PID: 5096, PI: Rieke), and archive spectra.
    The data reduction and processing are described in detail in
    \cite{Shi2014}. Here we only provide the reader with a summary. All spectra
    but one (see later) are the combination of the short-low ($5-14\,\rm \mu m$;
    SL) and long-low ($14-40\,\rm \mu m$; LL) resolution observations by
    $Spitzer$/IRS. For a given source, the SL spectrum was aligned to the LL
    spectrum and, in turn, the result was aligned to the $Spitzer$ 24$\,\rm \mu m$
    photometric point. This procedure was necessary due to slight calibration
    differences; \cite{Shi2014} report that the scaling factors are close to unity in the majority of cases (their Table 2). We
    confirm that the \cite{Shi2014} spectra are in good agreement with the
    $Spitzer$ 24$\,\rm \mu m$ photometric points that we have downloaded from
    the NASA/IPAC Infrared Science Archive. 

In order to extract the photometric
    information required for our SED fitting procedure, we consider the following three wavelength bands: $6.75-7.25\,\rm \mu m$, $20.75-21.25\,\rm \mu m$
     and $29.75-30.25\,\rm \mu m$ (all rest-frame). These were chosen to
    be approximately equally spaced across the spectrum, as well as avoid strong
    emission and absorption lines. Our SED fitting procedure uses the median fluxes in these regions. We have not considered the $Herschel$/SPIRE data (provided by \citealt{Petric2015}) in our SED
    fitting procedure (\S \ref{sec:SF assuming M12}) as they are mostly upper
    limits, especially in the higher redshift (z>0.2) sources. When possible,
    however, we consider the $Herschel$/SPIRE $250\,\rm \mu m$ information in
    our PAH-based calculation of the median intrinsic AGN SED (\S \ref{sec: obtaining intrinsic AGN SED}). The PAH-based SF luminosities
    considered in \S \ref{sec: measuring SF using PAHs} and \ref{sec: obtaining intrinsic AGN SED} are taken from \cite{Shi2007}. They were preferred to
    those presented in \cite{Shi2014} which are based on a combination of
    information from the $11.3\mu m$ PAH feature (solely due to SF) and FIR 
    photometry (including emission from both SF and AGN-heated dust), 
    therefore potentially introducing another source of uncertainty in our analysis. 
    We note that adopting the \cite{Shi2014} results would lead to larger SF luminosities and lower FIR AGN continua than derived with the \cite{Shi2007} results.
    Lower intrinsic AGN continua would also be a consequence of adopting a PAH to SF scaling such as that in \cite{Lyu2017} instead of that from \cite{Shi2007}.
 All BH masses used in this work are based on the most recent
    reverberation mapping results of \cite{Bentz2013} and the FWHM of the $H\beta$ line.
   We considered the following equation:
   \begin{equation}
       M_{\rm BH}=1.4625\times10^{5}\,f_{\rm BLR}\,\left( \frac{FWHM\left(H\beta\right)}{10^{3}\,km/s}\right) ^2\,R_{\rm BLR},
   \end{equation}
where $FWHM\left(H\beta\right)$ is the full width half maximum intensity of $H\beta$. $R_{\rm BLR}$ is the
 size of the broad line region and can be approximated by $K\,\left( \frac{L_{5100}}{10^{44}}\right) ^{\alpha}$;
  following \cite{Bentz2013} we set $K$ and $\alpha$ to 33.65 and 0.53 respectively. Finally, $f_{\rm BLR}$,
   which contains information about the geometry and dynamics of the broad line region gas, was set to 1.
   
    Regarding the bolometric luminosity $L_{\rm AGN}$, we used the expression
    from \cite{Netzer2013} which gives the following bolometric correction
    factor, {\it bol}, that multiplies $L_{\rm 5100}$: $\rm  bol=53-\log(L_{5100})$. This is in very good agreement with other estimates
    for objects in the general luminosity range of the PG sample but tends to
    over-estimate $L_{\rm AGN}$ for more luminous sources (see discussion in
    \citealt{Netzer2016}). 

 Our final sample consists of 85 PG QSOs; of these 69 were detected in all
    the three $Herschel$/PACS bands (though 2 suffer from blending at $160\mu \rm m$), 12 are undetected in one and/or two of
    the $Herschel$/PACS bands, 4 are undetected in all $Herschel$/PACS bands. The
    original PG QSOs sample \citep{Schmidt1983} contained 87 quasars, but
    \cite{Petric2015} only obtained $Herschel$ data for 85 of them. We note that PG0003$-$199 lacks the LL
    $Spitzer$/IRS observation and we find a large discrepancy between its
    spectrum and the full SED. For this source we therefore chose not to
    consider the information from $Spitzer$/IRS.

    \section{Measuring SF luminosities and intrinsic AGN SEDs}
    \label{sec: intro to SFR and AGN SED}
\begin{figure}
    \psfrag{y}[][][1.5][0]{normalised $\rm \lambda\,L_{\rm \lambda}$}
    \psfrag{x}[][][1.5][0]{$\rm \lambda_{\rm rest}$ ($\mu m$)}
    \psfrag{p1}[][][0.9][0]{\hspace{1.25cm}2MASS}
    \psfrag{p2}[][][0.9][0]{\hspace{1.04cm}WISE}
    \psfrag{p3}[][][0.9][0]{\hspace{2.35cm} PACS detections}
    \psfrag{p4}[][][0.9][0]{\hspace{3.07cm} PACS $3\sigma$ upper limits}
    \psfrag{p5}[][][0.9][0]{\hspace{2.47cm} SPIRE detections}
    \psfrag{p6}[][][0.9][0]{\hspace{3.20cm} SPIRE $3\sigma$ upper limits}
    \psfrag{t1}[][][0.9][0]{\hspace{1cm} EM12}
    \psfrag{t2}[][][0.9][0]{\hspace{0.80cm}S16}
    \includegraphics[scale=0.23,trim={13cm 1cm 12cm   0cm},clip]{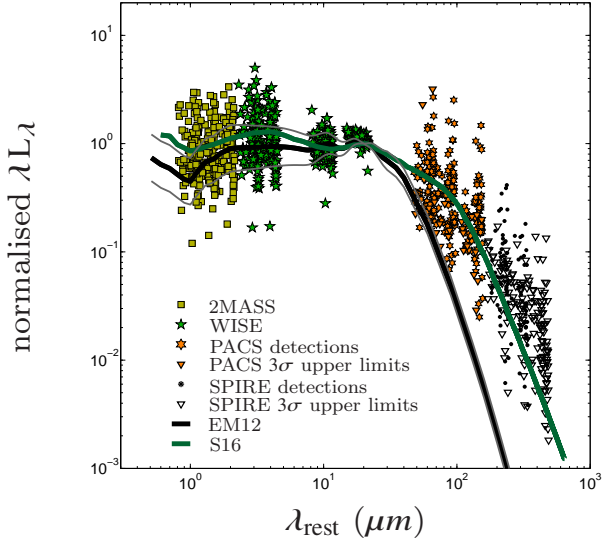} 
        \caption{Two different intrinsic AGN SEDs: in green is that found by S16,
        and in black the extended version of the \protect\citet[][EM12]{Mor2012} SED with corresponding $25^{\rm th}$ and $75^{\rm th}$ percentiles (light grey).
        Over-plotted are the normalised luminosities for our sample which include both AGN and host emission.}
        \label{fig: AGN_SEDs+points}
     \end{figure}

    The goals of this work are: {\it (i)} accurately measuring SF luminosities for all PG
    quasars, {\it (ii)} defining the intrinsic AGN SED of these sources, and
    {\it (iii)} investigating the growth rate of the stellar mass and BH mass in
    the sample. Our SF luminosity measurements are based on two independent methods. The first rests on the
    assumption that the intrinsic AGN SED of all sources is well-described by the extended version of the empirical median template
    identified by \cite{Mor2012}. This, together with the relevant $25^{\rm th}$ and $75^{\rm th}$ percentiles, was
    obtained by considering a sample of $\sim100$ local AGNs. The \cite{Mor2012} approach was
    based on a combination of PAH spectroscopy from $Spitzer$, with NIR and FIR
    photometry of different origin (see their paper for more details). They
    subtracted, from the overall spectrum, one of 40 FIR SF templates, ensuring
    that the only flux left around the $7.7\rm\,\mu m$ aromatic feature was
    consistent with the noise level. This resulted in three SEDs
    covering the rest frame wavelength $\lambda_{\rm rest}=1-30\rm\,\mu m$: a median SED and two others representing the $25^{\rm th}$ and $75^{\rm th}$ percentiles from it. In
    later works \citep[e.g.][]{Netzer2014,Netzer2016}, this was extended to
    longer wavelengths based on various clumpy torus models. The procedure
    involved adding a $100\rm\,K$ modified blackbody, with $\beta=1.5$, to the
    \cite{Mor2012} templates. Hereafter we will refer to this template as the extended \citet[][EM12]{Mor2012} SED. Although \cite{Mor2012} considered PAH features in
    their analysis, they took a different approach from \citet[][see \S\ref{sec: measuring SF using PAHs}]{Shi2007}, {\it requiring no assumption for the
    conversion from PAH to SF luminosity.} %
    The resulting median intrinsic AGN
    SED and the corresponding $25^{\rm th}$ and $75^{\rm th}$ percentiles are
    showed in Figure \ref{fig: AGN_SEDs+points} (black and grey lines).
    It is important to note that these SEDs are meant to represent the
    combination of dust emission from the central torus and that from the narrow
    line region (NLR). Because of the modified blackbody extrapolation beyond $30\, \mu \rm m$, they do not consider the possibility of AGN heated dust
    at a distance far beyond that of the NLR, and hence a temperature below $100\rm\,K$. In the reminder of this manuscript, when we mention ``torus" (e.g. torus luminosity, torus covering factor) in the context of the EM12 SED we actually mean ``torus$+$NLR".
    
 There is mounting observational evidence in favour of the EM12
    intrinsic AGN SED, especially if one assumes a narrow range in SED shape (a
    ``universal" SED). \cite{Netzer2016} used a sample of 66 {\it
    Herschel}/SPIRE-non-detections to provide constraints on the intrinsic FIR AGN SED. They found the EM12 intrinsic AGN SED to be
    consistent with the information provided by all the non-detections in their
    sample. They also showed that given a universal SED, the
    \cite{Mullaney2011} set of templates is consistent with a significant number of
    Herschel upper limits, and the \cite{Polletta2006,Polletta2007} SED, as used by \cite{Tsai2015}, is inconsistent with the
    majority of the observations. We note, however, that the samples considered by \cite{Mor2012}, \cite{Netzer2016} and the current work, are significantly more luminous than the sources in \cite{Mullaney2011}, thus making a direct comparison of their results problematic. In addition, and as already pointed out in the literature (e.g. \citealt{Lyu2017}), unlike the current work and the others mentioned above, \cite{Polletta2006, Polletta2007} do not explicitly correct for the star formation contribution in the FIR, making the comparison irrelevant. For these reasons, when in the remainder of this manuscript we examine the central issue of intrinsic AGN SED, we will mostly contrast the EM12 SED with that determined on a sub-sample of the sources considered here \citep{Symeonidis2016}. Finally the only direct, high resolution sub-mm observations of the nuclear obscuring structure to date (in NGC 1068, e.g.
    \citealt{GarciaBurillo2016,Gallimore2016}) are consistent with the EM12 intrinsic AGN SED.
    This remains the case even allowing for small differences due to aperture size and the fact that
    NGC 1068 is a type-II AGN.

     Figure \ref{fig: AGN_SEDs+points} shows
    the location of the $20\rm \mu m-$normalised luminosities calculated from
    the data in hand for our parent sample of PG QSOs, with respect to two
    intrinsic AGN SEDs. In solid black we show the
    EM12 SED and in green the mean SED suggested by \citet[][hereafter S16]{Symeonidis2016}.
    There is good agreement between the data and both intrinsic AGN
    templates in the near- and mid- IR. If we focus on the longer
    wavelengths, i.e. $Herschel$/PACS and $Herschel$/SPIRE, the situation looks
    different. We find (see Appendix) that the different shape of the S16 SED at this wavelengths is mostly due to the
    way it was calculated. In
    their work S16 start by calculating the {\it mean} observed quasar SED
    without normalising the individual SEDs at a given wavelength. This returns
    a mean observed quasar SED which is heavily weighted towards the more
    luminous sources (their PG sample spans nearly two orders of
    magnitude in luminosity). They later obtain the {\it mean} SF template for this ensemble, made from all
     the individual SF templates,  and subtract it from the mean observed spectrum to derive the {\it mean} intrinsic AGN SED. From 
Figure \ref{fig: AGN_SEDs+points} it is apparent that, even prior to
    subtracting the appropriate emission due to SF from the $Herschel$/PACS
    data, the suggestion that the S16 AGN SED is universal is inconsistent with
    the observations because $\sim50$ per cent of our PG QSOs already lie below
    this. The EM12 median intrinsic AGN SED seems a better representation
    of the general AGN population if, indeed, emission heated by the AGN is dominated by warm dust in the torus and NLR, and can be considered universal. In \S\ref{sec: revisiting AGN SED},
     we attempt to repeat the S16 procedure, with several important modifications, and discuss the meaning of the mean intrinsic AGN SED obtained by them.
     
     The first set of SF luminosities we consider in this work was
    obtained by adopting the EM12 median SED and the two
    accompanying SEDs representing the $25^{\rm th}$ and $75^{\rm th}$
    percentiles, as our best representation of the intrinsic AGN SED (\S \ref{sec:SF assuming M12}).
    The second, independent set of SF luminosities was 
    that estimated by \cite{Shi2007}. This
    method does not assume a particular intrinsic AGN SED, but a relation
    between the luminosity of PAH features as measured from the $Spitzer$ spectroscopy, 
     and the total infrared ($8-1000 \rm \mu m$) luminosity. This method is the only one used by S16 in their recent
    analysis of the PG sample. We note that in their paper S16 focus on those PG
    quasars at redshift $z<0.18$ in order to cap the number of upper limits in
    their analysis. In the current work, however, we focus on the full sample,
    but present the analysis of the same sample as that consider by  S16 in the
    Appendix.

    \subsection{Measuring SF luminosities assuming the EM12 intrinsic AGN SED}
    \label{sec:SF assuming M12}
    Here we estimate the SF activity for 81 PG QSOs (those with at least one $3\sigma$-$Herschel$/PACS
    detection) by assuming the extended \cite{Mor2012} intrinsic AGN SED
    discussed above, and employing a SED fitting method.
    
 First we estimate the torus luminosity by fitting the three intrinsic AGN
    SED models (median and $25^{\rm th}$ and $75^{\rm th}$ percentiles) to the near- and mid-infrared points ($2.15\rm\,\mu m\leq\lambda_{\rm rest}\leq30\rm\,\mu m$)
    of the object's SED. Given the redshift range of our sources, this includes the data from WISE and the photometric
    points extracted from the $Spitzer$/IRS spectra. This window in wavelength-space is
    where the intrinsic AGN emission dominates over that from the host. We do not consider the emission
    short-ward of $\lambda_{\rm rest}=2\rm\,\mu m$ as this is where the contamination from the
    host's old stellar population can be significant; furthermore, the
    accretion disk emission at longer wavelengths is negligible \citep[see e.g.][]{HernanCaballero2015}. 
    We avoid combining the MIR and FIR into a single $\chi^2$ fit, and instead use a
     least square procedure for the MIR for two reasons. First, the formal uncertainties 
     on the $Spitzer$ and $WISE$ points are much smaller than the scatter in the MIR data.
      Second, the uncertainties on the $Herschel$ FIR data are much larger and combining the two into a single $\chi^2$ procedure 
      will heavily bias the results.
    
 Upon determining the best fitting EM12 SED, we know its
    contribution to the total SED at wavelengths $\lambda_{\rm rest}\geq30\rm\,\mu m$. This quantity is then subtracted from the composite
    AGN$+$host SED in the wavelength range covered by {\it Herschel}/PACS
    ($70\,\rm \mu m\leq\lambda_{\rm obs}\leq160\rm\,\mu m$), such that the
    remaining luminosity can be associated with star formation alone. The
    star-formation luminosity ($L_{\rm SF}$) is estimated using the
    \citet[][CE01 hereafter]{Chary2001} templates for local star-forming galaxies. 
    There is evidence \citep[e.g.][]{Elbaz2010} for the CE01 templates underestimating
     the SF emission at wavelength $\lambda_{\rm rest}>150\rm \mu m$. We note, however,
      that for all our sources 2 ($70\mu \rm m$ and $100\mu \rm m$) out of the 3 PACS points considered
       in the SED fitting are at wavelength $\lambda_{\rm rest}<150\rm \mu m$, and for $\sim70$ per 
       cent of our sources the last PACS point ($160\mu \rm m$) is short-ward of $\lambda_{\rm rest}=150\rm \mu m$.
        Furthermore, the errorbars accompanying the $160\mu \rm m$ point are typically larger than the those at shorter wavelengths. 
        The effect of underestimating the SF emission at long wavelengths by the CE01 templates, albeit present, is therefore within this uncertainty.
        As a test investigating the impact that the consideration of the alternative \cite{Dale2002} set of SF templates has on the main conclusions
         of our work we refer the reader to \S\ref{sec: obtaining intrinsic AGN SED}. In our routine we fit
    both for the vertical normalisation (i.e. $L_{\rm SF}$) and the template
    shape (dust temperature). This is achieved by normalising each CE01 template
    by its total ($\lambda=8-1000\,\rm \mu m$) IR luminosity 
    prior to considering them in the fitting procedure. In this part of the
    fitting we minimise the $\chi^2$, searching for
    the best fitting combination of $L_{\rm SF}$ and template.
 Our fitting is comprised of two phases. First we look for the CE01 SF template nearest to the torus-subtracted FIR luminosity,
  and then we search the CE01 library within $\pm0.5\rm\, dex$ of the SF luminosity
 of such template to improve the match in shape. The uncertainty in $L_{\rm SF}$ is determined by
    marginalising over the best fitting CE01 template shape, and identifying the
    lowest and highest $L_{\rm SF}$ among those that satisfy the equation
    $\chi^2<\chi_{min}^2+1$. For further details on our method, as well as on the treatment of uncertainties, we refer the reader to \cite{Netzer2016}.
    
\begin{figure*}
    \begin{minipage}[c]{\textwidth}
    \begin{center}
    \begin{tabular}{c c}
    {\psfrag{y}[][][1.5][0]{$\rm \log\left( \lambda\,L_{\rm \lambda}/ erg\,s^{-1}\right) $}
    \psfrag{x}[][][1.5][0]{$\rm \lambda_{\rm rest}$ ($\mu m$)}
    \psfrag{p1}[][][0.9][0]{\hspace{1.34cm}2MASS}
    \psfrag{p2}[][][0.9][0]{\hspace{1.13cm}WISE}
    \psfrag{p4}[][][0.9][0]{\hspace{1.02cm} PACS}
    \psfrag{p3}[][][0.9][0]{\hspace{2.36cm} IRS photometry}
    \psfrag{p5}[][][0.9][0]{\hspace{1.13cm} SPIRE}
    \psfrag{t1}[][][0.9][0]{\hspace{3cm}intrinsic AGN SED}
    \psfrag{t2}[][][0.9][0]{\hspace{1.47cm}SF SED}
    \psfrag{t3}[][][0.9][0]{\hspace{2.43cm}composite SED}
    \hspace{-2.0cm}\includegraphics[trim=9cm 0cm 16cm 1cm, clip=true, width=0.6\textwidth]{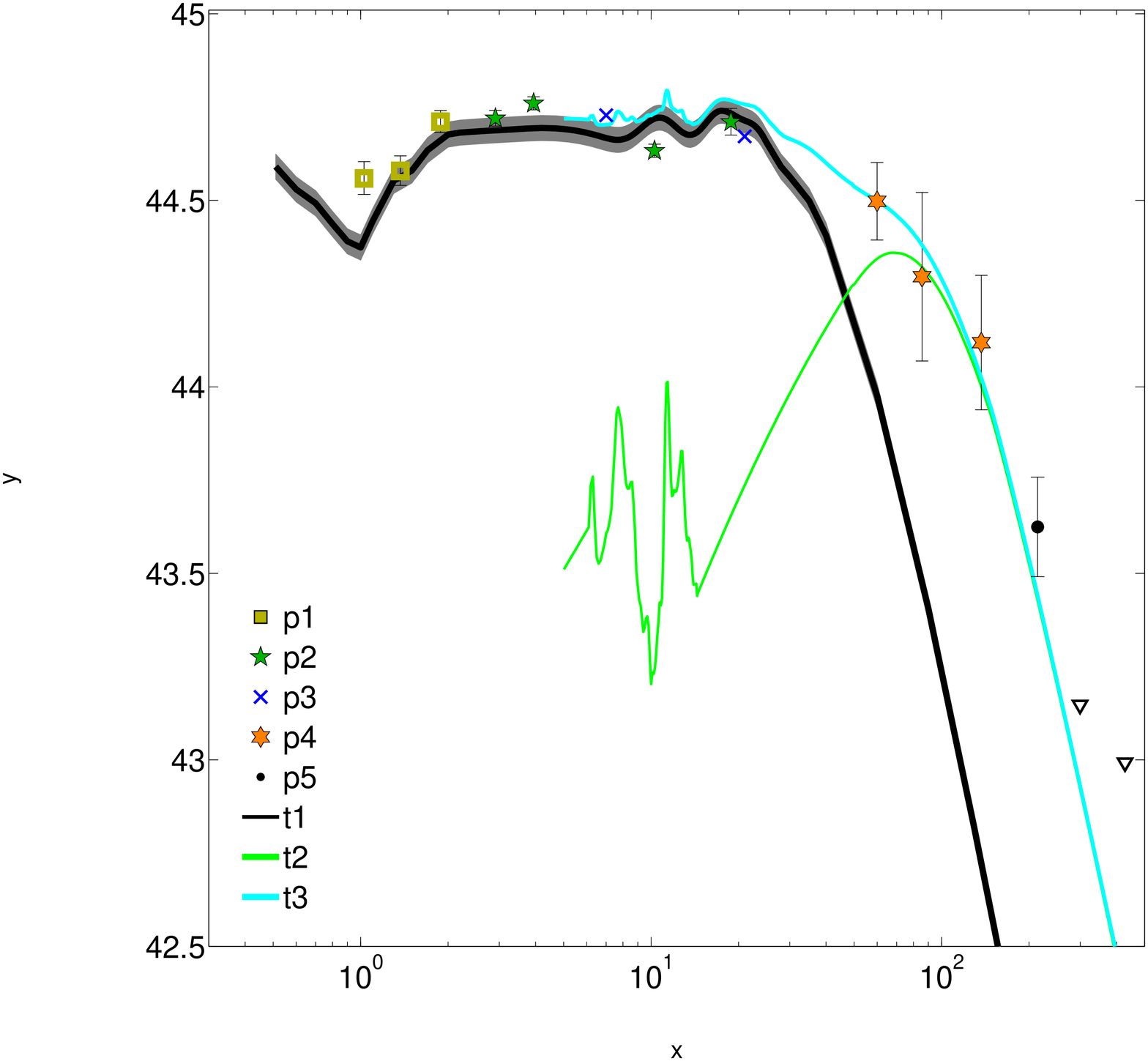}} &
    {\psfrag{y}[][][1.5][0]{$\rm \log\left( \lambda\,L_{\rm \lambda}/ erg\,s^{-1}\right) $}
    \psfrag{x}[][][1.5][0]{$\rm \lambda_{\rm rest}$ ($\mu m$)}
    \psfrag{p1}[][][0.9][0]{\hspace{1.34cm}2MASS}
    \psfrag{p2}[][][0.9][0]{\hspace{1.13cm}WISE}
    \psfrag{p4}[][][0.9][0]{\hspace{1.02cm} PACS}
    \psfrag{p3}[][][0.9][0]{\hspace{2.36cm} IRS photometry}
    \psfrag{p5}[][][0.9][0]{\hspace{1.13cm} SPIRE}
    \psfrag{t1}[][][0.9][0]{\hspace{3cm}intrinsic AGN SED}
    \psfrag{t2}[][][0.9][0]{\hspace{1.47cm}SF SED}
    \psfrag{t3}[][][0.9][0]{\hspace{2.43cm}composite SED}
    \hspace{-.8cm}\includegraphics[trim=9cm 0cm 16cm 1cm, clip=true, width=0.6\textwidth]{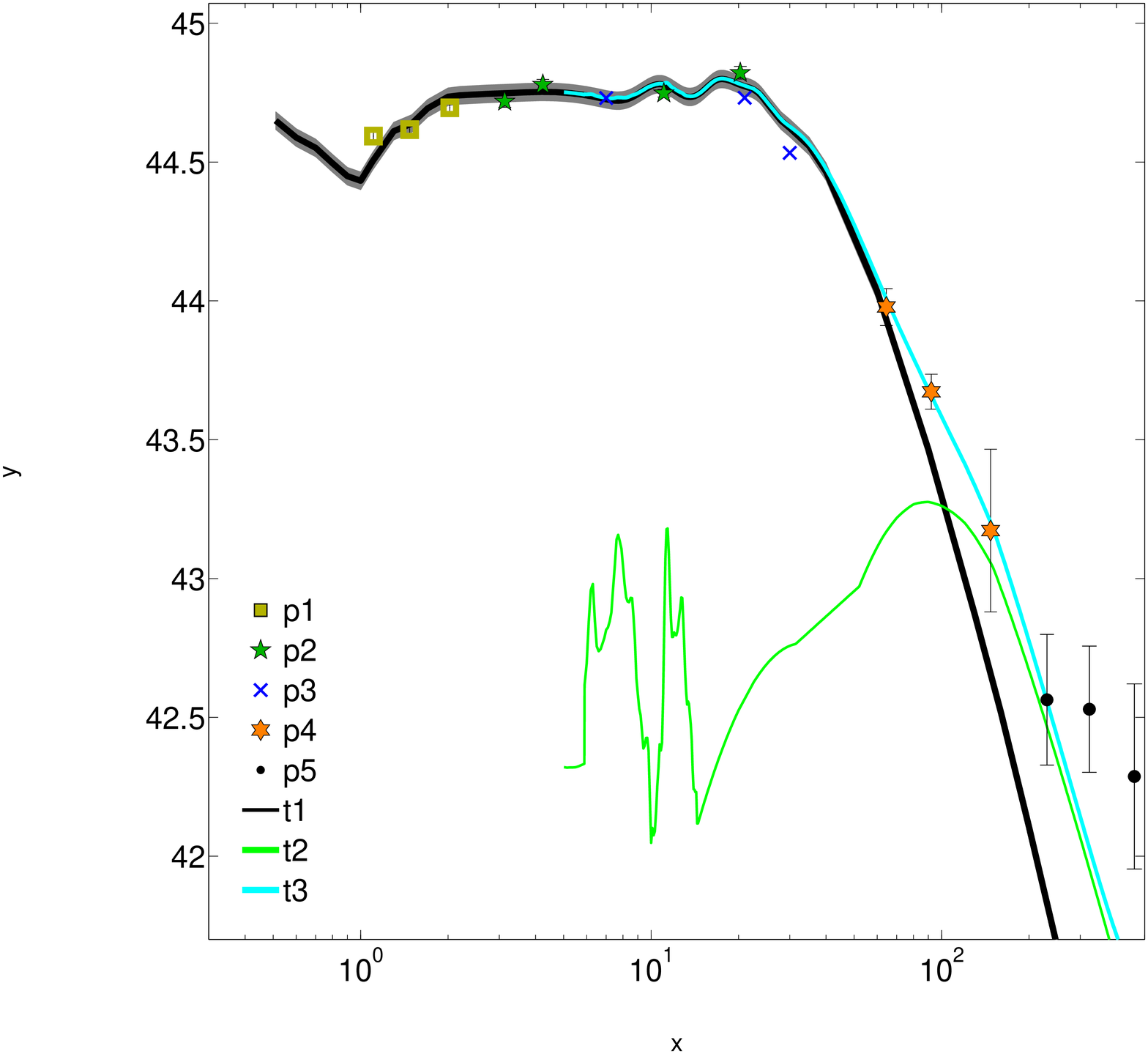}} \\
    \vspace{1cm}
    \end{tabular}
       \end{center}
    \end{minipage}
      \caption{Examples of SED fits; PG1322$+$659 (left panel) and PG1211$+$143
      (right panel). The point-like symbols refer to the same instruments as
      those in Figure \ref{fig: AGN_SEDs+points}, and the downward pointing triangles represent $3\sigma$ upper limits. 
      The black line corresponds
      to the best fitting EM12 template, and the grey shaded area
      corresponds to its uncertainty (see \S\ref{sec:SF assuming M12}). The green line
      represents the best fitting SF template from the CE01 set. The PG QSO
      in the left panel shows significant SF, while that in the right panel does not
      and most of its FIR emission is consistent with dusty torus$+$NLR emission.}
    \label{fig:SED_fit_examp}
    \end{figure*}

 In Figure \ref{fig:SED_fit_examp} we show two examples of SED fits,
    one where there is significant star formation (left panel) and one where the majority of the  FIR emission can be attribute to torus$+$NLR emission (right
    panel). We note that 4 sources (PG1022$+$519, PG1244$+$026, PG1351$+$236, and PG2304$+$042) have unsuccessful
    fits, due to peculiar NIR--MIR SED shapes. These were excluded from the remainder of our work.
       
    \subsection{Estimating star formation luminosities using PAH features}
    \label{sec: measuring SF using PAHs}
    As previously mentioned, this method relies on the existence of a widely-used
    relation between PAH luminosity and SF luminosity \citep[e.g.][]{Roussel2001,ForsterS2004,Wu2005},
    which remains uncertain due to the large intrinsic scatter and dependence on SF luminosity and metallicity \citep[e.g.][]{Calzetti2007,Smith2007}.
    \cite{Shi2007} measured PAH luminosities for the PG quasars and
    then determined the $L_{\rm SF}$ values later considered by S16 in their work, and that we consider here.

   \newpage 
      \begin{table*}
\caption{Quantities for our PG QSOs sample.}
\hspace*{-1cm}
\resizebox{19.5cm}{!}{
\begin{tabular}{cccccccccc}
\hline
$Name$  & $RA$ & $Dec$  & $z$ & $\log\, L_\text{5100}$\textsuperscript{(a)} 
& $\log\, M_\text{BH}$ & $\log L_\text{5$\mu$\rm m}$ & $\log L_\text{SF}$ &  location w.r.t. SFMS\textsuperscript{(b)} & choice of EM12 template\textsuperscript{(c)} \\
 & (deg) & (deg) &  & (erg$s^{-1}$) & (M$_{\odot}$) & (erg$s^{-1}$) & (erg$s^{-1}$) & & \\

\hline
PG0003+199\,\textsuperscript{$\epsilon$} & 1.5813 & 20.2029 & 0.025 & 43.748 & 7.222 & 43.914$\pm$0.019 & 43.213$\pm$0.1 & 0 & 3\\
PG0923+129  & 141.5137 & 12.7343 & 0.029 & 43.081 & 7.857 & 43.607$\pm$0.004 & 43.990$\pm$0.1 & 0 & 2\\
PG1534+580 & 233.9682 & 57.9026 & 0.030 & 43.404 & 7.706 & 43.518$\pm$0.092 & 43.208$\pm$0.1 & 0 & 1\\
PG1310-108 & 198.2741 & -11.1284 & 0.035 & 43.277 & 7.456 & 43.491$\pm$0.046 & 43.127$\pm$0.1 & 0 & 1\\
PG0921+525 & 141.3036 & 52.2863 & 0.035 & 43.098 & 6.979 & 43.729$\pm$0.022 & 43.216$\pm$0.1 & 0 & 3\\
PG1501+106 & 226.0050 & 10.4378 & 0.036 & 43.759 & 7.989 & 43.968$\pm$0.111 & 43.608$\pm$0.1 & 0 & 1\\
PG1535+547 & 234.1598 & 54.5592 & 0.038 & 43.580 & 6.874 & 43.784$\pm$0.093 & 43.064$\pm$0.1 & 0 & 3\\
PG2304+042\,\textsuperscript{$\beta$}\textsuperscript{$\alpha$} & 346.7621 & 4.5492 & 0.042 & 43.592 & 8.270 & 43.325$\pm$0.086 & $-$ & $-$ & --\\
PG1022+519\,\textsuperscript{$\alpha$} & 156.3803 & 51.6763 & 0.045 & 43.529 & 6.911 & 43.529$\pm$0.122 & $-$ & $-$ & --\\
PG1244+026\,\textsuperscript{$\alpha$} & 191.6469 & 2.3691 & 0.048 & 43.269 & 6.640 & 43.641$\pm$0.107 & $-$ & $-$ & --\\
PG1149-110 & 178.0147 & -11.3734 & 0.049 & 43.658 & 7.485 & 43.728$\pm$0.097 & 43.889$\pm$0.1 & 0 & 1\\
PG1119+120 & 170.4463 & 11.7384 & 0.049 & 43.876 & 7.281 & 44.033$\pm$0.084 & 44.061$\pm$0.1 & 1 & 1\\
PG0934+013 & 144.2543 & 1.0954 & 0.050 & 43.357 & 6.818 & 43.565$\pm$0.104 & 43.936$\pm$0.2 & 1 & 1\\
PG1351+236\,\textsuperscript{$\alpha$} & 208.5268 & 23.4303 & 0.055 & 43.681 & 8.090 & 43.663$\pm$0.051 & $-$ & $-$ & --\\
PG1011-040 & 153.5862 & -4.3112 & 0.058 & 43.632 & 7.095 & 43.852$\pm$0.047 & 43.895$\pm$0.1 & 1 & 1\\
PG1126-041 & 172.3194 & -4.4021 & 0.060 & 43.819 & 7.317 & 44.452$\pm$0.025 & 44.436$\pm$0.1 & 1 & 2\\
PG2130+099 & 323.1159 & 10.1387 & 0.061 & 44.262 & 7.684 & 44.569$\pm$0.003 & 44.316$\pm$0.1 & 1 & 2\\
PG0050+124 & 13.3956 & 12.6934 & 0.061 & 44.296 & 7.808 & 44.865$\pm$0.004 & 45.069$\pm$0.1 & 2 & 1\\
PG1229+204 & 188.0150 & 20.1581 & 0.064 & 44.134 & 7.830 & 44.091$\pm$0.000 & 43.923$\pm$0.1 & 0 & 1\\
PG0844+349 & 131.9269 & 34.7512 & 0.064 & 43.969 & 7.396 & 44.097$\pm$0.069 & 43.694$\pm$0.1 & 0 & 2\\
PG0049+171\,\textsuperscript{$\beta$} & 12.9782 & 17.4329 & 0.064 & 43.433 & 7.843 & 43.592$\pm$0.059 & 42.868$\pm$0.2 & 0 & 3\\
PG1448+273 & 222.7865 & 27.1575 & 0.065 & 43.986 & 6.687 & 43.988$\pm$0.051 & 43.934$\pm$0.1 & 1 & 1\\
PG2214+139 & 334.3011 & 14.2391 & 0.067 & 44.406 & 8.455 & 44.410$\pm$0.138 & 43.576$\pm$0.1 & 0 & 3\\
PG2209+184 & 332.9745 & 18.6972 & 0.070 & 43.703 & 8.164 & 43.867$\pm$0.169 & 43.839$\pm$0.1 & 0 & 3\\
PG1440+356 & 220.5311 & 35.4397 & 0.077 & 44.218 & 7.345 & 44.578$\pm$0.086 & 44.832$\pm$0.1 & 2 & 2\\
PG1211+143 & 183.5736 & 14.0536 & 0.085 & 44.479 & 7.663 & 44.750$\pm$0.032 & 43.474$\pm$0.1 & 0 & 2\\
PG1426+015 & 217.2774 & 1.2851 & 0.086 & 44.444 & 8.540 & 44.664$\pm$0.009 & 44.456$\pm$0.1 & 1 & 2\\
PG1351+640 & 208.3160 & 63.7627 & 0.087 & 44.393 & 8.551 & 44.738$\pm$0.099 & 44.708$\pm$0.1 & 1 & 1\\
PG1341+258 & 205.9864 & 25.6466 & 0.087 & 44.019 & 7.711 & 44.033$\pm$0.029 & 43.782$\pm$0.1 & 0 & 2\\
PG1411+442 & 213.4514 & 44.0039 & 0.089 & 44.311 & 7.867 & 44.709$\pm$0.017 & 44.028$\pm$0.1 & 0 & 3\\
PG0007+106 & 2.6292 & 10.9749 & 0.089 & 44.404 & 8.285 & 44.544$\pm$0.057 & 44.295$\pm$0.1 & 1 & 2\\
PG1404+226 & 211.5912 & 22.3962 & 0.098 & 44.165 & 7.071 & 44.104$\pm$0.040 & 43.971$\pm$0.1 & 1 & 3\\
PG0804+761 & 122.7442 & 76.0451 & 0.100 & 44.612 & 8.203 & 45.032$\pm$0.105 & 43.799$\pm$0.1 & 0 & 3\\
PG1617+175\,\textsuperscript{$\beta$}\textsuperscript{$\varphi$} & 245.0470 & 17.4077 & 0.114 & 44.283 & 8.477 & 44.643$\pm$0.419 & 42.000$\pm$0.3 & 0 & 3\\
PG1415+451 & 214.2534 & 44.9351 & 0.114 & 44.219 & 7.721 & 44.436$\pm$0.056 & 44.394$\pm$0.1 & 1 & 2\\
PG1552+085 & 238.6857 & 8.3726 & 0.119 & 44.432 & 7.573 & 44.357$\pm$0.071 & 43.670$\pm$0.1 & 0 & 3\\
PG1613+658 & 243.4882 & 65.7193 & 0.129 & 44.700 & 8.910 & 44.999$\pm$0.036 & 45.299$\pm$0.1 & 2 & 1\\
PG1435-067 & 219.5673 & -6.9724 & 0.129 & 44.388 & 8.118 & 44.502$\pm$0.070 & 43.586$\pm$0.1 & 0 & 3\\
PG1416-129 & 214.7659 & -13.1791 & 0.129 & 44.119 & 8.387 & 44.292$\pm$0.082 & 44.103$\pm$0.2 & 0 & 3\\
PG1612+261 & 243.5550 & 26.0712 & 0.131 & 44.425 & 7.806 & 44.602$\pm$0.114 & 44.694$\pm$0.1 & 1 & 1\\
PG0838+770 & 131.1888 & 76.8860 & 0.131 & 44.155 & 7.655 & 44.451$\pm$0.054 & 44.547$\pm$0.1 & 1 & 1\\
PG1626+554\,\textsuperscript{$\gamma$} & 246.9838 & 55.3754 & 0.133 & 44.449 & 8.327 & 44.317$\pm$0.243 & $-$ & $-$ & 3\\
PG1519+226 & 230.3094 & 22.4622 & 0.137 & 44.102 & 7.604 & 44.750$\pm$0.049 & 44.354$\pm$0.1 & 1 & 3\\
PG0026+129 & 7.3071 & 13.2678 & 0.142 & 44.665 & 7.809 & 44.678$\pm$0.077 & 43.725$\pm$0.1 & 0 & 3\\
PG1114+445\,\textsuperscript{$\varphi$} & 169.2767 & 44.2259 & 0.144 & 44.366 & 8.243 & 44.929$\pm$0.099 & 42.000$\pm$0.3 & 0 & 2\\
PG1115+407 & 169.6261 & 40.4317 & 0.154 & 44.208 & 7.672 & 44.676$\pm$0.005 & 44.931$\pm$0.2 & 2 & 3\\
PG1307+085 & 197.4458 & 8.3301 & 0.155 & 44.875 & 8.610 & 44.732$\pm$0.001 & 44.025$\pm$0.1 & 0 & 2\\
PG0052+251 & 13.7172 & 25.4275 & 0.155 & 44.782 & 8.581 & 44.858$\pm$0.052 & 44.558$\pm$0.2 & 1 & 3\\
PG1352+183\,\textsuperscript{$\beta$}\textsuperscript{$\varphi$} & 208.6487 & 18.0882 & 0.158 & 44.322 & 8.169 & 44.586$\pm$0.016 & 42.000$\pm$0.3 & 0 & 3\\
PG1001+054 & 151.0839 & 5.2168 & 0.161 & 44.253 & 7.653 & 44.703$\pm$0.028 & 44.073$\pm$0.1 & 0 & 3\\

\\
\cline{2-10}
\end{tabular}}
\label{tab:master table}
 \begin{tablenotes}
\item $(a)$ uncorrected for host galaxy contamination which, in our sample, we estimate to be typically $\sim20$ per cent.
 \item $(b)$ 0 refers to sources below the SFMS; 1 refers to sources on the SFMS; 2 refers to potential starbursts.
  For a description for how we determined the three groups we refer the reader to section \S\ref{sec: LSF vs LAGN}.
 \item $(c)$  1 refers to the 25$^{\rm th}$ percentile of the median EM12 template; 2 refers to the median EM12 template;
  3 refers to the 75$^{\rm th}$ of the median EM12 template. These have negative,
   flat and positive gradient in the optical-NIR (see grey and black lines in Figure \ref{fig: AGN_SEDs+points}), and are characteristic by $L_{\rm torus}/L_{5\mu\rm m}=[4.27,3.58,3.18]$ respectively.
 \item $\epsilon$ Source missing LL spectral information from $Spitzer$/IRS.
 \item $\beta$ Upper limit in one or two $Herschel$/PACS bands (70 and $160\mu\rm m$; with 3$\sigma$ being the requirement for a detection).
    \item  $\gamma$  Upper limit in all three $Herschel$/PACS bands (with 3$\sigma$ being the requirement for a detection).
    \item $\alpha$ Bad torus fit.
       \item $\delta$ Source with $Herschel$/PACS $160\mu\rm m$ observation affected by blending with a nearby source.
   \item $\varphi$ Source showing SF activity consistent with zero upon visual inspection; its $\log L_\text{SF}$ was arbitrarily set to $42.0\pm0.3$.
  
 \end{tablenotes}    
\end{table*}

\newpage
\addtocounter{table}{-1}
     \begin{table*}
\caption{Table \ref{tab:master table} continued.}
\hspace*{-1cm}
\resizebox{19.5cm}{!}{
\begin{tabular}{cccccccccc}
\hline

$Name$  & $RA$ & $Dec$  & $z$ & $\log\, L_\text{5100}$\textsuperscript{(a)} & $\log\, M_\text{BH}$ & $\log L_\text{5$\mu$\rm m}$ & $\log L_\text{SF}$
 &  location w.r.t. SFMS\textsuperscript{(b)} & choice of EM12 template\textsuperscript{(c)} \\
 & (deg) & (deg) &  & (erg$s^{-1}$) & (M$_{\odot}$) & (erg$s^{-1}$) & (erg$s^{-1}$) & & \\

\hline
PG1402+261 & 211.3176 & 25.9261 & 0.164 & 44.318 & 7.613 & 45.115$\pm$0.043 & 44.964$\pm$0.1 & 2 & 3\\
PG0157+001 & 29.9592 & 0.3946 & 0.164 & 44.671 & 8.631 & 45.223$\pm$0.210 & 46.043$\pm$0.1 & 2 & 1\\
PG1202+281 & 181.1755 & 27.9033 & 0.165 & 44.574 & 8.377 & 44.697$\pm$0.027 & 44.520$\pm$0.1 & 1 & 1\\
PG1048+342 & 162.9329 & 33.9907 & 0.167 & 44.064 & 7.823 & 44.413$\pm$0.010 & 44.174$\pm$0.2 & 1 & 2\\
PG1322+659 & 200.9563 & 65.6967 & 0.168 & 44.508 & 7.834 & 44.691$\pm$0.034 & 44.556$\pm$0.1 & 1 & 2\\
PG1151+117\,\textsuperscript{$\beta$} & 178.4553 & 11.4751 & 0.176 & 44.729 & 8.478 & 44.552$\pm$0.004 & 43.709$\pm$0.2 & 0 & 2\\
PG1116+215 & 169.7862 & 21.3217 & 0.177 & 45.126 & 8.557 & 45.294$\pm$0.121 & 44.301$\pm$0.1 & 0 & 3\\
PG1309+355 & 198.0740 & 35.2559 & 0.184 & 44.811 & 8.253 & 44.959$\pm$0.004 & 44.336$\pm$0.1 & 1 & 1\\
PG1012+008 & 153.7288 & 0.5604 & 0.185 & 44.631 & 8.346 & 44.830$\pm$0.003 & 44.631$\pm$0.1 & 1 & 2\\
PG0923+201\,\textsuperscript{$\varphi$}\textsuperscript{$\beta$} & 141.4780 & 19.9014 & 0.190 & 44.897 & 8.036 & 45.0779$\pm$0.177 & 42.000$\pm$0.3 & 0 & 3\\
PG0947+396\,\textsuperscript{$\delta$} & 147.7016 & 39.4474 & 0.206 & 44.657 & 8.437 & 44.925$\pm$0.011 & 44.502$\pm$0.2 & 1 & 2\\
PG1427+480 & 217.4295 & 47.7906 & 0.221 & 44.441 & 7.811 & 44.707$\pm$0.008 & 44.637$\pm$0.1 & 1 & 1\\
PG1121+422\,\textsuperscript{$\gamma$} & 171.1633 & 42.0292 & 0.234 & 44.975 & 8.162 & 44.709$\pm$0.115 & $-$ & $-$ & 3\\
PG0953+414 & 149.2183 & 41.2562 & 0.239 & 45.114 & 8.361 & 45.233$\pm$0.197 & 44.081$\pm$0.1 & 0 & 3\\
PG1004+130 & 151.8588 & 12.8156 & 0.240 & 45.232 & 8.864 & 45.097$\pm$0.033 & 44.759$\pm$0.1 & 1 & 1\\
PG1545+210\,\textsuperscript{$\beta$} & 236.9314 & 20.8713 & 0.266 & 45.074 & 8.878 & 45.110$\pm$0.043 & 44.283$\pm$0.2 & 0 & 3\\
PG1302-102 & 196.3876 & -10.5554 & 0.286 & 45.164 & 8.396 & 45.335$\pm$0.002 & 45.095$\pm$0.1 & 1 & 1\\
PG1700+518 & 255.3533 & 51.8222 & 0.292 & 45.671 & 8.753 & 45.785$\pm$0.023 & 45.723$\pm$0.1 & 2 & 2\\
PG1354+213\,\textsuperscript{$\beta$} & 209.1367 & 21.0646 & 0.300 & 45.202 & 8.714 & 44.983$\pm$0.027 & 44.322$\pm$0.2 & 0 & 2\\
PG1100+772 & 166.0570 & 76.9828 & 0.313 & 45.245 & 8.961 & 45.373$\pm$0.033 & 44.915$\pm$0.1 & 1 & 3\\
PG2251+113 & 343.5433 & 11.6106 & 0.323 & 45.625 & 8.885 & 45.448$\pm$0.107 & 44.594$\pm$0.1 & 0 & 3\\
PG2233+134 & 339.0320 & 13.7320 & 0.325 & 45.085 & 7.836 & 45.244$\pm$0.002 & 44.577$\pm$0.1 & 1 & 1\\
PG1216+069\,\textsuperscript{$\beta$}\textsuperscript{$\varphi$} & 184.8372 & 6.6440 & 0.334 & 45.844 & 9.195 & 45.257$\pm$0.099 & 42.000$\pm$0.3 & 0 & 3\\
PG1048-090\,\textsuperscript{$\gamma$}\textsuperscript{$\delta$} & 162.8747 & -9.3028 & 0.344 & 45.450 & 9.096 & 45.182$\pm$0.033 & $-$ & $-$ & 3\\
PG1049-006 ($\equiv$ PG1049$-$005) & 162.9643 & -0.8549 & 0.357 & 44.971 & 8.729 & 45.615$\pm$0.010 & 45.611$\pm$0.1 & 2 & 1\\
PG1425+267 & 216.8984 & 26.5374 & 0.366 & 45.453 & 9.257 & 45.314$\pm$0.011 & 44.891$\pm$0.1 & 1 & 1\\
PG1704+608 & 256.1724 & 60.7418 & 0.371 & 45.587 & 9.286 & 45.762$\pm$0.047 & 45.507$\pm$0.1 & 1 & 2\\
PG1512+370 & 228.6795 & 36.8473 & 0.371 & 45.065 & 8.872 & 45.263$\pm$0.018 & 44.708$\pm$0.1 & 1 & 2\\
PG0043+039 & 11.4467 & 4.1731 & 0.384 & 45.412 & 8.697 & 45.349$\pm$0.018 & 44.979$\pm$0.1 & 1 & 3\\
PG1543+489 & 236.3760 & 48.7692 & 0.400 & 45.269 & 8.108 & 45.696$\pm$0.054 & 45.837$\pm$0.1 & 2 & 1\\
PG1103-006\,\textsuperscript{$\beta$}\textsuperscript{$\varphi$} & 166.6324 & -0.8812 & 0.425 & 45.202 & 8.924 & 45.435$\pm$0.054 & 42.000$\pm$0.3 & 0 & 2\\
PG2308+098\,\textsuperscript{$\gamma$} & 347.8240 & 10.1376 & 0.432 & 45.389 & 9.237 & 45.455$\pm$0.040 & $-$ & $-$ & 3\\
PG0003+158\,\textsuperscript{$\beta$}\textsuperscript{$\varphi$} & 1.4968 & 16.1636 & 0.450 & 45.706 & 9.021 & 45.388$\pm$0.029 & 42.000$\pm$0.3 & 0 & 2\\
PG2112+059 & 318.7191 & 6.1285 & 0.466 & 45.710 & 8.931 & 45.978$\pm$0.093 & 45.423$\pm$0.1 & 1 & 3\\
PG1259+593\,\textsuperscript{$\beta$}\textsuperscript{$\varphi$} & 195.3039 & 59.0352 & 0.472 & 45.572 & 8.705 & 45.650$\pm$0.143 & 42.000$\pm$0.3 & 0 & 3\\

\\
\hline
\end{tabular}}

\end{table*}

    \begin{figure}     
       \psfrag{all}[][][0.9][0]{\hspace{4.7cm} All non non-upper limits in $L_{\rm SF, PAHs}$}
    \psfrag{t1s}[][][0.9][0]{\hspace{2.4cm}``Top $1\sigma$" sources }
        \psfrag{l1}[][][0.9][0]{\hspace{1.7cm} 1:1 relation}
 \psfrag{x2}[][][1.2][0]{\hspace{-0.5cm} $\log \left( L_{\rm SF,\,  PAHs}/ \rm erg\,s^{-1}\right)$}
    \psfrag{y2}[][][1.2][0]{$\log \left( L_{\rm SF,\,fit}/  \rm  erg\,s^{-1}\right)$}
    \includegraphics[scale=0.25,trim={18cm 0cm 12cm  0cm},clip]{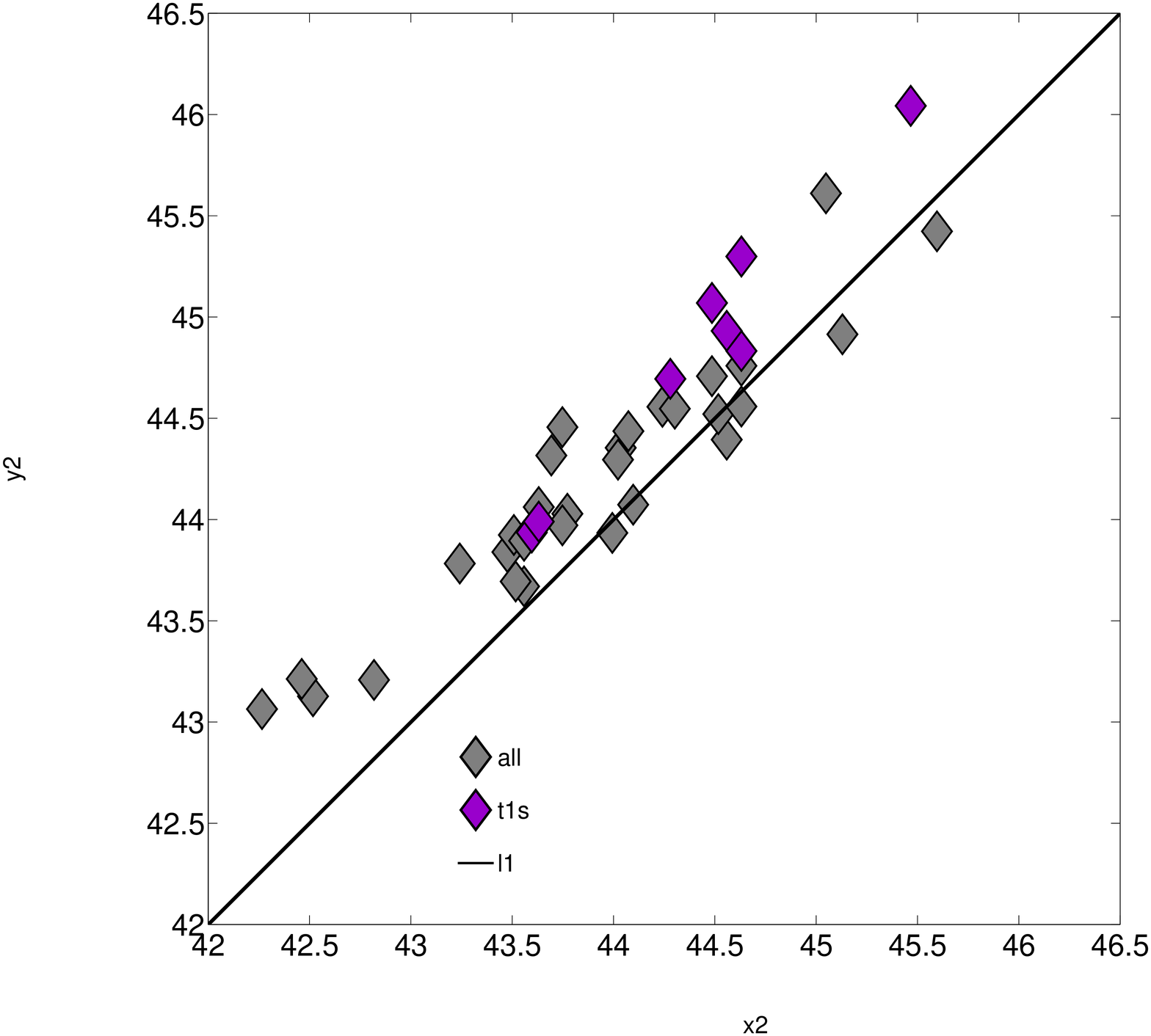} \\ 
                  \caption{Comparison of star formation luminosities as measured from aromatic features, along the x-axis, and from our SED fitting, along the y-axis, for PGs that are not upper limits in  $L_{\rm SF,\,PAHs}$ \citep{Shi2007}. The solid line is the 1:1 relation, while the purple points highlight those sources that we have defined as ``top $1\sigma$'' (\S\ref{sec: obtaining intrinsic AGN SED}). For most sources,  $L_{\rm SF,\,fit}$ is larger than the corresponding  $L_{\rm SF,\,  PAHs}$. It also appears that the majority of the ``top $1\sigma$''  sources lies in the region where $L_{\rm SF, fit}$ is $\sim3$ times $L_{\rm SF,\,  PAHs}$. }
         \label{fig: hist LSF_diff} 
      \end{figure}   
  
 Figure \ref{fig: hist LSF_diff} compares the star formation luminosities
    obtained from the two methods (SED fitting and aromatic features). For clarity we omit the PGs that, based on the work by \cite{Shi2007}, are upper limits in PAH-based SF luminosities. We find a median difference of
    $0.34\rm \,dex$ between the two sets of SF luminosities. As shown in the next section, this has important
    implications on the estimated intrinsic AGN SED shape.

    \subsection{The PAH-based intrinsic AGN SED}
    \label{sec: obtaining intrinsic AGN SED}
    The dataset considered here is superior to that considered in
    \cite{Mor2012}. First, all objects have high quality $Spitzer$/IRS spectra.
    Second, all objects have $Herschel$/PACS data, and third the sample is much
    better defined and hence represents better the population of luminous, local AGNs. As explained,
     though, the EM12 SED has been derived by arbitrarily adding a single temperature ($T=100\, \rm K$) 
     modified blackbody to the already-known SED at short wavelength. As we show below, this is found to be consistent 
     with the SED of many sources. The method, however, is limited in its ability to investigate the {\it range} 
     of intrinsic SEDs characterizing the entire population. For this reason one needs an independent way to estimate
      SF luminosities. Despite its limitations, the PAH-based method indeed constitutes such a tool, hence {\it we use it here keeping in mind the large associated uncertainties. }
  
    \begin{figure*}
     \psfrag{MN12}[][][0.9][0]{\hspace{0.2cm} EM12}
    \psfrag{GB}[][][0.9][0]{\hspace{1.3cm} S16 (mean)}
    \psfrag{TW}[][][0.9][0]{\hspace{2.17cm} This work, median}
    \psfrag{M12+GB}[][][0.9][0]{\hspace{2.64cm} Median of ``top $1\sigma$" sources}
    \psfrag{y}[][][1.4][0]{SF-subtracted, normalised $\rm \lambda\,L_{\rm \lambda}$}
    \psfrag{x}[][][1.4][0]{$\rm \lambda_{\rm rest}$ ($\mu m$)}
    \includegraphics[scale=0.3,trim={13cm 0cm 12cm  0cm},clip]{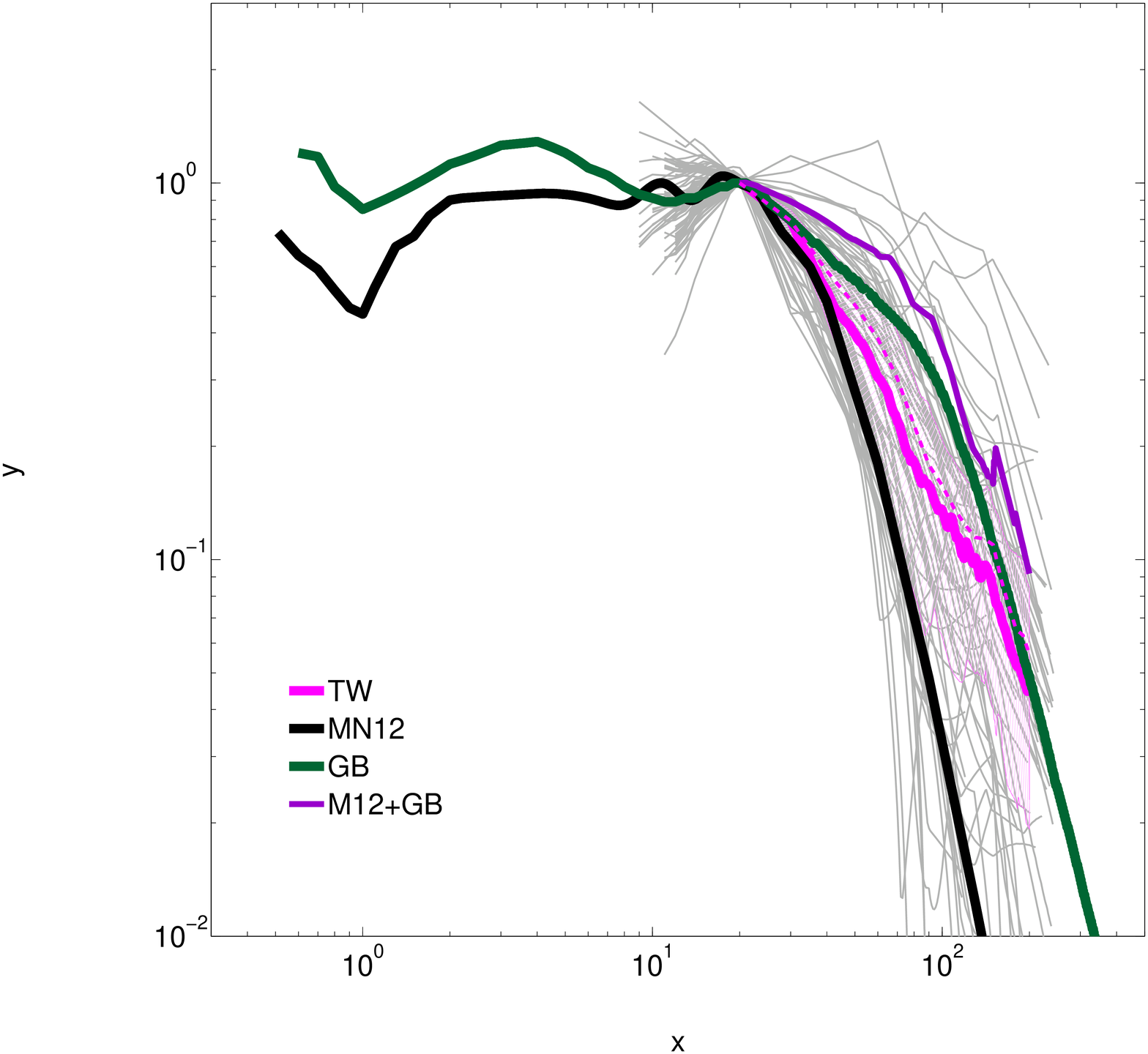} 
        \caption{Normalised SF-subtracted SEDs for our sample of PG QSOs (grey
        lines) when using the PAH-based method (updated where relevant as described in \S\ref{sec: obtaining intrinsic AGN SED}). The black and green
        lines correspond to EM12 and S16 intrinsic AGN SEDs, respectively. The
        solid magenta line is the median PAH-based intrinsic AGN SED obtained in the
        current work. The purple line is the intrinsic AGN SED obtained
        when considering those 12 sources that we defined as ``top $1\sigma$" in \S\ref{sec: obtaining intrinsic AGN SED}, and that we discuss in \S\ref{sec: revisiting AGN SED}. The dashed magenta line corresponds to
         the median intrinsic AGN SED we obtain when only considering PG quasars that are not upper limits in the PAH-based SF luminosities of \protect \cite{Shi2007}. }
        \label{fig: SF_subtracted_SEDs}
     \end{figure*}

  In this part of the analysis we consider the PAH-based SF luminosities provided by
    \cite{Shi2007} when these are not upper limits. If the \cite{Shi2007} estimates are upper limits we consider the $L_{\rm SF}$ values from our SED fitting as long as they are not larger. We followed this approach as we believe our $L_{\rm SF}$ to be the most reliable/accurate estimates available in these cases. For those 11 upper limits where our estimated SF luminosites are larger than those provided by  \cite{Shi2007} we follow the same procedure as S16, thus subtracting the SF templates corresponding to 1/2 of the \cite{Shi2007} upper limit.
    
    Our analysis of the PAH-based intrinsic AGN SED consists of a simple subtraction of a SF template from the overall
    SED, on an object by object basis. Specifically, for a given value of $L_{\rm SF}$, we find the template with the closest SF luminosity $L_{\rm SF,\,nearest}$ in the CE01 library. We then scale the chosen template by the ratio $L_{\rm SF}/L_{\rm SF,\,nearest}$. The result is then subtracted from the overall, observed AGN SED which has been previously interpolated between $5\mu\rm m$ and the longest wavelength where we have a flux detection.  
    
    There are four main 
    differences between the current work and that of S16.
    First, we use $Herschel$/PACS photometry in all three bands, while S16 used the less reliable $Spitzer$ and $AKARI$
    photometry in the wavelength range $\lambda=70-100\,\mu m$ (and $Herschel/PACS$ at $\lambda=160\,\mu m$). A direct comparison between their $Spitzer$ $70\,\mu m$ and our
    $Herschel$ $70\mu m$ photometries shows that the latter is higher for
    $\sim60$ percent of the sources. In addition,  several
    sources that are upper limits in the $Spitzer$ bands, are detections in the
    corresponding $Herschel$ bands. We emphasize that $Herschel$ provides much improved
    photometric accuracy, and better sampling of the $40-250\rm\, \mu m$ range,
    compared to $Spitzer$ and $AKARI$. Second, we
    calculate our mean and median SEDs in a different way compared to S16.
    We first subtract the SF emission, and then calculate the mean and median spectra after
normalising the individual
    intrinsic AGN SEDs at a chosen wavelength ($\lambda_{\rm rest}=20\mu \rm m$). Below we discuss the implications of this different procedure,
and in the Appendix we demonstrate how the
    choice of method leads to significantly different 
    SEDs. Third, we consider the entire 
    sample of PG quasars (cf. only those with $z<0.18$ in S16), making the results more applicable to the overall population.
     Lastly, the SF template we consider here are from CE01 while those
    used by S16 are the \citet[][hereafter DH02]{Dale2002} templates. To quantify the
    impact of this difference on our results, we repeated our analysis with the
    DH02 SF templates. We recovered a consistent result (shown in the Appendix) as when we use the CE01 library, thus the
    choice of SF templates does not significantly affect the findings discussed below.
    When following the subtraction procedure outlined above, for 11 sources we are left with SF-subtracted SEDs that are negative in
     any of the $Herschel$/PACS bands. For these we repeat the SF subtraction after replacing their previously considered $L_{\rm SF}$ 
     with the lower limit ($L_{\rm SF}-\delta L_{\rm SF}$ ), on this luminosity, taken from the relevant work (ours/S16).
    
 In Figure \ref{fig: SF_subtracted_SEDs} we show the individual
    SF-subtracted SEDs for our sample of PG QSOs (grey lines), normalised to $20\rm \mu m$, and the resulting
    median intrinsic AGN SED (solid magenta line). It is apparent
    that, when using the PAHs-based SF luminosities, there are many SEDs that are considerably
    colder than the EM12 SEDs, as is the resulting {\it median}. This (which we take to be the best representation of the population in this exercise) is also considerably below the 
    {\it mean} SED presented in S16 (green line). This has important consequences for the 
    characterization of the entire population, and the suggestion that (SF-unrelated) dust is present at large distances
from the BH. Quantitative estimates for the contribution of
such large-scale dust to the mean FIR luminosity are given in \S~\ref{sec:non-torus covering factor etc.}. In Figure \ref{fig: SF_subtracted_SEDs} we also show 
the median intrinsic AGN SED (dashed magenta line) we would obtained if we only considered
 those PG quasars that are not upper limits in the PAH-based SF luminosities of \cite{Shi2007}.
 This line indicates that even when we consider those PG quasars which have the more robust SF luminosities we obtain a median intrinsic AGN SED that is consistent with that of the full sample.

    We then considered those 12 sources that represent the top 16 per cent with the largest difference between their normalised SF-subtracted $70 \mu\rm m$ luminosity and that of the PAH-based median intrinsic AGN SED (``top $1\sigma$" sources hereafter). Their PAH-based median intrinsic AGN SED is given by the purple line, and is significantly different from all the other SEDs shown in this figure. \\

\begin{figure*}
    \begin{minipage}[c]{\textwidth}
    \begin{center}
    \begin{tabular}{c c c}
    {\psfrag{y}[][][1][0]{$\rm \lambda\,L_{\rm 70\mu m}/ \rm \lambda\,L_{\rm 12\mu m}$}
    \psfrag{x}[][][1][0]{$\rm L_{\rm SF,\,fit}$(erg/s)}
    \psfrag{blah1}[][][0.9][0]{sample's range in optical-luminosity}
    \hspace{-1.5cm}\includegraphics[trim=11cm 0cm 16cm 1cm, clip=true, width=0.33\textwidth]{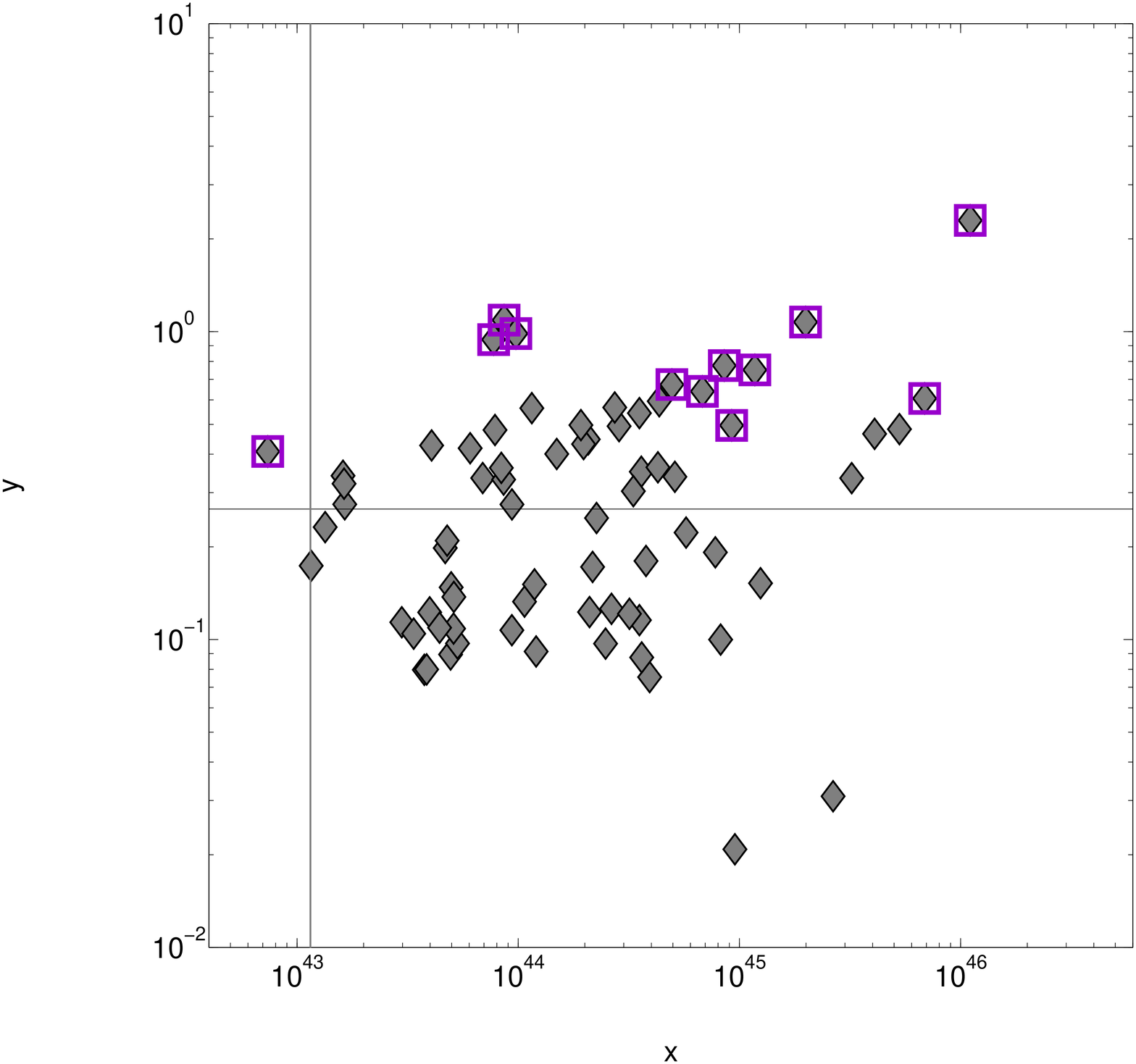}} 
     &
    {\psfrag{y}[][][1][0]{$\rm \lambda\,L_{\rm 70\mu m}/ \rm \lambda\,L_{\rm 12\mu m}$}
    \psfrag{x}[][][1][0]{$\rm L_{\rm SF,\,PAHs}$(erg/s)}
        \psfrag{y}[][][1][0]{$\rm \lambda\,L_{\rm 70\mu m}/ \rm \lambda\,L_{\rm 12\mu m}$}

    \includegraphics[trim=11cm 0cm 16cm 1cm, clip=true,
    width=0.33\textwidth]{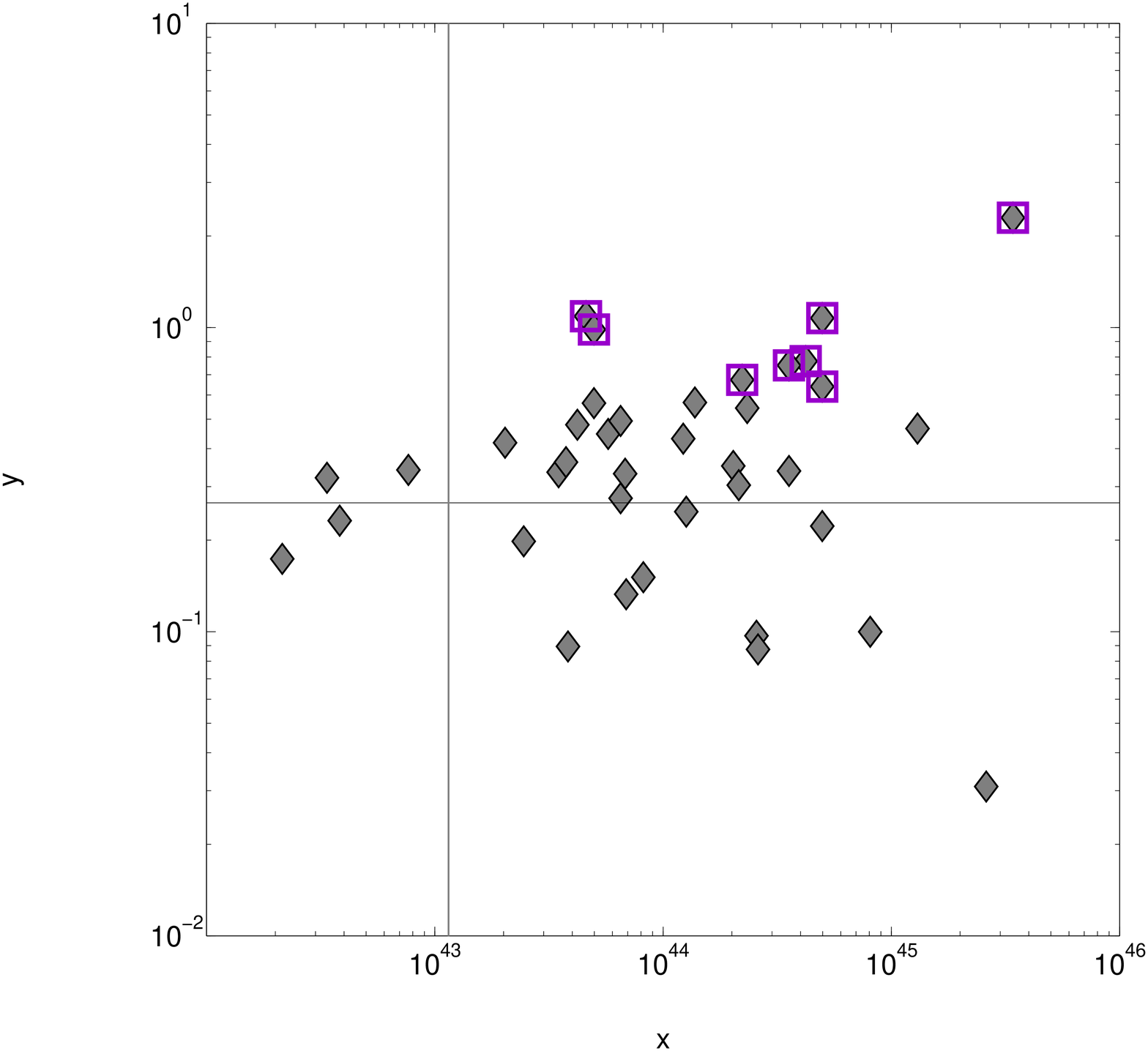}}
    &
    {\psfrag{x}[][][1][0]{$\rm \lambda\,L_{\rm 5\mu m}$ (erg/s)}
   \psfrag{y}[][][1][0]{$\rm \lambda\,L_{\rm 70\mu m}/ \rm \lambda\,L_{\rm 12\mu m}$}
    \psfrag{blah3}[][][0.9][0]{sample's range in black hole mass}
    \includegraphics[trim=11cm 0cm 16cm 0cm, clip=true,
    width=0.33\textwidth]{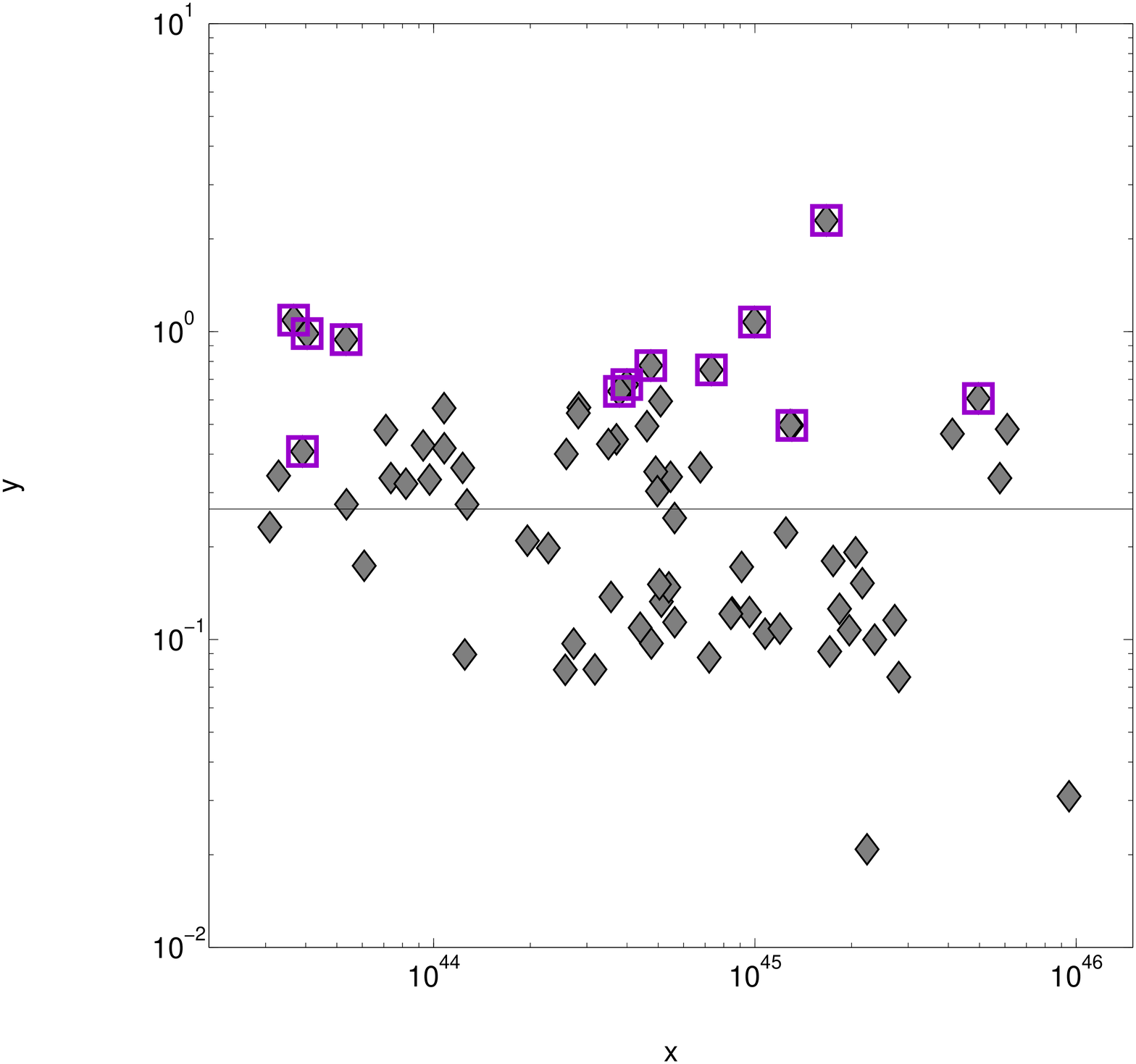}} \\ 
     \end{tabular}
       \end{center}
    \end{minipage}
      \caption{Ratio of the SF-subtracted luminosity measured at $70\rm \mu m$
      and that measured at $12\rm \mu m$ for our PG QSOs, as a function of SF and
      AGN luminosities. The purple squares highlight the ``top $1\sigma$'' sources. The left panel considers the SF estimated from our SED
      fitting that assumes the EM12 SEDs, while the middle panel that from
      \protect\cite{Shi2007}, based on PAHs, for non-upper limits only and 
      updated, if required, as previously mentioned (\S\ref{sec: obtaining intrinsic AGN SED}). No significant correlation 
      appears between the shape of the intrinsic AGN SED of the full population, and the three quantities considered. The ``top $1\sigma$" sources, however, mostly appear at the high end of the distributions. In the left panel the vertical line indicates the SF luminosity corresponding to star formation rate $SFR=0.1\, \rm M_{\odot}\, yr^{-1}$.} 
    \label{fig: ratios_LSF_L5um}
    \end{figure*}

 We also investigated whether the different SED shapes 
    derived with PAH-based luminosities were driven by SF or torus luminosity, and the results are
    summarized in Figure \ref{fig: ratios_LSF_L5um}. Here we show the ratio
    between the SF-subtracted luminosity measured at $70\mu m$ and that measured
    at $12\mu \rm m$ (i.e. a proxy for shape in this region of the SED) as a
    function of SF luminosity from our SED fitting (left-hand panel), SF
    luminosity from PAHs (middle panel), and torus
    luminosity (right panel). It is clear that in all three panels there is no
    correlation for the general population, and we can therefore conclude that the differences in shape do
    not depend on  $L_{\rm SF}$ and  $L_{\rm AGN}$ ($\propto L_{\rm 5\mu m}$). On the other hand, if we focus on the ``top $1\sigma$" sources, highlighted by purple squares, they mostly appear at the high end of the $L_{\rm SF}$ and $L_{\rm 5\mu m}$ distributions. \\

      We note that for the results discussed in \S\ref{sec:torus covering factor} and \S \ref{sec:SFR BHAR} we assume the EM12 template, as explained below and in the relevant sections.

    \section{Discussion}
    \label{sec:results}
Thus far we have compared and contrasted the results obtained when employing different methods to estimate the SF activity and the typical intrinsic AGN SED in low redshift AGNs. We found that the consideration of aromatic features yields star formation luminosities that are consistently smaller than those obtained from MIR-FIR SED fitting. Furthermore, we found that the methodology followed to constrain the typical FIR shape of the intrinsic AGN SED in PG quasars strongly impacts on the outcome. 

In this section we discuss and interpret our findings. We revisit the issue of ``intrinsic AGN SED'' 
and we explore the properties of the cooler dust, referred to here as ``large-scale AGN-heated dust"
 (see \S\ref{sec:non-torus covering factor etc.} on the possibility of this dust being part of the torus/being SF-related), 
 which were identified when considering PAH-based SF luminosities. Furthermore we also want to investigate the torus covering factor and the instantaneous
relative growth rate of BHs and their host galaxies, when assuming the EM12 SED to describe the torus and the NLR emission in the considered quasars. Part of the reason why we consider the EM12 SED, rather than that obtained in \S\ref{sec: obtaining intrinsic AGN SED}, is that we believe it to be more robust because determined on a larger sample. Our choice, however, is further motivated in the remainder of this section.

\subsection{Comparison of the PAH-based and SED-fitting-based SF luminosities}
\label{sec: L_SFus vs L_SFshi}
    The SF estimates from SED fitting, assuming the EM12 SED,
    are systematically larger than those obtained from the PAH-based method (see Figure \ref{fig: hist LSF_diff}). It
    is, however, hard to conclusively determine which method is more reliable.
    On the one hand, fitting SF templates using {\it Herschel} data is far
    superior in terms of data quality, and does not suffer from the
    uncertainties due to fitting the PAH spectrum and the large, intrinsic,
    luminosity-dependent scatter due to the conversion from PAH luminosity to SF
    luminosity. On the other hand, the PAH-based method is independent of the assumed torus SED. 

Finally we compared our SED fitting-based SF luminosities with those calculated by \citet[][their Table 4]{Petric2015}. Similarly to the current study, they also considered the full sample of PG quasars and estimated their SFR from FIR $Herschel$ photometry.
We find that, when considering their $L_{\rm FIR\, \left[40-500\,\mu \rm m\right] }$, they are consistent with a typical difference of $0.07\,\rm dex$. We note that \cite{Petric2015} also finds SF luminosities that are approximately a factor of 2 larger than those quoted in \cite{Shi2007}.

    \subsection{Re-visiting the intrinsic AGN SED}
    \label{sec: revisiting AGN SED}
    \subsubsection{Observational constrains on the intrinsic AGN SED}
    \label{sec: obs constrain on AGN SED}
\begin{figure*}
    \begin{minipage}[c]{\textwidth}
    \begin{center}
    \begin{tabular}{c c}
    {\psfrag{y}[][][1.5][0]{$\rm \log\left( \lambda\,L_{\rm \lambda}/
    erg\,s^{-1}\right) $}
  \psfrag{p1}[][][0.9][0]{\hspace{3.3cm}PACS $3\sigma$ upper limits}
    \psfrag{p2}[][][0.9][0]{\hspace{4.45cm}SPIRE$-250\,\rm \mu m$ $3\sigma$ upper limits}
    \psfrag{p3}[][][0.9][0]{\hspace{1.12cm}EM12}
    \psfrag{p4}[][][0.9][0]{\hspace{0.83cm}S16}
        \psfrag{p5}[][][0.9][0]{\hspace{1.66cm}This work}
         \psfrag{y}[][][1.4][0]{normalised $\rm \lambda\,L_{\rm\lambda}$}
    \psfrag{x}[][][1.4][0]{$\rm \lambda_{\rm rest}$ ($\mu m$)}
    \psfrag{PG quasars}[][][0.7][0]{upper limits for PG quasars}
  \hspace{-2.0cm}\includegraphics[trim=9cm 0cm 16cm 1cm, clip=true,
    width=0.6\textwidth]{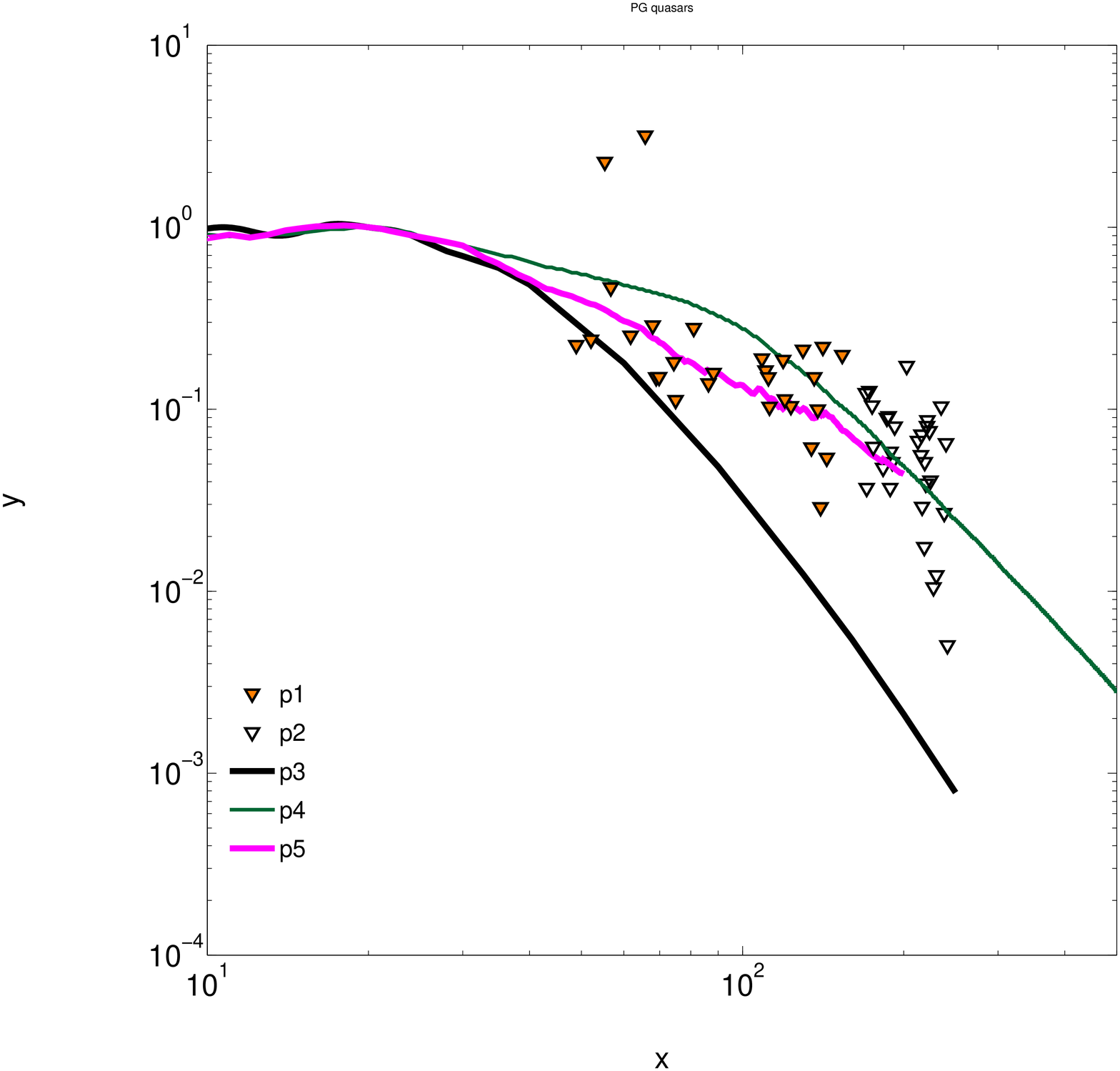}} & 
  {   \psfrag{p1}[][][0.9][0]{\hspace{4.45cm}SPIRE$-250\,\rm \mu m$ $3\sigma$ upper limits}
    \psfrag{p2}[][][0.9][0]{\hspace{5.39cm}median stacks of SPIRE non-detections}
    \psfrag{t1}[][][0.9][0]{\hspace{1.15cm}EM12}
    \psfrag{t2}[][][0.9][0]{\hspace{0.83cm}S16}
        \psfrag{t3}[][][0.9][0]{\hspace{1.67cm}This work}
         \psfrag{y}[][][1.4][0]{normalised $\rm \lambda\,L_{\rm\lambda}$}
    \psfrag{x}[][][1.4][0]{$\rm \lambda_{\rm rest}$ ($\mu m$)}
     \psfrag{high z sample}[][][0.7][0]{upper limits/median stacks for $z=2-3.5$ quasars \citep{Netzer2016}}
    \hspace{-.8cm}\includegraphics[trim=9cm 0cm 16cm 1cm, clip=true,
    width=0.6\textwidth]{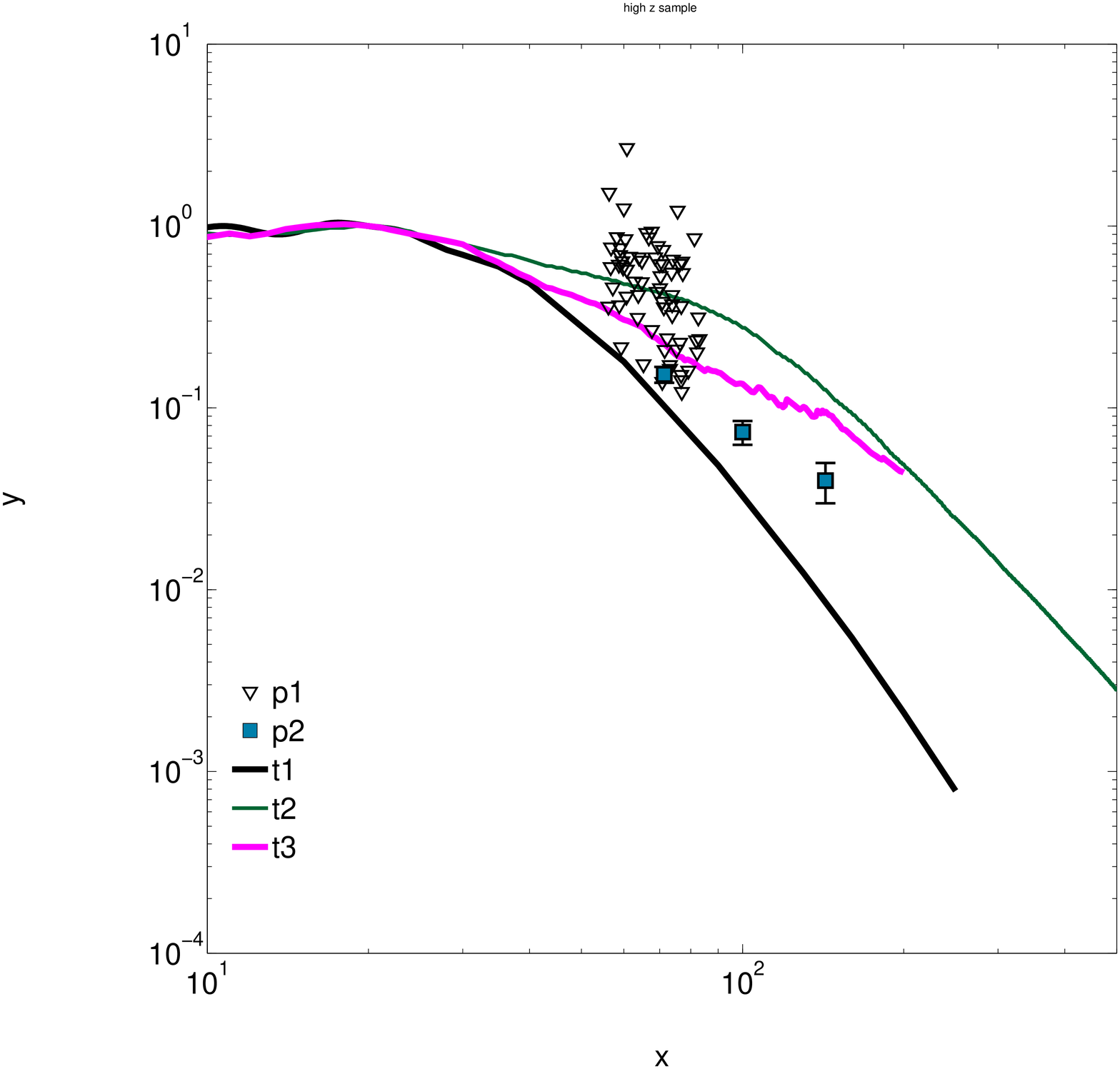}} \\ 
    \vspace{1cm}
    \end{tabular}
       \end{center}
    \end{minipage}
      \caption{{\bf Left panel.} Zoom-in on the long wavelength region ($\lambda_{\rm rest}>10\mu\rm m$) of the 
      3 intrinsic AGN SEDs considered in this work. The triangles represents $Herschel$/PACS and SPIRE $3\sigma$ upper limits for our PG QSOs. 
     Approximately 80 per cent of the $Herschel$/PACS upper limits lie below the S16 (mean) intrinsic torus SED
     (cf. 50 per cent below the median intrinsic AGN SED identified in the current work). {\bf Right panel.} The lines have the same 
      meaning as the adjacent panel, however here the triangles represent  $Herschel$/SPIRE $250\mu \rm m$ $3\sigma$ upper limits of the
       $z=2-3.5$ \protect\cite{Netzer2016} sample. The squares are median stacks of the $Herschel$/SPIRE non-detection for the
        same sample. In this case $\sim40$ per cent of the triangles lie below the S16 SED and, the latter is significantly above the stacked points.}
    \label{fig: ULs comparison}
    \end{figure*}

 Figure \ref{fig: ULs comparison} shows the comparison of the EM12
    template (median), the S16 {\it mean} intrinsic AGN template and the {\it median} intrinsic AGN SED identified in this work,
    to the $Herschel$ $3\sigma$ upper limits from the current work (left panel) and those presented in Figure 7
in \citet[][right panel]{Netzer2016}. Here we only consider FIR upper limits as these are the best candidates for cases where the FIR emission is least contaminated by SF activity, thus being AGN-dominated. As previously mentioned, among the PG QSOs, there are 4 upper limits in all three $Herschel$/PACS bands ($\sim5$ per cent of the total), and 12 upper limits ($\sim14$ per cent of the total) in one or two $Herschel$/PACS bands (see Table \ref{tab:master table}). The EM12 template is consistent with all $3\sigma$ upper limits, both at high and low redshift. In contrast $\sim80$ per cent of the $Herschel$/PACS $3\sigma$ upper limits in the PG sample, and $\sim40$ per cent of the $Herschel$/SPIRE $3\sigma$ upper limits from \cite{Netzer2016},
 fall below the mean S16 SED. For the PG sample, we would expect a smaller fraction of the upper limits to fall below the S16 mean, 
if this SED indeed represented the mean population properties. This is most noticeable in the range $\lambda_{\rm rest}=70-120 \mu \rm m$,
 where nearly all $3\sigma$ upper limits fall below the S16 line. This contradiction is even more apparent in the high-redshift sample since, if the
S16 mean SED were representative of this population, all sources below the line would instead be within the $Herschel$ detection limit, appearing as real
$Herschel$ detections rather than upper limits.

Moreover, the median stacks obtained from the $250 \mu \rm m$ upper limits in \citet[][66 sources]{Netzer2016}
that are shown in the diagram, mark the locations that divide the population in equal parts (i.e. 50 per cent of
the luminosities in these undetected sources should lie below these stacks). This is in clear contradiction with the assumption that the mean intrinsic AGN SED
 proposed by S16 represents this population since, in such a case, the calculated stacks (median or mean) would
lie close to the S16 SED rather than a factor of $\sim 3$ below.
On the other hand, the median PAH-based intrinsic SED calculated for the PG sample in the current work lie much 
closer to the high redshift median stacks, as should be the case if the ``typical'' intrinsic SED shape is independent
of AGN luminosity. 

 All the above considerations suggest that there is no simple way to distinguish between EM12 SED and the median, 
 PAH-based SED obtained here. Conversely, even allowing for the intrinsic scatter expected around an averaged quantity,
  the mean S16 SED seems to be in conflict with many observations, most noticeable involving $Herschel$ upper limits. 
  As shown in the Appendix, we believe that the mean S16 SED represents a few more luminous sources rather than the entire population.

\subsubsection{Large-scale dust and its impact on the estimate of SF luminosities}
\label{sec: large-scale dust and its impact on the estimate of SF luminosities}
\begin{figure*}
    \begin{minipage}[c]{\textwidth}
    \begin{center}
    \begin{tabular}{c c}
    {\psfrag{MN12}[][][0.9][0]{\hspace{0.5cm}EM12}
    \psfrag{TW}[][][0.9][0]{\hspace{1.3cm}This work}
    \psfrag{GB}[][][0.9][0]{\hspace{2.1cm} mod. blackbody}
    \psfrag{M12+GB}[][][0.9][0]{\hspace{2.46cm}EM12$+$mod. blackbody}
    \psfrag{y}[][][1.5][0]{normalised $\rm \lambda\,L_{\rm \lambda}$}
    \psfrag{x}[][][1.5][0]{$\rm \lambda_{\rm rest}$ ($\mu m$)}
    \psfrag{y}[][][1.5][0]{normalised $\rm \lambda\,L_{\rm \lambda}$}
    \psfrag{x}[][][1.5][0]{$\rm \lambda_{\rm rest}$ ($\mu m$)}
    \hspace{-2.5cm}\includegraphics[trim=9cm 0cm 16cm 1cm, clip=true,
    width=0.6\textwidth]{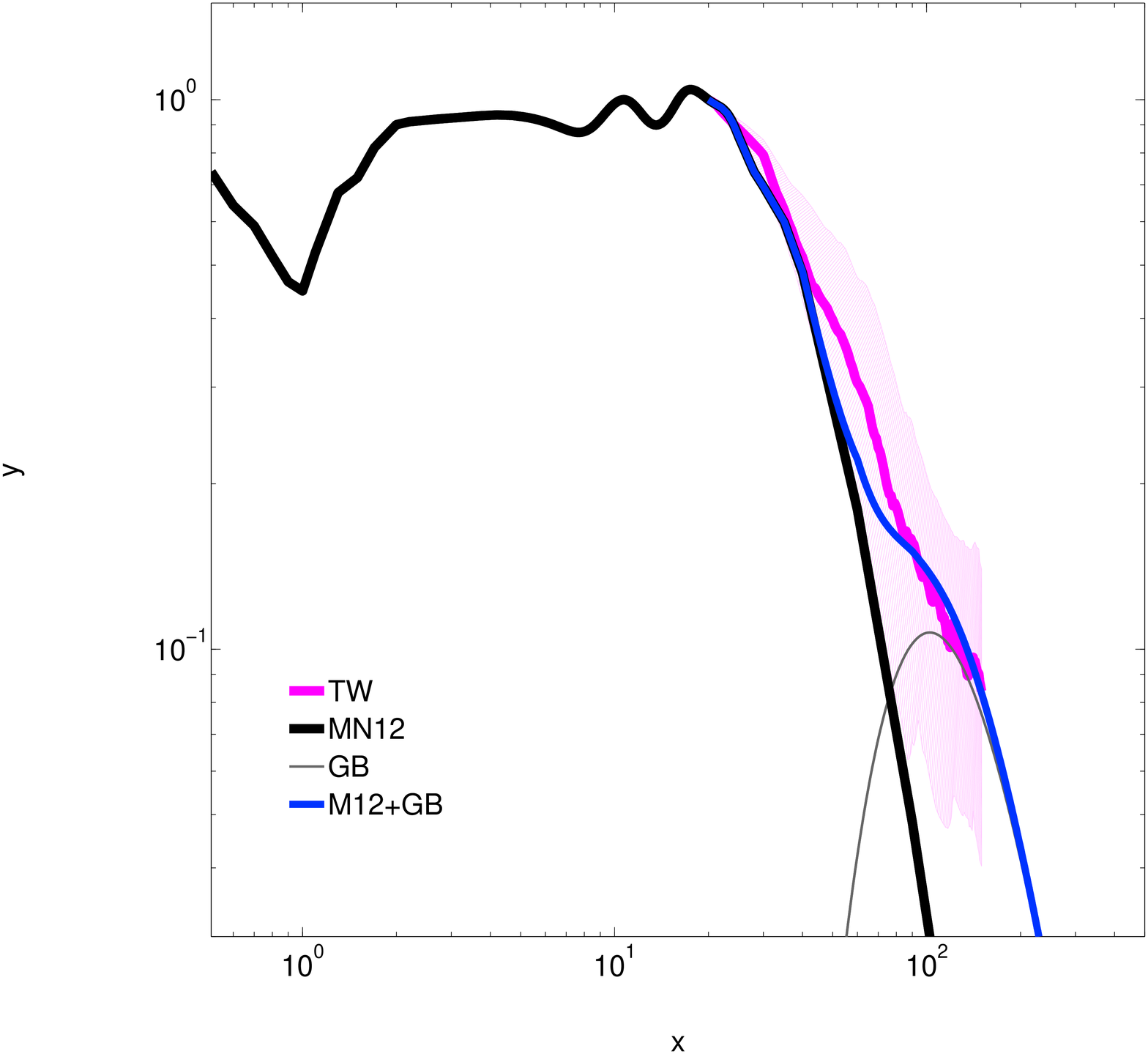}} & 
    {\psfrag{MN12}[][][0.9][0]{\hspace{0.5cm}EM12}
    \psfrag{TW}[][][0.9][0]{\hspace{4.22cm}This work, ``top $1\sigma$" AGN SEDs}
    \psfrag{GB}[][][0.9][0]{\hspace{3.5cm} sum of 2 mod. blackbodies}
    \psfrag{M12+GB}[][][0.9][0]{\hspace{2.65cm}EM12$+$mod. blackbodies}
    \psfrag{y}[][][1.5][0]{normalised $\rm \lambda\,L_{\rm \lambda}$}
    \psfrag{x}[][][1.5][0]{$\rm \lambda_{\rm rest}$ ($\mu m$)}
    \psfrag{y}[][][1.5][0]{normalised $\rm \lambda\,L_{\rm \lambda}$}
    \psfrag{x}[][][1.5][0]{$\rm \lambda_{\rm rest}$ ($\mu m$)}
   \hspace{-0.5cm} \includegraphics[trim=10cm 0cm 16cm 1cm, clip=true,
    width=0.61\textwidth]{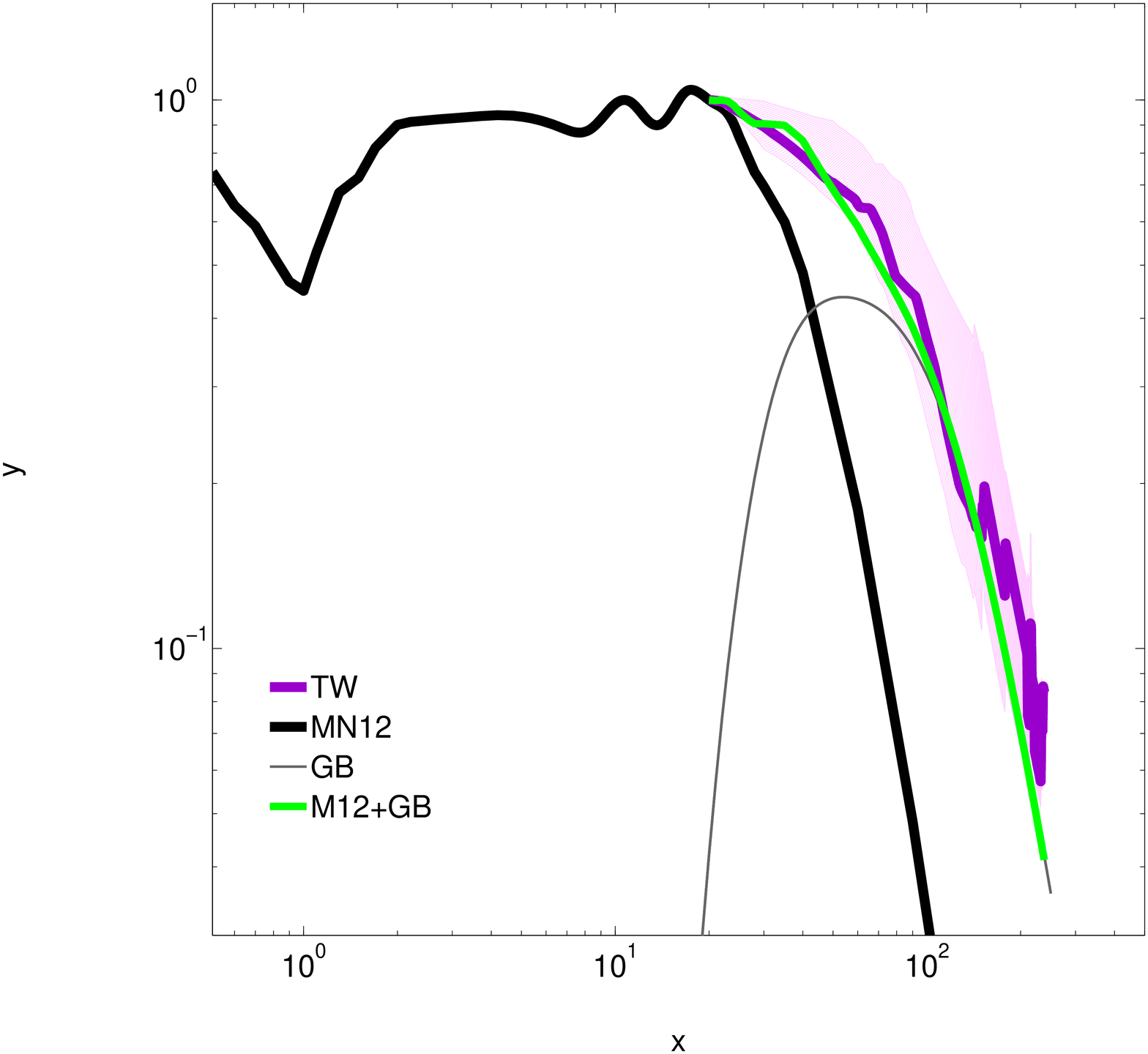}} 
    \\
    \vspace{1cm}
    \end{tabular}
       \end{center}
    \end{minipage}
    \caption{{\bf Left panel.} In magenta we show the median intrinsic AGN SED as
    found in this work (with pink hatching representing the $25^{\rm
    th}$ and $75^{\rm th}$ percentiles) when considering our PG sample. In blue we show, for illustration only, the composite SED obtained by summing a modified blackbody with temperature of $T_{\rm
    cold\,\,dust}=25.5\rm\, K$ and $\beta=1.5$ (grey line) to the EM12 median intrinsic AGN
    SED (black line). {\bf Right panel.} The black line has the same meaning as
    in the adjacent panel, while the purple line is the median intrinsic AGN
    SED obtained when only considering the 12 ``top $1\sigma$"
    sources (see \S\ref{sec: obtaining intrinsic AGN SED}). In this case one modified blackbody ($\beta=1.5$) does 
    not provide a satisfactory fit, and here we show the sum of two modified blackbodies (grey line) with temperatures $T=29\rm\, K$ 
    and $T=60\rm\, K$ {\it for illustration}.}
    \label{fig: GB fits}
    \end{figure*}

 Figure \ref{fig: SF_subtracted_SEDs} also unveils that some SEDs obtained with the PAH-based method
lie far above the median and the mean SEDs of the sample. As previously mentioned, the purple
    line represents the median intrinsic AGN SED obtained when considering ``top $1\sigma$" sources (as defined in \S\ref{sec: obtaining intrinsic AGN SED}).
      We also show this median SED in the right panel of Figure \ref{fig: GB fits}. {\it If} the PAH-based method provides an accurate way to derive the intrinsic
AGN-heated dust emission, these sources may have the strongest contribution from
    the potential large-scale dust component. {\it Alternatively}, these sources could be
those with the least reliable PAH-based SF estimates. Figure \ref{fig: hist LSF_diff} shows that these sources (marked in purple) 
clearly lie where the discrepancy in the SF measurement between 
the two methods is large. 
We cannot, however, say whether this deviation from the median is due to the subtraction of an erroneous amount of PAH-determined SF,
 or because of extra contribution by large-scale AGN-heated dust which invalidates the consideration of the EM12 template.\\
    \\
   \indent The left panel of Figure \ref{fig: GB fits} shows the PAH-based median intrinsic AGN
    SED, and its $25^{\rm
    th}-75^{\rm th}$ percentiles range (pink hatching). The dark blue line in this figure was obtained by summing
    the median EM12 SED to a modified blackbody ($\beta=1.5$) with a temperature of $T_{\rm
    cold\,\,dust}=25.5\,\rm K$. While this provides the best single temperature fit (through $\chi^{2}$ minimisation), we caution that it should not be considered as the ultimate value given the large associated uncertainties. In fact, the excess over the EM12 SED can be fitted, reasonably well, with a range of temperatures, from $20\,\rm K$ to about $30\,\rm K$.
    In the right panel of Figure \ref{fig: GB fits} we show, in purple, the median intrinsic AGN SED obtained when considering the
     ``top $1\sigma$" sources, with the corresponding
  $25^{\rm th}-75^{\rm th}$ percentiles (pink hatching). In this case, there is no way to fit the difference between 
  this SED and EM12 with a single modified blackbody. Upon experimenting we find that the required range of temperatures is between $10\,\rm K$ 
  and $60\,\rm K$. Given the large uncertainties associated with the PAH-based method, we choose not to attempt a full 
  fit of this difference, and only to measure the differences in the total emitted radiation (see \S\ref{sec:non-torus covering factor etc.}). 
  We also repeated this analysis replacing the set of modified blackbodies with the CE01 SF templates. 
  We found that we can fit the residual emission with a single SF template, both for the SED based on the full sample and that based on the ``top $1\sigma$" sources. 
  Although the resulting additional SF is within the uncertainties accompanying the best fitting SF luminosities,
   it may be indicative of star formation that was not properly captured by the PAH-based estimates (and consequently by the subtraction).

\vspace{0.3cm} Finally, we estimated the uncertainties in the measured SF luminosities introduced by the various intrinsic AGN SEDs, as a function of AGN luminosity. These numbers can also be used to estimate the contribution of cold, large-scale AGN dust, if real, to the total SF luminosity. The natural way to do such a test is to fix $L_{\rm SF}$\ and then compute the
  ratio of SF luminosity and AGN-heated dust luminosity at a pre-chosen wavelength, for every considered value of $L_{\rm torus}$. Here 
  we chose this fiducial wavelength to be $70\mu\rm m$ which is close to the peak in $\rm \nu L_{\nu}$ of most SF templates
   used in this work. 
   We examined two cases, the EM12 SED and the new PAH-based median intrinsic AGN SED obtained in this work.
    For illustration, we fix the SFR to be $10\,\rm M_{\odot}/yr$ and show the results for various ratios of
$L_{\rm torus}/L_{SF}$ in Table \ref{tab:SF_uncertainty}. We note that the consideration of different SF SEDs,
 other than that arbitrarily picked here, would yield different numbers. The differences, however, are minor and would
  not impact on the main conclusions of this test. The uncertainty on $L_{\rm SF}$ listed in the table are calculated
   assuming no subtraction of the AGN contribution at $70\mu \rm m$. As shown in the table, for the EM12 SED, significant uncertainties, approaching a factor 2, require 
 $L_{\rm torus}/L_{SF}>20$. For the new PAH-based median intrinsic AGN SED, this number is $L_{\rm torus}/L_{SF}\sim10-20$.
  Thus, the SFR uncertainties are small unless $L_{\rm AGN}>>L_{\rm SF}$. These numbers are very different from those presented 
 in S16, where they conclude that the intrinsic AGN SED can be neglected when the observed total host luminosity 
 (SF$+$AGN contribution) at $60\mu \rm m$ is twice the AGN luminosity at $5100\, \AA$. 
 The results of our test show that the original $L_{\rm SF}-L_{\rm AGN}$ correlation presented in \citet[][their Figure 4]{Rosario2012} is still valid, unlike what stated by S16 (their Figure 16).

\begin{table*}
\hspace{-1.0cm}
\resizebox{15cm}{!}{
\begin{tabular}{cccc}
\hline
 $L_{\rm torus}$/$L_{\rm SF}$  & $L_{\rm SF\,\, 70\, \mu \rm m}$/$L_{\rm EM12\,\, 70 \mu\rm m}$ & $L_{\rm SF}$ uncertainty EM12  & $L_{\rm SF}$ uncertainty PG median    \\
            &                            & no torus subtraction   & no AGN-dust subtraction         \\

\hline
 1                   & 18.7                  &  5\%                         &  10\%                  \\
 5                   & 3.74                  &  21\%                  &  36\%                    \\
10                   & 1.87                  &  36\%                      &  53\%                \\
20                   & 0.93                  & 52\%                        &  69\%                    \\
30                   & 0.62                  & 62\%                        &  77\%               \\
50                   & 0.37                  & 73\%                   &  85\%                  \\
\hline
\end{tabular}}
\caption{Uncertainty on SF luminosity due to AGN-heated dust for various assumptions about the dust SED. 
EM12 is the Mor \& Netzer (2012) SED (torus$+$NLR dust only).
For this SED, $L_{\rm torus}=30.5\times L_{70 \, \mu \rm m}$.
``PG median'' is the median SED derived in this work using PAH-based SF luminosities (see \ref{sec: obtaining intrinsic AGN SED}). 
At 70\mic, this SED is a factor of $\sim 2.1$ more luminous than the EM12 SED. For the PG quasar sample  \LAGN$\approx 2 \times L_{\rm torus}$.
The uncertainties are calculated at 70\mic\ and are assumed to be also the uncertainties on the total SFR. We assume no subtraction of the AGN-dust contribution at $70\mu \rm m$. The numbers in the table can also be used to estimate the $70\mu \rm m$ emission by extended, AGN-heated dust (if real) relative to the SF emission at this wavelength. In our sample the median $L_{\rm SF\,\, 70\, \mu \rm m}$/$L_{\rm EM12\,\, 70 \mu\rm m}$ is $\sim 2.6$.
}
\label{tab:SF_uncertainty}
\end{table*}

        \subsection{Torus covering factor}
    \label{sec:torus covering factor}
 Although it does not constitute a main point of our work, we investigated the covering factor of PG QSOs as they are a significantly more complete sample than others previously considered in the literature for this matter. We follow the methods and discussion in \cite{Netzer2015} and
    \cite{Netzer2016}, and consider the torus$+$NLR component based on the
    chosen EM12 template. After dividing our sample in 5 equally spaced $L_{\rm AGN}$ bins we find that the median covering factor for the anisotropic case ($b=0$ in Equation 3 from
    \citealt{Netzer2016}) varies between $\sim0.3$ and $\sim0.4$ (with a typical error of $\sim0.07$) across the 5 AGN luminosity bins. For the isotropic case  ($b=1$ in Equation 3 from \citealt{Netzer2016}) we find the median covering factor to range between $\sim0.4$ and $\sim0.6$ (with a typical error of $\sim\,0.2$) across the considered AGN luminosity bins. We also find no
    statistically significant trend for the change of covering factor with AGN
    luminosity, confirming the findings of \cite{Mateos2016,Mateos2017}, \cite{Netzer2016}
    and \cite{Stalevski2016}. Lastly we report a lack of correlation between covering factor and Eddington ratio.\\
         {\indent We emphasize that the results mentioned here are independent of the choice of intrinsic AGN SED (EM12 vs. the PAH-based median identified in \S\ref{sec: obtaining intrinsic AGN SED}). As we show in the next section, the difference between the integrated EM12 luminosity and the integrated PAH-based template luminosity is negligible.}

    \subsection{Non-torus dust: temperature, covering factor and location}
    \label{sec:non-torus covering factor etc.}
    Here we estimate the covering factor of the potential additional, colder dust
    component (see \S  \ref{sec: large-scale dust and its impact on the estimate of SF luminosities} and Figures \ref{fig: SF_subtracted_SEDs}, \ref{fig: GB fits})
    relative to the covering factor of the torus$+$NLR emission. We do this by simply
    comparing the integrated luminosity of the PAH-based median intrinsic AGN SED identified in this 
    work ($L_{\rm cold\,\,dust}$) to that of the EM12 SED ($L_{\rm torus+NLR\,\,dust}$), 
    which represents emission from the torus and the NLR only. We do so in
    two wavelength ranges: $\lambda_{\rm rest}\sim 1-243\rm\, \mu m$ and
    $\lambda_{\rm rest}\sim20-243\rm\, \mu m$. The wider wavelength range provides
    us with information about the ``full" covering factor, while the narrower
    range gives us an idea of the relative contributions in the region where the
    extra cold dust component is more noticeable. We note that short-ward of $20\mu\rm m$ we assume the new PAH-based intrinsic SED to be identical to the median EM12. Here we do not consider the
    case of an-isotropic dust emission, since the large-scale, colder dust is
    most likely optically thin.  For the PAH-based median SED of the entire sample, we find
    $L_{\rm cold\,\,dust}/L_{\rm torus+NLR\,\,dust}=1.26$ in the narrower wavelength
    range, and $L_{\rm cold\,\,dust}/L_{\rm torus+NLR\,\,dust}=1.06$ for
    the wider wavelength range. For comparison these numbers are $\sim58$ per cent and $\sim12$ per cent when considering the S16 mean SED in place of our median, PAH-based SED. For the group of 12 sources with potentially the largest
    contribution from the cold dust component (i.e. the ``top $1\sigma$" sources), we have $L_{\rm
    cold\,\,dust}/L_{\rm torus+NLR\,\,dust}=1.90$ for the narrower range and $L_{\rm
    cold\,\,dust}/L_{\rm torus+NLR\,\,dust}=1.19$ in the wider range. The contributions with respect to the bolometric
    AGN luminosity is about 50 per cent of the contributions calculated above
    because the torus covering factor is $\sim1/2$. This means that in the
    majority of the sources the cold dust covering factor is of order $\sim2-3$
    per cent, which we consider consistent with zero. We note that this number is
    based on total energy considerations and the geometrical covering factor can
    be larger if the dust is optically thin to some of the AGN radiation.
    
 Moreover we can use the derived temperature ($T_{\rm cold\,\,dust}$,
    \S\ref{sec: revisiting AGN SED}) together with some basic assumption (e.g.
    the SED of the optically thin dust is well described by a single $\beta$), to
    estimate the distance between the AGN and the extra dust component. The first possibility
     is optically thin dust, where the distance ($r$) from an emitting
    source of energy $L_{\rm AGN}$ is given by the following equation:
    \begin{equation}
    \frac{r}{pc}=\left(\frac{1500}{T_{\rm
    cold\,\,dust}}\right)^{\frac{4+\beta}{2}}\,\left( \frac{L_{\rm
    AGN}}{10^{46}}\right) ^{\frac{1}{2}}\, ,
    \end{equation}
    where $\beta\sim1.5$. For the median based on the ``top $1\sigma$" sources we cannot estimate $r$, as we could not find a satisfactory 
    fit with a single-temperature modified blackbody. On the other hand, for the generic median the temperature is $T\sim25.5\,\rm K$ and  $L_{\rm
    AGN}=2.3\times10^{45}\rm\, erg/s$, yielding $r\sim35\,\rm kpc$. We
    caution that this estimate may not be entirely meaningful because many sources within
    the full sample do not show a clear extra dust component (see Figure
    \ref{fig: SF_subtracted_SEDs}). This number ($r=35\,\rm kpc$) is very large, and puts the extra-dust component outside the host galaxy of most sources. The recent
    Symeonidis (2016) work applies the same considerations to the sources from
    Netzer et al (2016) that are typically 50 times more luminous than our current sample. The
    corresponding distances in this case are about 7 times larger, i.e.
    $\sim245\,\rm kpc$. Again this is not plausible, as these scales are significantly beyond the size of any galaxy at these redshifts \citep[e.g.][]{vanDerWel2014}. 
        
   The second possibility is that the dust is optically
    thick at some wavelengths, and hence can be located closer to the central
    AGN. This  raises a different issue related to the column density
    required to produce the extra emission. Rough calculations show that for a gas of solar metallicity with a column density of $N_{\rm H}\sim10^{21.3}\rm\, cm^{-2}$, the
    range of temperatures inside a dust shell, whose location is $4\,\rm kpc$
    from a source with $L_{\rm AGN}=2\times10^{45}\,\rm erg/sec$, is $T=25-45\,\rm K$ 
    (depending on the gas density and ionizing continuum shape). This results in a spectrum which is not too different from the
    single temperature SED assumed for optically thin dust. However, this column
    is large enough to attenuate, considerably, the optical UV continuum of the
    source, if located along the line of sight. This is not observed in the
    optical-UV SED of the sources in question. Moreover, if the covering factor of this component is of order a few percent,
     this gas will produce strong emission lines, of low or high ionization (depending on the density), 
     that can outshine the NLR emission. Regarding the possibility of unusually cold torus dust, energy considerations, combined with the assumed temperature, 
     suggest a minimal radius of approximately $160\,\rm pc$. 
     Thus, the location and physical properties of the component responsible for the additional dust emission (if any) must be carefully modeled. 
     
  Within this context, we note that  \cite{Baron2016} recently found evidence for
    large-scale dust in luminous SDSS type-I AGN, responsible for a typical
    reddening $E_{\rm B-V}\sim0.08\,\rm mag$ ($N_{\rm H}\sim10^{20.7}\rm\, cm^{-2}$  for a solar metallicity gas). 
    This is roughly in agreement with the necessary additional dust emission but not with the low temperature deduced here.
    
 Finally, we caution that these geometric scale considerations for an AGN-heated cold dust
  component will not apply if the non-torus cold dust is related to star formation that was
   not properly captured by the PAH-based SF estimate, as for a PAH to SFR scaling following \cite{Lyu2017} rather than \cite{Shi2007}. If this were the case, the SF-related dust could be placed
    at any location in the galaxy that is not directly exposed to AGN radiation, for example in the large solid angle shielded by the torus. 

    \subsection{Black hole and stellar mass growth}
    \label{sec:SFR BHAR}
    In this section we want to investigate how the black hole, and host's stellar
    mass grow, in the PG quasars. Specifically we want to compare these growths to those found in
    other samples, in particular those with higher redshift and significantly
    more massive BHs. The median BH mass for the PG sample is $M_{\rm BH}\sim 3\times10^{8}\,\rm M\odot$, while the sample on
    which we will focus for the comparison (i.e. \citealt{Netzer2016}) is
    characterised by a typical black hole mass at $z\sim3$ of $M_{\rm
    BH}\sim10^{10}\,\rm M\odot$.

    \subsubsection{$L_{\rm SF}$ vs. $L_{\rm AGN}$} 
    \label{sec: LSF vs LAGN}
    The topic of $L_{\rm SF}$ vs. $L_{\rm AGN}$ has been considered in many
    earlier works using several parallel approaches and, hence, showing
    seemingly different results (see below). This
    issue has been discussed at length in \cite{Netzer2016} and we refer the
    reader to their work for more details. Here we only discuss the points most
    relevant to the current study.
    
 The first important issue is how the sample is selected -- either by
    using its AGN properties \citep[e.g][]{Netzer2009, Rosario2012, Stanley2015}
    or its IR properties \citep[e.g][]{Chen2013,Delvecchio2015}. Given the combination of colour
    and flux used for the selection of the PG sample, we only discuss samples
    selected by their AGN properties (X-ray or optical-UV). The second issue is
    whether the analysis refers to the mean, median or individual IR
    luminosities. Samples analyzed by their mean IR (mostly FIR) luminosity give
    more weight to sources of higher luminosity. On the other hand, flux limited
    UV or X-ray samples contain many more low luminosity sources, close to the
    flux limit of the sample. 
    
    Using mean observed properties is most useful when studying the cosmic SFR of the
    entire population. This approach, however, is accompanied by the loss of the information about the distribution of SFR in a
    certain bin of redshift and $L_{\rm AGN}$. For example, in the
    \cite{Stanley2015} sample only 10--20 per cent of the AGNs in a given
    $L_{\rm AGN}$ bin were individually detected by $Herschel$. \cite{Mullaney2015} also showed that, given a distribution of star formation rates, the use of its linear-mean rather than the individual values can lead to the results being biased by the more luminous outliers. This highlights also another
    issue: the consideration of median properties vs. mean properties in the analysis.
    The great advantage of the sample under consideration here is the possibility to
    use all the above methods: individually measured $L_{\rm SF}$ (for  $\sim
    95$ per cent of the sources), and mean $L_{\rm SF}$ and median $L_{\rm SF}$ for
    the entire sample.
    
\begin{figure*}
    \begin{minipage}[c]{\textwidth}
    \begin{center}
    \begin{tabular}{c c}
    {
    \psfrag{us med}[][][0.7][0]{\hspace{1.67cm} This work, median}
    \psfrag{us mean}[][][0.7][0]{\hspace{1.3cm} This work, mean}
    \psfrag{stan}[][][0.7][0]{\hspace{1.2cm} Stanley+ 15}
    \psfrag{R121}[][][0.7][0]{\hspace{1.8cm} Rosario+ 12, z$\sim 0$}
    \psfrag{R122}[][][0.7][0]{\hspace{2.7cm} Rosario+ 12, z$=0.2-0.5$}
    \psfrag{z}[][][0.7][90]{z}
    \psfrag{x}[][][1.3][0]{$\rm L_{\rm AGN}$ (erg/s)}
    \psfrag{y}[][][1.3][0]{$\rm L_{\rm SF,\, fit}$ (erg/s)}
    \hspace{-4.0cm}\includegraphics[trim=7cm 0cm 15.5cm 1cm, clip=true,
    width=0.45\textwidth]{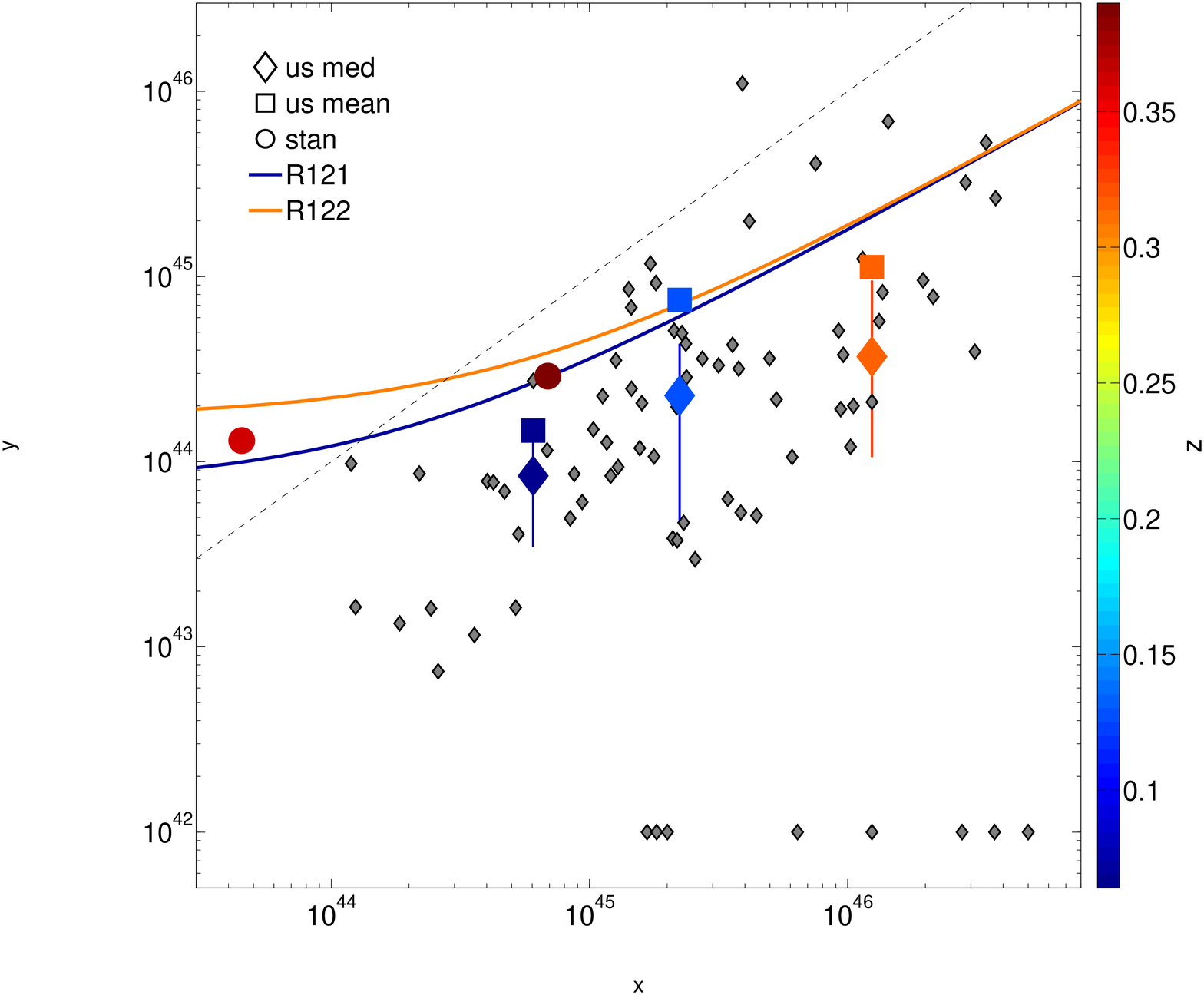}} 
     &
     {\psfrag{p1}[][][0.7][0]{\hspace{3.2cm} Sources below the SFMS}
    \psfrag{p2}[][][0.7][0]{\hspace{2.72cm} Sources on the SFMS}
    \psfrag{p3}[][][0.7][0]{\hspace{1.25cm} Starbursts}
    \psfrag{l2}[][][0.7][0]{\hspace{1.42cm} 1:1 relation}
    \psfrag{l1}[][][0.7][0]{\hspace{3.35cm} Best fit for SFMS sources}
     \psfrag{x}[][][1.3][0]{$\rm L_{\rm AGN}$ (erg/s)}
    \psfrag{y}[][][1.3][0]{$\rm L_{\rm SF,\, fit}$ (erg/s)}
     \hspace{-4cm}\includegraphics[trim=7cm 0cm 16cm 1cm, clip=true,
    width=0.475\textwidth]{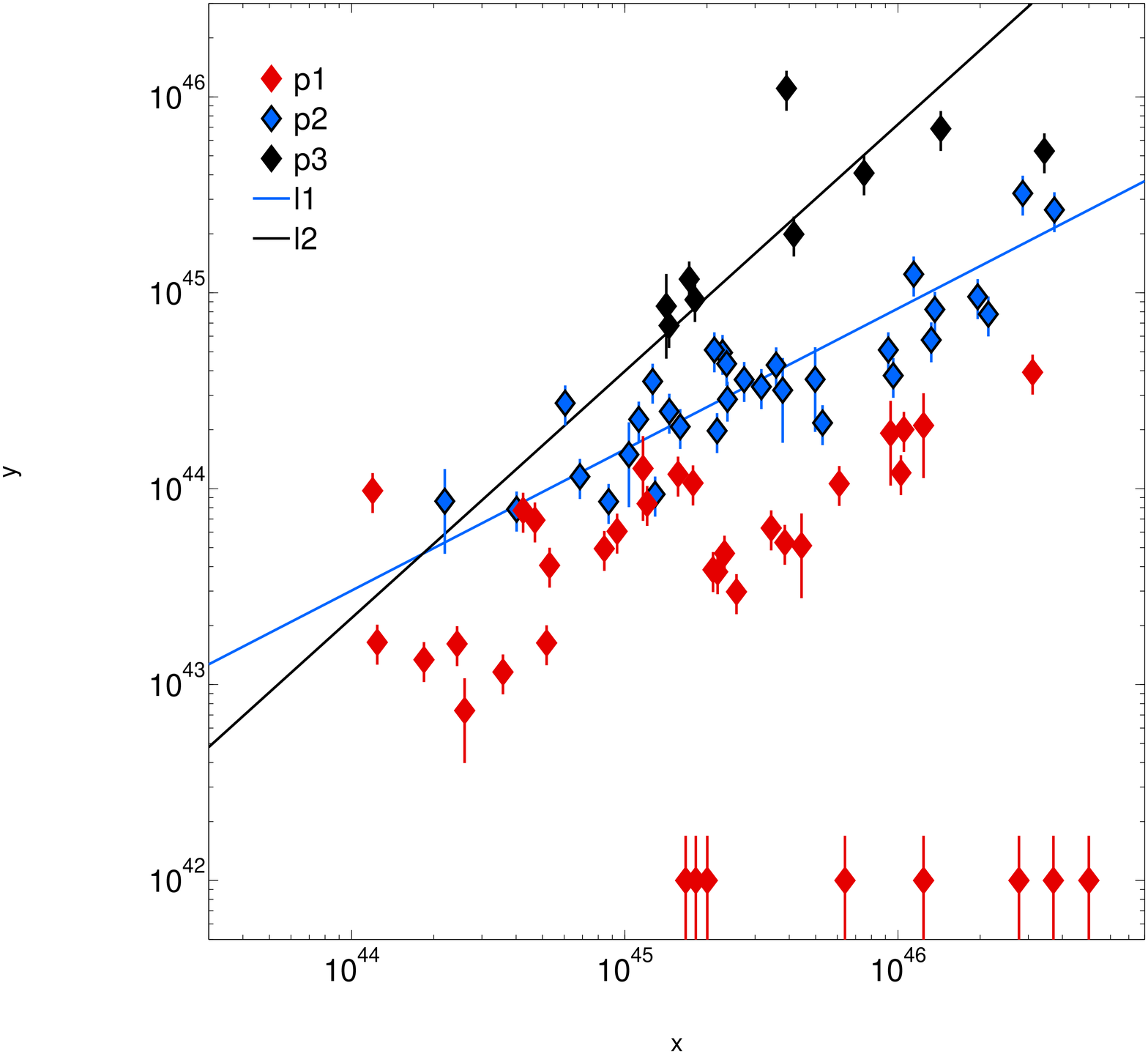}}\\
    
     {
       \psfrag{R121}[][][0.7][0]{\hspace{1.8cm} Rosario+ 12, z$\sim 0$}
    \psfrag{R122}[][][0.7][0]{\hspace{2.7cm} Rosario+ 12, z$=0.2-0.5$}
    \psfrag{x}[][][1.3][0]{$\rm L_{\rm AGN}$ (erg/s)}
    \psfrag{y}[][][1.3][0]{$\rm L_{\rm SF, PAHs}$ (erg/s)}
    \hspace{5cm}\includegraphics[trim=7cm 0cm 16cm 1cm, clip=true,
    width=0.45\textwidth]{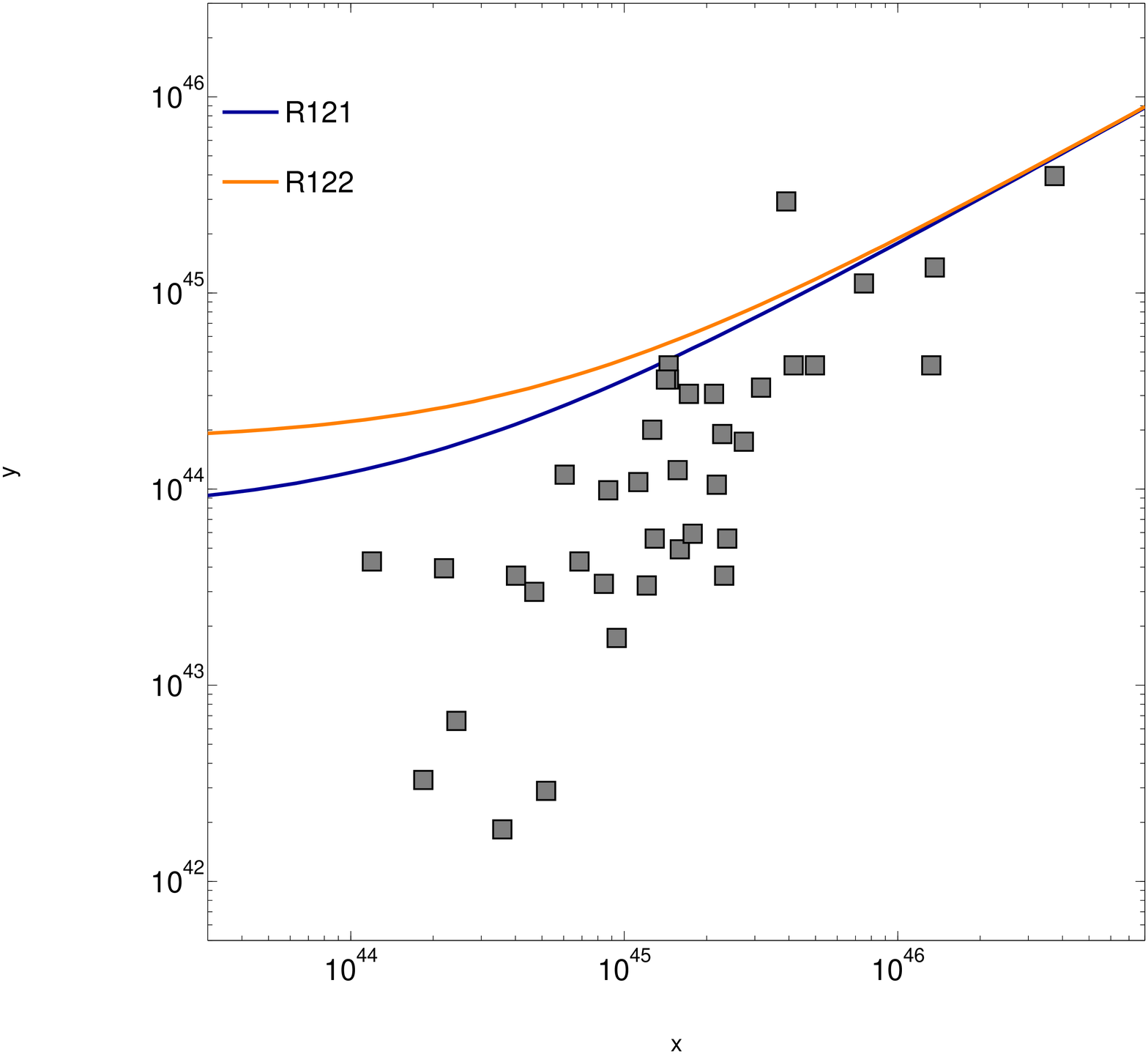}} 
    \\
    \end{tabular}
       \end{center}
    \end{minipage}
    \caption{{\bf Top left panel.} Median (diamonds) and mean (squares) $L_{\rm SF}$ in 3 bins of
    $L_{\rm AGN}$  as found in our sample of PG QSOs, when assuming the EM12 AGN SED. 
    The uncertainties on the median points
    represent the $25^{\rm th}$ and $75^{\rm th}$ percentiles of the
    distribution in each $\rm L_{\rm AGN}$ bin. The grey points represent the
    individual sources, while the colour-coding is by redshift and facilitates the comparison with the modified
    mean relations of  \protect\citet[][solid lines, see \S\ref{sec: LSF vs LAGN} for details]{Rosario2012}
    and the mean $L_{\rm SF}$ from \protect\citet[][circles]{Stanley2015}. 
    {\bf Top right panel.} $L_{\rm SF}$ vs. $L_{\rm AGN}$ for the PG sample using the EM12-based method to derive SF luminosities.
Objects below, on and above the SFMS are colored with green, blue and black respectively. The regression line for
the sources on the SFMS is shown with the corresponding colour
(see \S\ref{sec: LSF vs LAGN} for details). The black line represents the 1:1 relation and appears to be in good agreement with the $L_{\rm SF}- L_{\rm AGN}$
 relation observed for the sources above the SFMS.
    {\bf Bottom panel.}
    $L_{\rm SF}$ vs. $L_{\rm AGN}$ for SF luminosities based on PAH features. Regardless of
    the method to estimate SFRs, we find a trend: $\rm L_{\rm SF}$ increases
    as $\rm L_{\rm AGN}$ increases. For clarity we omitted those sources that
    are upper limits in \protect\cite{Shi2007}. } 
    \label{fig:LSF vs LAGN}
    \end{figure*}

 In the top left panel of Figure \ref{fig:LSF vs LAGN} we show the
    mean and median star formation luminosities, as found in our sample of PG
    QSOs (large squares and large diamonds respectively) when assuming the EM12 AGN
    template. The colour-coding is based on the average redshift of the
    respective bins, while the grey points show the underlying distribution of
    individual sources. It is apparent that in our sample there exists a
    trend between AGN luminosity and SF luminosity. In the same panel, we also show
    two curves obtained from \cite{Rosario2012}, scaled up assuming $L_{\rm
    SF}=2\times L_{\rm 60\mu \rm m}$. There seem to be a consistent behaviour within
    the luminosity range where the two works overlap, regardless of whether we
    consider mean or median SF luminosities for our sample. Finally, in this figure the
    circles represent the findings of \cite{Stanley2015}, within a similar
    redshift range to that of our sample. 
 This diagram illustrates the steepening of the relationship, from
    basically no dependence in the SF-dominated region, to a logarithmic slope
    in the AGN-dominated region. 

To test this more quantitatively, we focus on individual sources, avoiding all methods based on mean properties. We want to investigate whether the relationship between $L_{\rm SF}$ and $L_{\rm AGN}$
depends on the SF properties of the host, i.e. the location of the host with respect to the 
SF main sequence (SFMS). This requires information about \mstar\ for all the host galaxies.
In our sample of type-I AGN, we do
not have measurements of \mstar. We can, however, use standard low redshift scaling relations 
to get an approximation for the total stellar mass, starting from the measured \MBH.
Such relations have been investigated in numerous papers, e.g. 
\cite{Kormendy2013,Lasker2014a, Lasker2014b, Reines2015, Shankar2016}.
There are large differences between the various studies linked to galaxy morphology 
(i.e. the differences between total mass and bulge mass in early- and late-type galaxies), the method used to measure the BH mass
(e.g. direct dynamical measurements, $\sigma^*$-based measurements, $L_{\rm K}$-based measurements,
reverberation mapping-based measurements, BHs in megamasers), and the sample selection method.
A clear trend emerging from all such studies is that \mstar/\MBH\ increases with decreasing
BH mass (compare e.g. the \citealt{Reines2015} ratio of $\sim4000$ for $10^7 < M_{\rm BH} < 10^8$ \msun\ with ratios
of 200--400 found in systems with \MBH$\sim 10^9$ \msun\, e.g. \citealt{Kormendy2013}). 
There are, however, large differences between all these results, some of which may be related to the uncertainties in estimating BH mass in systems that are not bulge-dominated.
We use a method which is similar to that used in \cite{Trakhtenbrot2010}. The BH mass is estimated
from the \MBH-$\sigma^*$ relationship presented in \cite{Kormendy2013}, and \mstar\ from broad-band SDSS
photometry. For the current study, we obtained such measurements for about 20,000 SDSS/DR7 type-II AGNs and fitted
them by, 
\begin{equation}
\log M_{*}/M_{\rm BH}=3.3-0.5\,(\log M_{\rm BH}-7.5) \, .
\end{equation}
 The numerical values obtained in this way are in agreement with the \cite{Reines2015} and \cite{Lasker2016} for low \MBH, and converge to the \cite{Kormendy2013} 
ratios for the most massive BHs in our sample, 
assuming $M_{*}=(1-2) \times M_{\rm bulge}$. Furthermore, these results in $M_{*}/M_{\rm BH}$ are similar to those suggested in \cite{Greene2016} for all (43 early type and 24 late type) galaxies in their
sample, over the range of \MBH\ considered here.
We have also experimented with a simpler relationship (Equation 10 from \citealt{Kormendy2013} assuming $M_{*}=2 \times M_{\rm bulge}$), finding values of $M_{*}$ which are not significantly different
from the previous approximation for the most massive BHs in the sample.
While the uncertainties in this simplified approach are large,
the above approximation still allows us to
constrain some important {\it average} properties related to SF in the hosts of PG quasars. We therefore take the black hole mass estimates for our sample, obtained as described in \S\ref{sec:Data} and listed in Table 1, and use Equation 3 to obtain stellar masses for every PG quasar.

                  We used our SF luminosities and the assumption that $1\,M_{\odot}/\rm yr=1\times 10^{10}\, L_{\odot}$, together with the assigned $M_{*}$ and
                   the parameterization of the 
                   SFMS given in \cite{Whitaker2012}, to determine the location of all
                   our PG QSOs with respect to
                   this sequence. We consider sources that are ``below the SFMS''
                   those with SFRs which are at least $0.3\,\rm dex$ below the
                   redshift dependent expression given by
                   \cite{Whitaker2012}. This yielded 31 objects on, and 37 objects below, the SFMS. 
                   There are 9 more sources which we consider to be above the main sequence and potential starbursts. 
                   Their  SFR is more than $0.5\,\rm dex$ above the line we use to define the SFMS at their redshift. 
                   These objects are suspected to be characterised by high density molecular
gas which leads to a much higher efficiency in SF, perhaps caused by mergers. The exact definition 
of the boundaries of the SFMS changes slightly from one study to the next, but this makes 
no significant difference to the results considered here and those in \S\ref{sec:SFR/BHAR}. The fraction of starbursting hosts found
here is considerably larger than that found in large samples of SF galaxies \citep{Rodighiero2011}. This is not surprising
given the suspected merger activity in many hosts of PG quasars
\citep[e.g.][]{Veilleux2009}. 
We note that our analysis may not necessarily work correctly on an object by object basis, 
due to the large uncertainties, yet we can still apply it to investigate average properties of large samples (like that considered here).

    The top right panel of Figure \ref{fig:LSF vs LAGN} shows all PG sources with
    reliable SF luminosities (77 objects) and the best linear fit calculated by
    the standard BCES regression procedure \citep{Arkitas1996}, that takes into account
    the uncertainties on both variables. For the error on $L_{\rm AGN}$ 
    we have assumed $0.15\, \rm dex$ for all sources. For the error on $L_{\rm SF}$, 
    we do not allow an uncertainty below $0.1\,\rm dex$. For some sources, the formal 
    uncertainty is in the range $0.1-0.2\,\rm dex$; to all these we assigned an uncertainty of $0.2\,\rm dex$. 
    Finally, 8 sources are consistent with no SF with a formal upper limit of $\log(L_{\rm SF}/\rm erg\,s^{-1} )<42.3$.
     All these were assigned SFRs corresponding to $\log(L_{\rm SF}/\rm erg\,s^{-1} )=42.0$ with an uncertainty of $0.3\,\rm dex$. 
     We calculated the regression for the three sub-samples
(on, below and above the SFMS), as well as for the entire sample. 
The correlations for the entire sample, and for the sources below/above the SFMS, are not significant. This
is due to the very large spread in $L_{\rm SF}$ for a given $L_{\rm AGN}$, caused by objects with negligible SFR,
and the small number of sources above the SFMS respectively. On the other
hand, the correlation for the objects on the SFMS is highly significant.
It is given by
$\log L_{\rm SF} = (0.72 \pm 0.08)\,\log L_{\rm AGN}+(11.80 \pm 3.2)$

The correlation of SFMS sources is consistent with that proposed in \cite{Netzer2009} for the population
of type-II AGN with $L_{\rm AGN}>L_{\rm SF}$ in the SDSS, and with the recent work by \cite{Ichikawa2016}. 
Interestingly, there appears a considerable difference in the $L_{\rm AGN}-L_{\rm SF}$ relation between the group of sources on the SFMS and that above the SFMS
 (that we caution only includes 9 sources). The latter seems to closely follow a 1:1 relation.

While the result of the correlation analysis depend on two well determined quantities, $L_{\rm AGN}$ and $L_{\rm SF}$,
 its main uncertainty is the definition of the SFMS group, as this in turn depends on the large error on the the stellar mass estimates.
  The best way to verify the reliability of this part, is to compare our stellar masses with those recently obtained by \cite{Zhang2016} that are well-determined. 
  They investigated the location of PG QSOs on the SFMS with photometrically-determined stellar masses. They report a tentative dependence of the
   specific SFR (sSFR) on $L_{\rm AGN}$, with the more luminous AGNs showing higher sSFRs. A direct comparison of their stellar masses to those obtained in our work shows good agreement, 
   with a typical difference of $\sim0.2\,\rm dex$. The scatter is distributed nearly symmetrically along the 1:1 line, showing no evidence for systematics in our estimates.
    We also note that the consideration of stellar masses based on alternative scaling relations \citep[e.g.][]{Kormendy2013, Reines2015} yields a less
     good agreement in the comparison to the \cite{Zhang2016} photometrically-determined stellar masses. 
     This is further support that our stellar mass estimates,  although not necessarily exact on a source by source basis,
      give a good representation for the typical properties of the AGN population.

When inspecting Figure \ref{fig:LSF vs LAGN}, it is interesting to note some real differences between our UV-selected sample and several X-ray selected samples,
 like those presented by \cite{Rosario2012}, \cite{Stanley2015} and \cite{Shimizu2016}. The former selection method (UV-selected) tends to pick only AGN-dominated
  sources while the latter (X-ray selected) contain many SF-dominated objects. In general, the X-ray selected samples reach lower AGN luminosity and also lower SFRs.
   In addition some of the samples provide only mean SFRs, neglecting the location of individual sources in the $L_{\rm AGN}-L_{\rm SF}$ plane.
   Nonetheless some of the differences, especially the lack of rise of $L_{\rm SF}$ with $L_{\rm AGN}$ at low $L_{\rm AGN}$ seen in several X-ray selected samples,
    seem to be in conflict with the properties of the PG QSO sample. 
    
For completion, we show in the bottom panel of Figure \ref{fig:LSF vs LAGN} the relationship 
between $L_{\rm SF}$
    estimates from the PAH-based method and $L_{\rm AGN}$.
    For clarity, we only plot those sources for which PAH-derived star formation
    luminosities are not upper limits.
   The same behaviour of $L_{\rm SF}$ and $L_{\rm AGN}$ is
    present regardless of the method used to measure SF luminosities. We consider this
    result less meaningful than that shown in the top left panel of Figure
    \ref{fig:LSF vs LAGN}, since $\sim50$ per cent of the sources could not be used (because
    upper limits) and hence the sample cannot be considered ``complete".
  
    Finally we note that very recently \cite{Shimizu2016} reported the lack of a correlation for $L_{\rm SF}-L_{\rm AGN}$,
     in a large sample of local X-ray selected AGNs, from the $Swift\,\, Burst\,\, Alert\,\, Telescope$ (BAT) 58 month catalogue.
      After obtaining the full data for their Table 1, and applying the bolometric correction factor they suggest ($L_{\rm AGN}=10.5\times L_{14-195\rm \, keV}$),
       we find a significant overlap for approximately half of the luminosity range covered by the PG quasars. The lack of
        correlation remains also when we only consider those BAT sources which are AGN-dominated.
            We suggest two possible reasons for the observed differences. 
            First, having more AGNs with high $L_{\rm SF}$ and low $L_{\rm AGN}$ in the \cite{Shimizu2016} sample,
             and second the much fewer high $L_{\rm AGN}$ sources ($L_{\rm AGN}>2\times10^{45}$) in the \cite{Shimizu2016} sample. 
             These are indeed the ones mostly driving the correlation observed in the current work. 

    \subsubsection{SFR/BHAR across cosmic time}
    \label{sec:SFR/BHAR}
                   The ratio SFR/BHAR can be interpreted as the relative
                   ``instantaneous growth rate'' of the stellar and BH mass. The term ``instantaneous" is correct for the AGN  but may be misleading for the SFR.
                    The FIR radiation we measure is emitted by dust which is heated by stars of different ages, occasionally without any on-going star formation.
                     It reflects the average SFR over a period of $\sim100\, \rm Myr$. 
                      Nevertheless, this is an appropriate terminology when comparing to the entire SF history of the galaxy. For $L_{\rm AGN}=L_{\rm SF}$, and for mass to radiation conversion
                   efficiency $\eta=0.1$, SFR/BHAR is $\sim140$. If the ratio
                   does not change with time, and if the duty cycles
                   of SF and BH accretion  (the fractions of time the
                   processes are ``on'')
                   are identical, at the end of the
                   process we would expect \mstar/\MBH=SFR/BHAR. At low
                   redshift, the duty cycles for SF and BH accretion are very
                   different, as evident, for example, by the much larger number
                   of high mass star-forming galaxies compared to the numbers of luminous
                   AGNs. This means that in the local Universe, the accumulation
                   of stellar mass during the time the BH is active is only a small fraction of the final
                   stellar mass. A rough estimate of the mean relative duty
                   cycle of the two processes is given  by
                   (\mstar/\MBH)/(SFR/BHAR). This is a ratio we can determine, with large uncertainty, using the values of \mstar\
obtained as described in \S~\ref{sec: LSF vs LAGN}. 

We now proceed to contrasting the low redshift PG quasars, and the higher redshift, highest luminosity $z=2-3.5$ AGNs in the \cite{Netzer2016} sample.
 We emphasize that we do not aim to evolutionally link these two samples, but rather to compare them in order to learn more about the stellar mass and BH mass growth for BHs of different masses.
  To facilitate this comparison, we note that the median of \mstar/\MBH\ in
                   the PG quasar sample is $\sim485$. On the other hand, the
                   high redshift sample contains
                   the most massive BHs at those epochs, that are probably on
                   their way to also become the most massive BHs in the local
                   Universe.  Their mean \MBH\ is unknown, but was estimated by
                   \cite{Netzer2016} to be $\sim10^{9.5-10} M_{\odot}$.  From
                   \cite{Kormendy2013} we find that the hosts of such BHs in the local Universe are
                   large elliptical galaxies and hence we assume
                   $M_{*}=M_{bulge}$. For these sources we therefore estimate $M_{*}/M_{BH}\approx
                   100$.

    \begin{figure*}
    \psfrag{p1}[][][0.9][0]{\hspace{6.2cm} This work ($z=0-0.5$), sources below the SFMS}
    \psfrag{p2}[][][0.9][0]{\hspace{5.76cm} This work ($z=0-0.5$), sources on the SFMS}
    \psfrag{p3}[][][0.9][0]{\hspace{4.32cm} This work ($z=0-0.5$), starbursts}
    \psfrag{p4}[][][0.9][0]{\hspace{6.40cm}Netzer+ 16 ($z=2-3.5$), {\it Herschel}
    non-detections}
    \psfrag{p5}[][][0.9][0]{\hspace{5.75cm} Netzer+ 16 ($z=2-3.5$), {\it Herschel} detections}
    \psfrag{x}[][][1.1][0]{Age of the Universe (Gyr)}
    \psfrag{y}[][][1.1][0]{SFR/BHAR}
    \includegraphics[scale=0.3,trim={13cm 1cm 12cm
    0cm},clip]{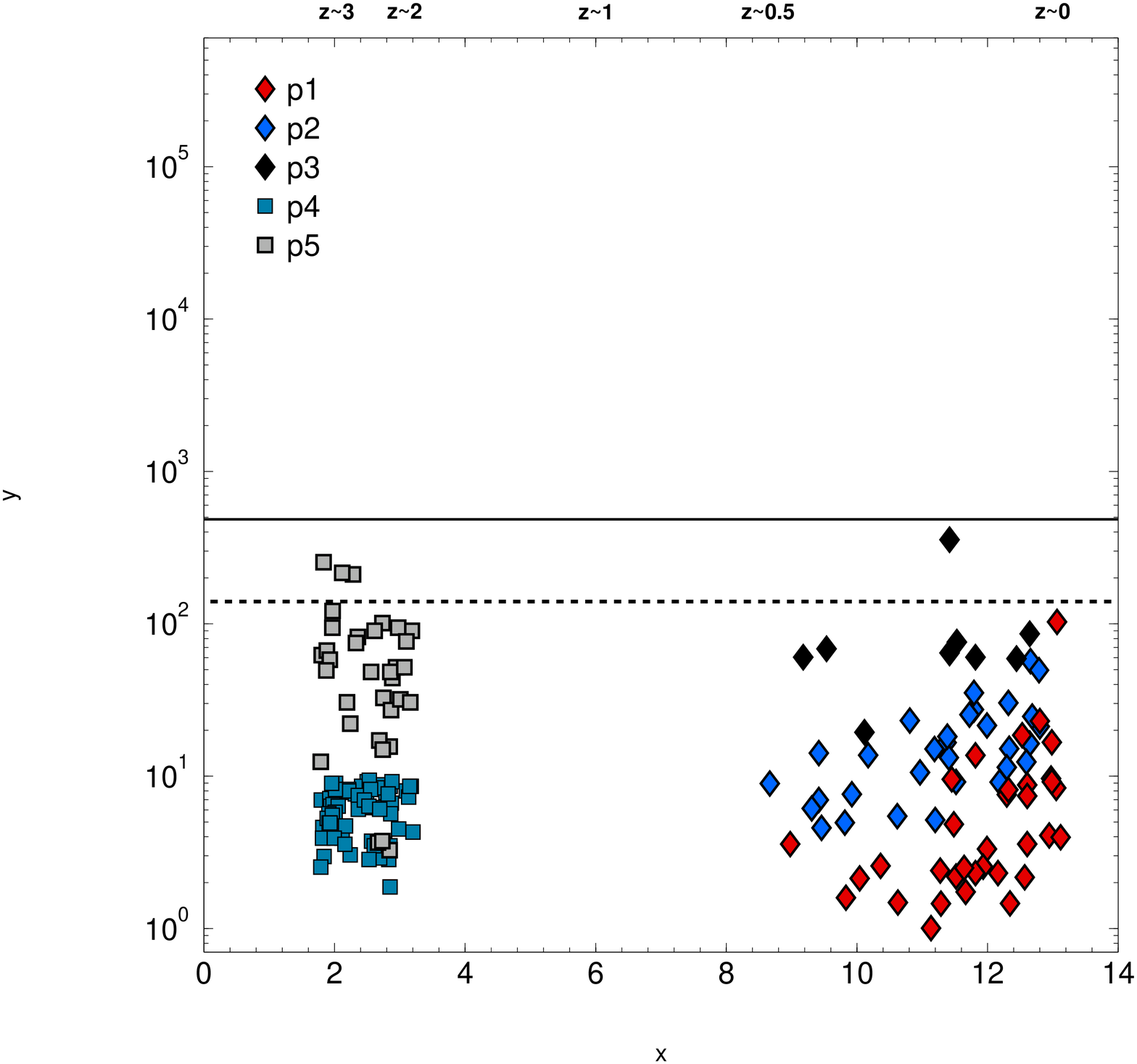} 
    \vspace{0.2cm}
      \caption{Instantaneous stellar mass to black hole growth rates as a function of
      cosmic time, for two samples. The blue diamonds represent the PG
      QSOs on the SFMS.
      The black diamonds represent PG QSOs that are
      candidates for starbursts (at least $0.5\,\rm dex$ above the
      SFMS) and the red diamonds are PG QSOs below the SFMS. The squares represent
      the sources from \protect\cite{Netzer2016}, with {\it Herschel} detections
      in grey and  {\it Herschel} non-detections in light blue. The horizontal lines
      are drawn at SFR/BHAR=140 (to mark the case where $L_{\rm AGN}=L_{\rm SF}$) and 485
      (the typical \mstar/\MBH\ of the PG QSOs). While we do not trust our division between below/above the SFMS on
       an individual object's basis, the well-defined colour segregation for the three groups shown here is confirmation that our division works overall.}
        \label{fig:SFR/BHAR}
    \end{figure*}             
    
 Figure \ref{fig:SFR/BHAR} shows the distribution
    of SFR/BHAR across cosmic time for two samples: our PG sample, the
   and the $z=2-3.5$ sample of
                   Netzer et al. (2016). The PG QSO sample has individual measurements of
                    both BHAR and SFR. As previously mentioned, the high redshift sources were observed by
                   {\it Herschel}/SPIRE, with 34 detections and 66 upper limits.
                   The individual measurements of the detected sources were used
                   to obtain the grey squares in Figure \ref{fig:SFR/BHAR}. The
                   66 upper limits were stacked to obtain a meaningful median flux
                   corresponding to $SFR\sim 100\, \rm M_{\odot}/yr$. For
                   these objects (shown as blue squares), we used the stack's
                   $L_{\rm SF}$ value and their individual $L_{\rm AGN}$, to calculate their SFR/BHAR ratios.
                 As explained in Netzer et al.
                   (2016), some of these sources are probably on the
                   SFMS and some,
                   perhaps the majority, below the SFMS in quenched hosts. We purposely avoid a comparison with those works that do not have individual measurements in $L_{\rm SF}$ but use stacks or mean values \citep[e.g.][]{Stanley2015} as, in such studies, the spread in SFR/BHAR at a particular redshift cannot be properly explored. 
                   
                   The histogram of SFR/BHAR for
                   PG quasars on and above the SFMS is shown in Figure
                   \ref{fig:SFR/BHAR_hist}; the median of this ratio is
                   $\sim17$.
                   For the $z=2-3.5$ sample, we consider sources on or above the SFMS
                   to be only the 34 sources detected by {\it Herschel}; their 
                   median SFR/BHAR ratio is $\sim50$.

                   The distribution of SFR/BHAR for sources on or above the SFMS,
                   combined with the same ratio for the sources in quenched
                   hosts, can be interpreted
                   as the progression of the duty cycles of SF and BH activity
                   along cosmic time, and as a function of BH mass. The explanation adopted here applies {\it only}
                    to AGNs that are hosted in SF galaxies on and above the SFMS. We compare two epochs: $2-3\,\rm Gyr$ after the Big Bang and the local
                   Universe. For the high redshift sources, the estimated
                   (\mstar/\MBH)/(SFR/BHAR) ratio is $\sim2$ meaning that in the
                   first $2-3\,\rm Gyr$, the most massive BHs were actively
                   growing during
                   about half the time their host galaxies grew by SF. These
                   numbers are consistent with the scenarios discussed in
                   \cite{Netzer2014} and \cite{Netzer2016} for
                   this population. For the local, lower mass objects the situation is very
                   different, since  (\mstar/\MBH)/(SFR/BHAR) is $\sim20-100$ with a median value of $\sim56$. This would imply that the BHs are actively growing
                   during about $2-5$ per cent of the time that their hosts are
                   forming stars on and above the SFMS. For example, if a typical
                   SF episode is about $150\rm\, Myr$, the
                   equivalent BH growth time is about $3\rm\, Myr$. According to
                   this scenario, we observe the PG quasars during this $3-5$ per
                   cent of the SF event, which may also be the last event of fast BH growth of 
                   this population. We emphasize, again, that all numbers used here are for sources on and above the SFMS. We attribute the very different ratios of
                   (\mstar/\MBH)/(SFR/BHAR) at these epochs as due to 
                   very different duty cycles. This may be
                   related to either different conditions in the Universe, e.g.
                   the relative fraction of molecular gas
                   in galaxies, or different population properties, particularly
                   BH mass. Finally, we caution that the AGN activity discussed here 
   is not necessarily a single
                   event, and the $5\rm\, Myr$ can be broken into several
                   shorter episodes. Also, there must have
                   been earlier significant SF episodes since the duration and
                   typical SFR measured here are not enough to explain the
                   stellar mass of these
             objects. It is very likely that earlier episodes have different SFR/BHAR.
                   
  The above explanation depends on several
                   assumptions. On the high luminosity side (i.e. the redshift
                   $z=2-3.5$ sample), we assume that the growth of BHs below the
                   SFMS does not add much
                   to their total mass. A simple scenario involving negative
                   feedback, operating both on SF and BH activity, is consistent
                   with this assumption.
                   On the low luminosity side (i.e. PG QSOs), we assume that the
                   active BHs in SF-dominated systems, like most sources
                   in the \citealt{Rosario2012} and \citealt{Stanley2015}
                   samples, do not influence the SF in their hosts. The prediction here is that most such
                   systems cannot be distinguished from SF galaxies on the SFMS
                   (for more discussion about the comparison of main sequence SF
                   galaxies and AGN hosts see \citealt{Santini2012}, \citealt{Rosario2013}, \citealt{Hicox2014} and \citealt{Mullaney2015}). This is because shallower X-ray surveys would miss the AGN X-ray emission, thus mistakenly classifying a significant fraction of these sources as
                   pure SF galaxies.
                   
  There are additional uncertainties and caveats in
                    the above scenario, mostly related to the lack of robust
                    stellar mass measurements and
                   the use of mean (or median) rather than individual
                   properties. This is the case for the $z=2-3.5$ sample where
                   2/3 of the sources do not have
                   individual SFR measurements, and none of the sources have
                   stellar mass measurements. In addition, we did not
                   distinguish SFMS galaxies from starburst galaxies, where SF
                   efficiency is high and the stellar mass growth time is
                   shorter. Given the lack of information on individual \mstar, we prefer
                   not to discuss these issues any further.

                 Lastly, we refer to several points in the new \cite{Lyu2017} paper that was already mentioned in \S\ref{sec:intro}. Their work addresses AGN SED, specifically that proposed in \cite{Elvis1994} and then modified in \citet[][ME15 hearafter]{Xu2015}, which is based on a sample of $\sim50$ high luminosity sources. The \cite{Lyu2017} analysis suggests ME15 to be the best representation of the intrinsic AGN SED. Our comparison reveals that this SED is consistent with that from EM12 at wavelengths longer that $\sim2\mu\rm m$ (see the Appendix). \cite{Lyu2017} also suggest that the prescription used by S16 to convert PAH-based luminosities to SF-based luminosities tends to under-estimate $L_{\rm SF}$, again in line with our conclusions. By following the same method as S16 but adopting a different PAH/SF scaling, \cite{Lyu2017} find a PG-based intrinsic AGN SED that significantly differs from that of S16 (their Figure 6). Their conclusions contrast the existence of a significant amount of large-scale, cold AGN heated dust as postulated in S16.

   \section{conclusions}
\label{sec:conc}
We presented a new analysis of the entire IR spectrum of the PG quasar sample
 taking advantage of the most updated  $Spitzer$/IRS (MIR) and 
$Herschel $ (FIR) observations. 
The analysis focuses on three major issues: the SED of AGN-heated
dust in these sources, the various correlations between $L_{\rm SF}$ and $L_{\rm AGN}$ in this and other
AGN samples, and the relative growth rates of BH (BHAR) and stellar mass (SFR).
The main results of our work can be summarized as follows:

    \begin{figure*}
    \begin{minipage}[c]{\textwidth}
    \begin{center}
    \begin{tabular}{c c}
    {\psfrag{y2}[][][1.1][0]{normalised fraction}
    \psfrag{x2}[][][1.1][0]{$\rm \log\left(SFR/BHAR\right) $}
    \hspace{-2.5cm}\includegraphics[trim=9cm 0cm 16cm 1cm, clip=true,
    width=0.45\textwidth]{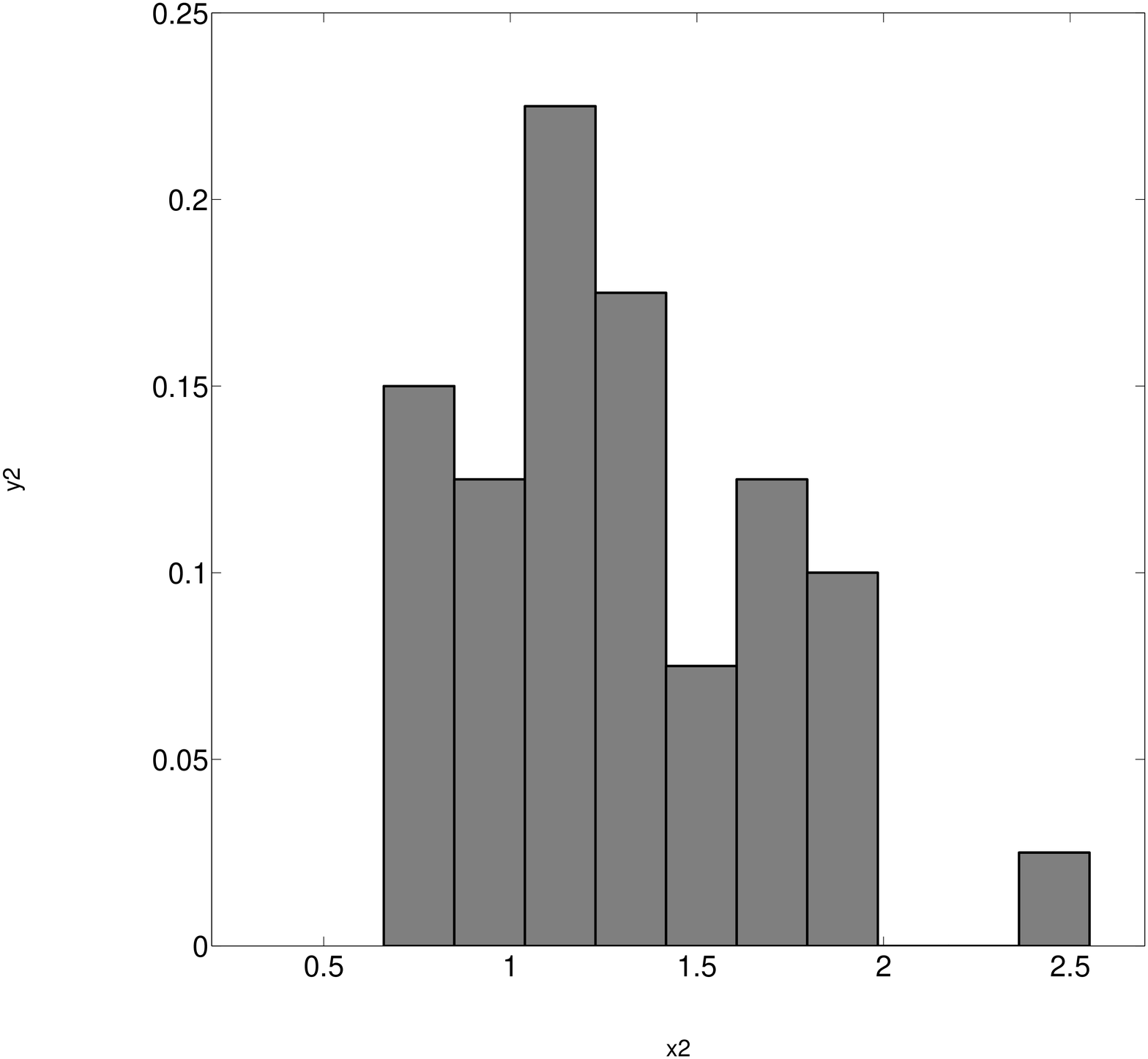}} & 
    {\psfrag{y2}[][][1.1][0]{normalised fraction}
    \psfrag{x2}[][][1.1][0]{$\rm \log\left(M_{*}/M_{\rm
    BH}\right)/\left(SFR/BHAR\right) $}
    \includegraphics[trim=9cm 0cm 16cm 1cm, clip=true,
    width=0.45\textwidth]{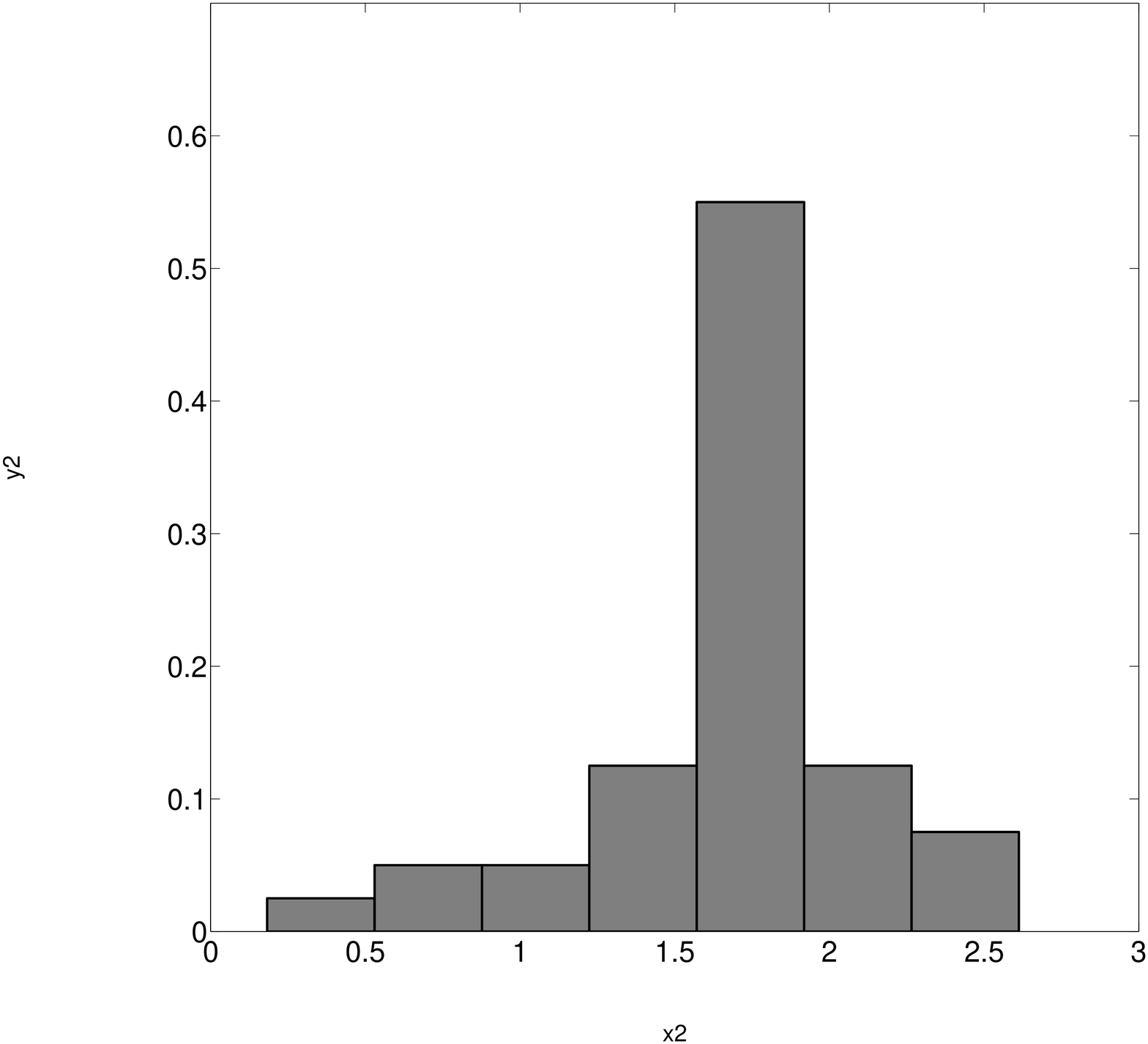}}\\ 
    \vspace{1cm}
    \end{tabular}
       \end{center}
    \end{minipage}
       \caption{{\bf Left panel.} Histogram of the ratio of SFR to BHAR, for PG
       QSOs on and above the SFMS;  the median ratio is $\sim17$. {\bf Right panel.}
       Histogram of the ``relative duty cycles'' of SF and BH activity, defined
       as $M_{*}/M_{\rm BH}/(SFR/BHAR)$. In this case the median value is
       $\sim56$, and is significantly
       different from the value of $\sim2$ which is our estimate of this
       quantity for the high redshift, high luminosity \protect\cite{Netzer2016}
       sample.}
        \label{fig:SFR/BHAR_hist}
    \end{figure*}

\begin{enumerate}
\item
We compared two different ways of measuring $L_{\rm SF}$ relying on different assumptions.
The first is based on fitting $\it Herschel$/PACS observations using the torus SED presented in \cite{Mor2012}, after
extending it to the FIR (EM12). This method assumes no emission from larger-scale AGN-heated dust.
The second method follows the work of S16 where the
SFRs are based on PAH luminosity, as measured from $\it Spitzer$/IRS spectroscopy by \cite{Shi2007}. We find that SF luminosities determined from SED fitting are consistently larger than those based on aromatic features (and the specific $L_{\rm PAH}$ to $L_{\rm SF}$ conversion assumed by \citealt{Shi2007}). The SF templates matching the PAH-based  SF luminosities are then
subtracted from the observed MIR-FIR SEDs of the PGs, and the remaining flux is interpreted as the entire (torus and large-scale) dust emission. Our results are at odds with what presented in S16, however there are several important differences between our analysis and theirs. Firstly we used the entire PG sample while S16 used only sources
with $z<0.18$. In addition we used $\it Herschel$/PACS 70$\, \mu \rm m$ and 100$\, \mu \rm m$,  data that were not used by S16. Most importantly,
 S16 calculated the mean SED of the AGN-heated dust without
normalization while we look for the SED shape after normalization at $20\, \mu \rm m$. 
The different procedures result in  
median
and mean SEDs that are much weaker and ``hotter'', in the FIR, than those proposed by S16. In particular (see Appendix)
 we show that removing the four AGNs that are most luminous at $70\mu\rm m$ from the S16 sample, reduces the mean S16 SED
 by a factor of $\sim2$ at this wavelength. Our median AGN SED derived with the PAH-based method 
represents only $\sim6$ per cent increase on the total ($1-243\, \mu \rm m$) luminosity of
the EM12 torus SED. The increase on the $20-243\, \mu \rm m$ range is $\sim26$ per cent. 
The covering factor required to explain this amount of extra emission by remote AGN-heated dust is only  $\sim3$ per cent.
 Moreover, the excess can also be explained by a systematic under-estimation of the SFR by the PAH-based method.
\item
Assuming that the PAH-based method is correct, we can estimate the properties of the additional AGN-heated dust ($T_{\rm cold\, dust}$).
 For example, for the full sample we find $T_{\rm cold\, dust}$ in the range $\sim 20-30\,\rm K$. On the other hand, for the group of 12 sources representing
the ``top $1 \sigma$" distribution of cold, large-scale dust, we estimate the required
temperature range to be $T_{\rm cold\, dust}=10-60\, \rm K$.
In both cases, the location of this dust, if optically thin, is outside the host galaxy. Optically thick dust can be located closer to
the central source but this requires large column densities and seems to be in conflict with several other observed properties of the sources in the
sample. We emphasize, again, that in both these cases, we could find a single SF template that explained the excess emission.  
\item
We obtained rough estimates of \mstar\ using a relation for \MBH\ and $M_{*}$ determined on a type-II AGN sample from the SDSS.
 We also verified that these are in reasonable agreement with the recent photometric $M_{*}$  estimates of \cite{Zhang2016}. 
We used our \mstar\ values to find the location of the host galaxies in the SFR--\mstar\ plane
and identify the SF main-sequence (SFMS) for our sample.
We examined the correlation between $L_{\rm SF}$ and  $L_{\rm AGN}$ and note a significant difference between objects that are
on the SFMS, below the SFMS, and above the SFMS (candidates for starbursting hosts). Assuming $L_{\rm AGN} \propto L_{\rm SF}^{\alpha}$, we
find $\alpha=0.72\pm0.08$ for objects on the SFMS (31 sources),  and no significant correlation for objects below and above the SFMS (37 and 9 sources respectively).
There is no correlation for the entire sample because of the sources that are below the SFMS. The potential starbursts, however, appear to cluster along the 1:1 line in the $\log L_{\rm SF}-\log L_{\rm AGN}$ plane. 
We investigated the differences between our results and those found
in earlier works, based on X-ray selected samples, like \cite{Rosario2012} and \cite{Stanley2015}.
 We suggest that part of the difference lies in the fact that those samples are dominated by
  objects that are undetected by $\it Herschel$. The corresponding studies must, therefore,
   use stacks or upper limits. This in turn forces the use of mean properties, rather than measured properties
of individual sources. However, some of the difference may be real and, thus far, without explanation. 
\item
We examined the relative instantaneous growth rates of BH and stellar mass (SFR/BHAR), and compared
them with those measured for the most luminous AGNs at $z=2-3.5$ \citep{Netzer2016}. For sources hosted by SF galaxies, the comparison
of SFR/BHAR with \mstar/\MBH\ gives an indication of the relative duty cycles of SF and BH accretion at the two epochs. The differences
in relative duty cycles are very large, ranging from about $20-100$ in the local Universe to about 2 at redshift $z=2-3.5$. We use
these numbers to suggest different scenarios for BH and SF growth rates in systems containing active BHs with a mass
of $\sim 3 \times 10^{8}$\msun\ in the local Universe, and those containing active BHs with a mass of $\sim 10^{10}$\msun\
at high redshift.
\end{enumerate}

    \section*{Acknowledgments}
The authors thank N\'uria Castell\'o-Mor for helpful discussions, and George Rieke, Taro Shimizu, Benny
 Trakhtenbrot, David Rosario and Myrto Symenodis for clarifications and important information. They also acknowledge very helpful comments by the anonymous referee.
 Funding for this work was provided by the Israel Science Foundation grant 284/13. PACS has been developed by a consortium of institutes led by MPE
(Germany) and including UVIE (Austria); KUL, CSL, IMEC (Belgium); CEA,
OAMP (France); MPIA (Germany); IFSI, OAP/OAT, OAA/CAISMI, LENS, SISSA
(Italy); IAC (Spain). This development has been supported by the funding
agencies BMVIT (Austria), ESA-PRODEX (Belgium), CEA/CNES (France),
DLR (Germany), ASI (Italy), and CICYT/MCYT (Spain).

    \bibliographystyle{mnras}
    \bibliography{PGs} 

    \appendix
    \section{Comparison to Symeonidis et al. (2016)}

    Here we focus on the lower redshift ($z<0.18$) sub-sample of PG QSOs
    considered in S16. A detailed description of the selection criteria for this
    sub-sample is given in S16. Briefly, as well as the redshift cut, they
    required radio-quietness and $L_{\rm 2KeV-1\mu m}>10\times L_{\rm 8-1000\mu
    m}$. This part of our analysis is based on 43 sources, identified from the
    names provided in Figure 3 of S16. Our sample size differs from that of S16 (47 QSOs)
    because we excluded 1 source due to having upper limits in all 3 $Herschel$/PACS bands, and 3 sources for having atypically-looking torus emission in the near- and mid-IR.
 
\begin{figure}
    \psfrag{MN12}[][][0.85][0]{\hspace{0.25cm} EM12}
    \psfrag{S16paper}[][][0.9][0]{\hspace{0.6cm} S16 (mean)}
        \psfrag{TWall}[][][0.85][0]{\hspace{3.79cm}This work, full sample (median)}
   \psfrag{TWS16}[][][0.85][0]{\hspace{3.5cm} This work, S16 sample (median)}
   \psfrag{meanS16}[][][0.85][0]{\hspace{3cm} This work, S16 sample (mean)}
    \psfrag{y}[][][1.4][0]{SF-subtracted, normalised $\rm \lambda\,L_{\rm
    \lambda}$}
    \psfrag{x}[][][1.4][0]{$\rm \lambda_{\rm rest}$ ($\mu m$)}
    \includegraphics[scale=0.25,trim={13cm 0cm 12cm  0cm},clip]{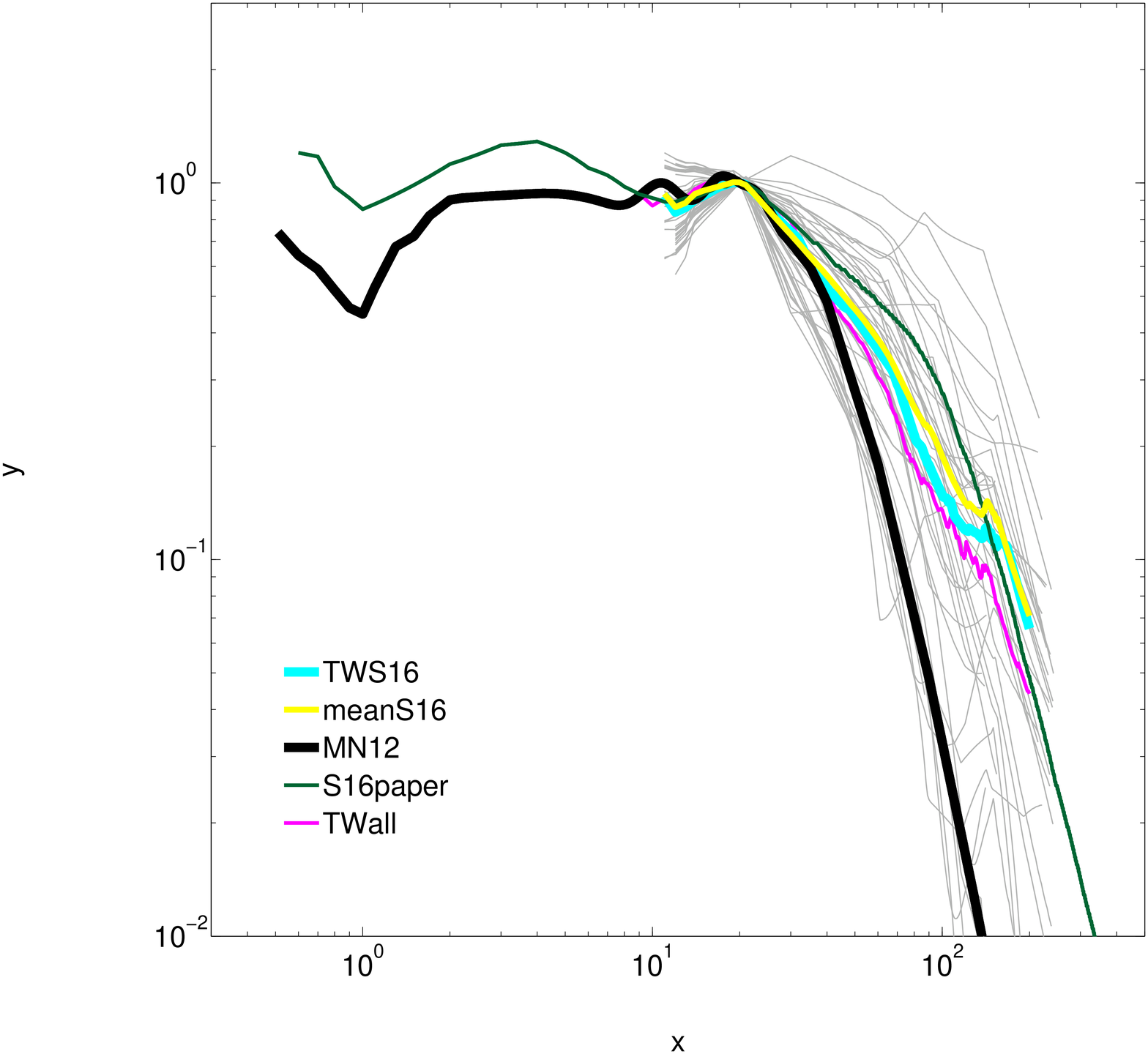}
        \caption{Normalised SF-subtracted SEDs for the $z<0.18$ sub-sample of PG QSOs
        considered in S16. This resulting median SED (cyan) is consistent with that derived from the full population (magenta),
         and differs from that found by S16 (green). We attribute this difference mostly to the lack of normalisation of the SF-subtracted SEDs
         prior to the calculation of the mean in S16, and the presence of a few sources with extremely high 
         $L_{\rm 70\,\mu \rm m}/L_{\rm 12\,\mu \rm m}$. The mean of the S16 sample calculated after normalisation
          (yellow line) is closer to our median than to the mean S16 SED.}
        \label{fig: SF_subtracted_SEDs_subsample}

     \end{figure}

    When considering the lower redshift PG QSOs sub-sample, we identify a colder
    median intrinsic AGN SED than the EM12 median, yet warmer than that presented in S16. This is shown in Figure \ref{fig: SF_subtracted_SEDs_subsample}, and is in agreement with
    the results presented in \S\ref{sec: obtaining intrinsic AGN SED} for the
    overall population. If, however, we follow the same procedure as S16, i.e. finding the mean observed AGN SED without normalization and subtracting 
    from it the mean SF template, we recover an intrinsic AGN SED that is consistent with that found in S16. We here stress the
     importance of normalising the considered SEDs before calculating their average, when seeking to study SED shapes. 
    
    To quantify the effect that the lack of normalisation has on the calculation of the mean SED, we calculated the 
    ratio between the average sample luminosity at $70\,\mu \rm m$ and the average
     sample luminosity at $12\,\mu \rm m$ ($\langle L_{70\,\mu \rm m} \rangle/\langle L_{12\,\mu \rm m} \rangle$). 
     We find that this quantity increases by $\sim35$ per cent when it is calculated on the un-normalised SF-subtracted SEDs,
      compared to the normalised SF-subtracted SEDs. Furthermore we find this ratio to be extremely sensitive
       to the most luminous sources ($L_{70\,\mu \rm m}>3 \times10^{44}\,\rm erg/s$ -- 4 PG quasars)
        when not normalizing the SF-subtracted SEDs. Upon discarding the 4 most luminous sources,
         $\langle L_{70\,\mu \rm m} \rangle/\langle L_{12\,\mu \rm m} \rangle$ decreases by
          $\sim46$ per cent in the un-normalised case, and by $\sim16$ per cent in the normalised case. 
          This indicates that the mean intrinsic AGN SED proposed by S16 represents mostly
           the sources with the largest $L_{70\,\mu \rm m}$ in the sub-sample of $z<0.18$ PG quasars, rather than the entire population.
    
\section{The PAH-based intrinsic AGN SED using an alternative set of SF templates}
\begin{figure}
     \psfrag{M12}[][][0.9][0]{\hspace{0.48cm} EM12}
    \psfrag{S16}[][][0.9][0]{\hspace{1.30cm} S16 (mean)}
    \psfrag{TW}[][][0.9][0]{\hspace{5.70cm} This work (median) with CE01 SF templates}
    \psfrag{DH}[][][0.9][0]{\hspace{5.75cm} This work (median) with DH02 SF templates}
    \psfrag{y}[][][1.4][0]{SF-subtracted, normalised $\rm \lambda\,L_{\rm \lambda}$}
    \psfrag{x}[][][1.4][0]{$\rm \lambda_{\rm rest}$ ($\mu m$)}
    \includegraphics[scale=0.25,trim={13cm 0cm 12cm  0cm},clip]{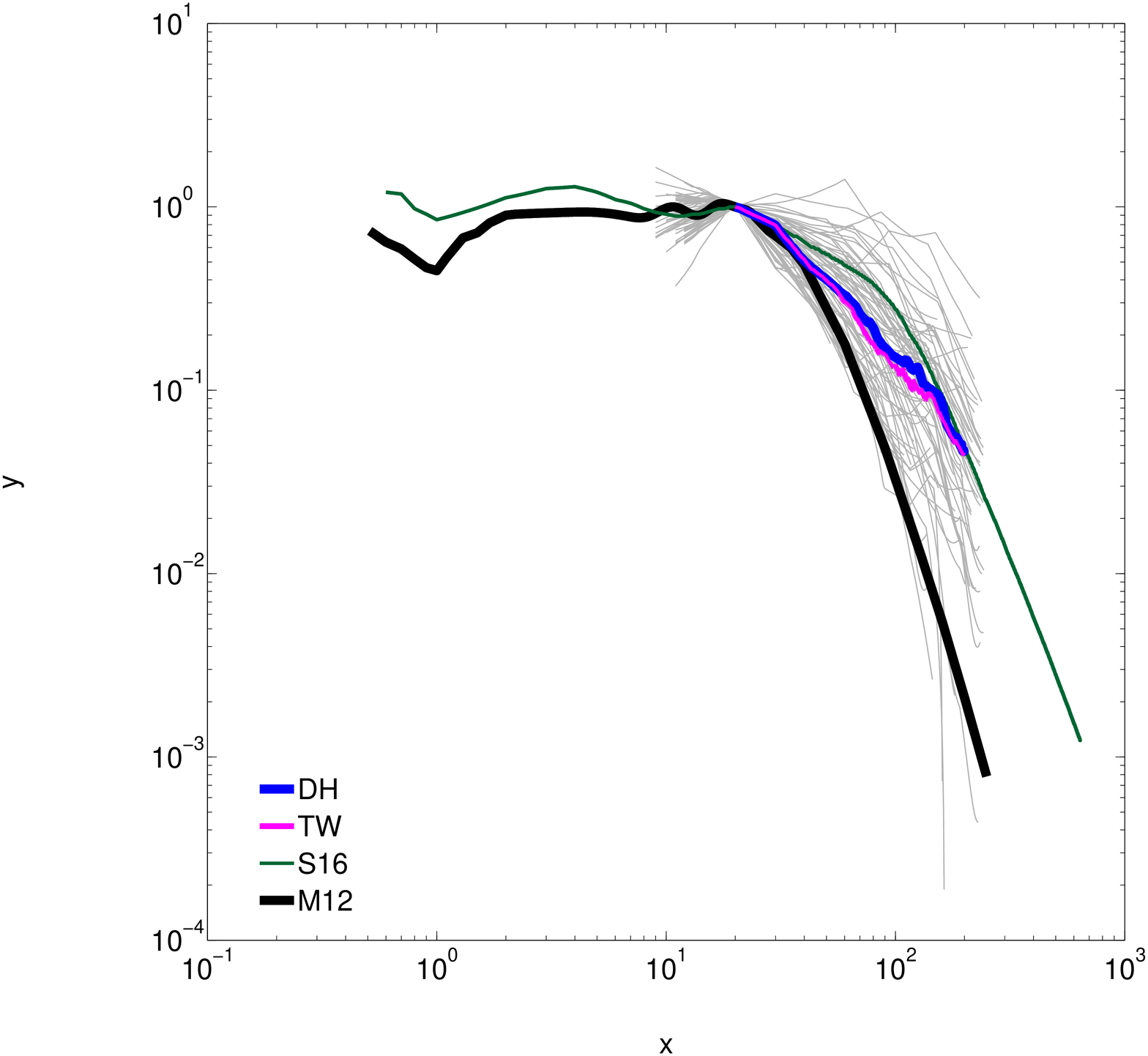} 
        \caption{Same as Figure \ref{fig: SF_subtracted_SEDs}, but here we have subtracted the DH02 SF templates to obtain SF-subtracted intrinsic AGN SEDs of our PGs. Regardless of the SF library considered we obtain consistent median intrinsic AGN SEDs}
        \label{fig: highlight_discrepancy_lambda_DH}
     \end{figure}

Here we check whether the choice of SF library in the derivation of the PAH-based intrinsic AGN SED impacts on the result. In this exercise we follow the same method as highlighted in \S\ref{sec: obtaining intrinsic AGN SED}, but we consider the DH02 library of SF templates. The result is showed in Figure \ref{fig: highlight_discrepancy_lambda_DH} and is consistent with that presented in \S\ref{sec: obtaining intrinsic AGN SED}.

\section{Comparison between EM12 and ME15}
\begin{figure}
     \psfrag{l1}[][][0.9][0]{\hspace{0.7cm} EM12}
    \psfrag{l2}[][][0.9][0]{\hspace{0.7cm} ME15}
        \psfrag{y}[][][1.5][0]{normalised $\rm \lambda\,L_{\rm \lambda}$}
    \psfrag{x}[][][1.5][0]{$\rm \lambda_{\rm rest}$ ($\mu m$)}
    \includegraphics[scale=0.25,trim={13cm 0cm 12cm  0cm},clip]{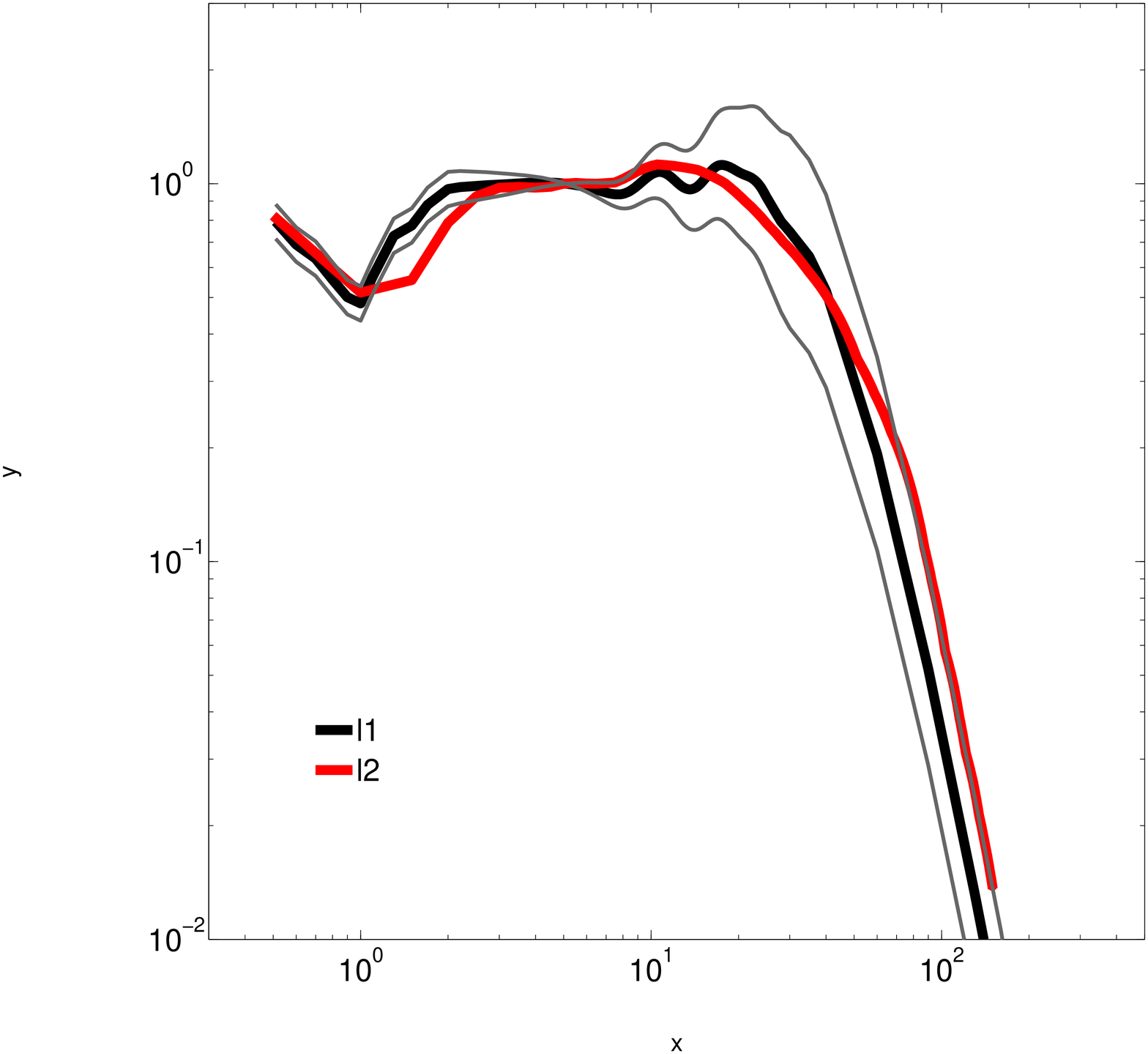} 
        \caption{Comparison between the EM12 intrinsic AGN SED (black, with $25^{\rm th}$ and $75^{\rm th}$ percentiles in grey) and 
        that from \protect \citet[][red]{Xu2015}, which we refer to as ME15.
         For wavelengths longer than $\lambda=2\mu\rm m$ the two SEDs appear in agreement within the uncertainties.
        }
        \label{fig: EM12 vs Xu}
     \end{figure}

Figure \ref{fig: EM12 vs Xu} shows in black the median EM12 intrinsic AGN SED (with $25^{\rm th}$ and $75^{\rm th}$ percentiles in grey), and in red the intrinsic AGN SED from \cite{Xu2015}, as shown in \cite{Lyu2016}. The two templates, which are both normalised at $5\mu\rm m$ agree long-ward of $\sim 2\mu\rm m$, with small differences.

             \section{Intrinsic AGN SEDs}

Below we provide the data for the intrinsic AGN SEDs considered in this work. We note that for the PAH-based SEDs we do not provide data short-ward of $20\mu\rm m$ as below this wavelength we assume the EM12 SED (e.g. in \S\ref{sec:non-torus covering factor etc.}). Furthermore, the EM12 template is identical to that given in \cite{Mor2012} for wavelength $\lambda>2.2\mu\rm m$, while at shorter wavelengths the interpolation considered here is slightly different from that in \cite{Mor2012}.
       
         \clearpage
 \begin{sidewaystable}
\resizebox{23cm}{!}{
\begin{tabular}{cccccccc}
\hline
Rest wavelength   & Extended EM12 median intrinsic AGN SED & Extended EM12 25$^{th}$ percentile & Extended EM12 75$^{th}$
 percentile & Median, PAH-based intrinsic AGN SED for full PG sample & 25$^{th}$ percentile for full PG sample (PAH-based) & 75$^{th}$ percentile for
  full PG sample (PAH-based) & Median, PAH-based intrinsic AGN SED for ``top 1$\sigma$" PGs \\

$\lambda_{\rm rest}$	  &  $\rm \lambda\,L_{\rm \lambda}$  & $\rm \lambda\,L_{\rm \lambda}$ 
	&	$\rm \lambda\,L_{\rm \lambda}$  &  $\rm \lambda\,L_{\rm \lambda}$ 		&		$\rm \lambda\,L_{\rm \lambda}$  &
		$\rm \lambda\,L_{\rm \lambda}$  &	$\rm \lambda\,L_{\rm \lambda}$ \\
$\left(\mu \rm m\right) $	 & 	(arbitrary units) & (arbitrary units) & (arbitrary units) & (arbitrary units) &
 (arbitrary units) & (arbitrary units) & (arbitrary units) \\

\hline
1.00 & 0.4487 & 0.2731 & 0.7372 &      &      &      &     \\
2.00 & 0.9010 & 0.5484 & 1.4802 &      &      &      &     \\
3.00 & 0.9262 & 0.5841 & 1.4687 &      &      &      &     \\
4.00 & 0.9370 & 0.6112 & 1.4364 &      &      &      &     \\
5.00 & 0.9322 & 0.6283 & 1.3832 &      &      &      &     \\
6.00 & 0.9111 & 0.6338 & 1.3098 &      &      &      &     \\
7.00 & 0.8821 & 0.6328 & 1.2296 &      &      &      &     \\
8.00 & 0.8752 & 0.6469 & 1.1841 &      &      &      &     \\
9.00 & 0.9220 & 0.7015 & 1.2118 &      &      &      &     \\
10.00 & 0.9838 & 0.7699 & 1.2571 &      &      &      &     \\
11.00 & 0.9971 & 0.8020 & 1.2396 &      &      &      &     \\
12.00 & 0.9537 & 0.7878 & 1.1545 &      &      &      &     \\
13.00 & 0.9080 & 0.7698 & 1.0709 &      &      &      &     \\
14.00 & 0.9033 & 0.7855 & 1.0388 &      &      &      &     \\
15.00 & 0.9418 & 0.8395 & 1.0567 &      &      &      &     \\
16.00 & 0.9998 & 0.9129 & 1.0950 &      &      &      &     \\
17.00 & 1.0389 & 0.9711 & 1.1114 &      &      &      &     \\
18.00 & 1.0406 & 0.9953 & 1.0879 &      &      &      &     \\
19.00 & 1.0226 & 1.0004 & 1.0453 &      &      &      &     \\
20.00 & 1.0000 & 1.0000 & 1.0000 & 1.0000 & 1.0000 & 1.0000 & 1.0000\\
21.00 & 0.9824 & 1.0038 & 0.9615 & 0.9837 & 0.9717 & 1.0049 & 1.0031\\
22.00 & 0.9675 & 1.0096 & 0.9272 & 0.9530 & 0.9405 & 0.9779 & 0.9930\\
23.00 & 0.9432 & 1.0047 & 0.8854 & 0.9285 & 0.9025 & 0.9627 & 0.9820\\
24.00 & 0.9049 & 0.9836 & 0.8325 & 0.9048 & 0.8637 & 0.9391 & 0.9636\\
25.00 & 0.8561 & 0.9492 & 0.7722 & 0.8847 & 0.8210 & 0.9199 & 0.9512\\
26.00 & 0.8139 & 0.9201 & 0.7200 & 0.8653 & 0.7888 & 0.9060 & 0.9390\\
27.00 & 0.7743 & 0.8921 & 0.6720 & 0.8470 & 0.7600 & 0.8906 & 0.9268\\
28.00 & 0.7391 & 0.8676 & 0.6296 & 0.8286 & 0.7328 & 0.8746 & 0.9153\\
29.00 & 0.7164 & 0.8567 & 0.5991 & 0.8116 & 0.7016 & 0.8600 & 0.9048\\
30.00 & 0.6951 & 0.8462 & 0.5710 & 0.7935 & 0.6716 & 0.8463 & 0.8943\\
31.00 & 0.6733 & 0.8197 & 0.5531 & 0.7531 & 0.6501 & 0.8225 & 0.8820\\
32.00 & 0.6528 & 0.7947 & 0.5362 & 0.7120 & 0.6311 & 0.7989 & 0.8699\\
33.00 & 0.6336 & 0.7713 & 0.5204 & 0.6780 & 0.6090 & 0.7753 & 0.8594\\
34.00 & 0.6154 & 0.7492 & 0.5055 & 0.6545 & 0.5858 & 0.7579 & 0.8491\\
35.00 & 0.5983 & 0.7284 & 0.4915 & 0.6298 & 0.5624 & 0.7358 & 0.8388\\
36.00 & 0.5721 & 0.6965 & 0.4700 & 0.6006 & 0.5335 & 0.7161 & 0.8286\\
37.00 & 0.5478 & 0.6668 & 0.4499 & 0.5785 & 0.5048 & 0.6997 & 0.8185\\
38.00 & 0.5250 & 0.6392 & 0.4313 & 0.5517 & 0.4832 & 0.6899 & 0.8084\\
39.00 & 0.5038 & 0.6133 & 0.4138 & 0.5331 & 0.4572 & 0.6805 & 0.7983\\
40.00 & 0.4839 & 0.5891 & 0.3975 & 0.5174 & 0.4322 & 0.6696 & 0.7884\\
41.00 & 0.4556 & 0.5546 & 0.3742 & 0.4958 & 0.4103 & 0.6580 & 0.7785\\
42.00 & 0.4295 & 0.5229 & 0.3528 & 0.4754 & 0.3905 & 0.6467 & 0.7687\\
43.00 & 0.4055 & 0.4936 & 0.3331 & 0.4586 & 0.3740 & 0.6352 & 0.7590\\
44.00 & 0.3833 & 0.4666 & 0.3149 & 0.4539 & 0.3557 & 0.6226 & 0.7493\\
45.00 & 0.3628 & 0.4417 & 0.2980 & 0.4418 & 0.3385 & 0.6103 & 0.7397\\
46.00 & 0.3438 & 0.4186 & 0.2824 & 0.4323 & 0.3223 & 0.5984 & 0.7302\\
47.00 & 0.3262 & 0.3971 & 0.2679 & 0.4250 & 0.3097 & 0.5868 & 0.7226\\
48.00 & 0.3098 & 0.3772 & 0.2545 & 0.4178 & 0.3070 & 0.5804 & 0.7160\\
49.00 & 0.2946 & 0.3586 & 0.2420 & 0.4062 & 0.2986 & 0.5760 & 0.7096\\
50.00 & 0.2804 & 0.3413 & 0.2303 & 0.3976 & 0.2860 & 0.5678 & 0.7063\\
51.00 & 0.2671 & 0.3252 & 0.2194 & 0.3866 & 0.2729 & 0.5591 & 0.7006\\
52.00 & 0.2547 & 0.3101 & 0.2092 & 0.3789 & 0.2616 & 0.5496 & 0.6947\\
53.00 & 0.2431 & 0.2960 & 0.1997 & 0.3746 & 0.2528 & 0.5498 & 0.6890\\
54.00 & 0.2323 & 0.2828 & 0.1908 & 0.3649 & 0.2429 & 0.5422 & 0.6835\\
55.00 & 0.2221 & 0.2703 & 0.1824 & 0.3555 & 0.2311 & 0.5348 & 0.6782\\
56.00 & 0.2125 & 0.2587 & 0.1745 & 0.3454 & 0.2194 & 0.5224 & 0.6731\\
57.00 & 0.2035 & 0.2477 & 0.1672 & 0.3332 & 0.2098 & 0.5094 & 0.6681\\
58.00 & 0.1950 & 0.2374 & 0.1602 & 0.3246 & 0.2000 & 0.4967 & 0.6634\\
59.00 & 0.1870 & 0.2277 & 0.1536 & 0.3129 & 0.1914 & 0.4843 & 0.6568\\
60.00 & 0.1795 & 0.2185 & 0.1474 & 0.3050 & 0.1842 & 0.4744 & 0.6457\\
61.00 & 0.1702 & 0.2071 & 0.1398 & 0.3011 & 0.1825 & 0.4696 & 0.6378\\
62.00 & 0.1614 & 0.1965 & 0.1326 & 0.2950 & 0.1722 & 0.4684 & 0.6365\\
63.00 & 0.1533 & 0.1866 & 0.1259 & 0.2890 & 0.1649 & 0.4645 & 0.6358\\
64.00 & 0.1457 & 0.1774 & 0.1197 & 0.2832 & 0.1561 & 0.4596 & 0.6355\\
65.00 & 0.1386 & 0.1687 & 0.1138 & 0.2768 & 0.1454 & 0.4514 & 0.6352\\
66.00 & 0.1319 & 0.1606 & 0.1083 & 0.2648 & 0.1395 & 0.4410 & 0.6318\\
67.00 & 0.1256 & 0.1530 & 0.1032 & 0.2533 & 0.1341 & 0.4305 & 0.6226\\
68.00 & 0.1198 & 0.1458 & 0.0984 & 0.2456 & 0.1289 & 0.4243 & 0.6118\\
69.00 & 0.1142 & 0.1391 & 0.0938 & 0.2422 & 0.1245 & 0.4133 & 0.6014\\
70.00 & 0.1091 & 0.1328 & 0.0896 & 0.2327 & 0.1200 & 0.4014 & 0.5913\\
71.00 & 0.1042 & 0.1268 & 0.0856 & 0.2284 & 0.1159 & 0.3917 & 0.5816\\
72.00 & 0.0996 & 0.1212 & 0.0818 & 0.2202 & 0.1102 & 0.3811 & 0.5699\\
73.00 & 0.0952 & 0.1159 & 0.0782 & 0.2122 & 0.1043 & 0.3710 & 0.5562\\
74.00 & 0.0911 & 0.1109 & 0.0749 & 0.2034 & 0.0995 & 0.3607 & 0.5427\\
75.00 & 0.0873 & 0.1062 & 0.0717 & 0.1962 & 0.0941 & 0.3499 & 0.5296\\
76.00 & 0.0836 & 0.1018 & 0.0687 & 0.1913 & 0.0919 & 0.3394 & 0.5168\\
77.00 & 0.0801 & 0.0976 & 0.0658 & 0.1902 & 0.0899 & 0.3297 & 0.5043\\
78.00 & 0.0769 & 0.0936 & 0.0631 & 0.1824 & 0.0862 & 0.3222 & 0.4921\\
79.00 & 0.0738 & 0.0898 & 0.0606 & 0.1845 & 0.0838 & 0.3150 & 0.4801\\
80.00 & 0.0708 & 0.0862 & 0.0582 & 0.1814 & 0.0809 & 0.3051 & 0.4751\\

\end{tabular}}
\caption{Data for the intrinsic AGN SEDs considered in our work.}
\label{tab:SEDs}
\end{sidewaystable}
%
%

\addtocounter{table}{-1}
 \begin{sidewaystable}
\vspace*{17cm}
\resizebox{23cm}{!}{
\begin{tabular}{cccccccc}
\hline
Rest wavelength   & Extended EM12 median intrinsic AGN SED & Extended EM12 25$^{th}$ percentile & Extended EM12 75$^{th}$
 percentile & Median, PAH-based intrinsic AGN SED for full PG sample & 25$^{th}$ percentile for full PG sample (PAH-based) & 75$^{th}$ percentile for
  full PG sample (PAH-based) & Median, PAH-based intrinsic AGN SED for ``top 1$\sigma$" PGs \\

$\lambda_{\rm rest}$	  &  $\rm \lambda\,L_{\rm \lambda}$  & $\rm \lambda\,L_{\rm \lambda}$ 	&	$\rm \lambda\,L_{\rm \lambda}$
  &  $\rm \lambda\,L_{\rm \lambda}$ 		&		$\rm \lambda\,L_{\rm \lambda}$  &	$\rm \lambda\,L_{\rm \lambda}$  &	$\rm \lambda\,L_{\rm \lambda}$ \\
$\left(\mu \rm m\right) $	 & 	(arbitrary units) & (arbitrary units) & (arbitrary units) & (arbitrary units) & (arbitrary units) & (arbitrary units) & (arbitrary units) \\

\hline
81.00 & 0.0680 & 0.0828 & 0.0559 & 0.1776 & 0.0789 & 0.2962 & 0.4714\\
82.00 & 0.0654 & 0.0796 & 0.0537 & 0.1726 & 0.0771 & 0.2880 & 0.4677\\
83.00 & 0.0629 & 0.0765 & 0.0517 & 0.1677 & 0.0746 & 0.2802 & 0.4642\\
84.00 & 0.0605 & 0.0736 & 0.0497 & 0.1634 & 0.0715 & 0.2744 & 0.4609\\
85.00 & 0.0582 & 0.0709 & 0.0478 & 0.1592 & 0.0694 & 0.2688 & 0.4576\\
86.00 & 0.0561 & 0.0682 & 0.0461 & 0.1641 & 0.0661 & 0.2700 & 0.4545\\
87.00 & 0.0540 & 0.0657 & 0.0444 & 0.1610 & 0.0630 & 0.2643 & 0.4515\\
88.00 & 0.0520 & 0.0634 & 0.0428 & 0.1594 & 0.0632 & 0.2648 & 0.4486\\
89.00 & 0.0502 & 0.0611 & 0.0412 & 0.1591 & 0.0649 & 0.2621 & 0.4457\\
90.00 & 0.0484 & 0.0589 & 0.0398 & 0.1566 & 0.0678 & 0.2575 & 0.4431\\
91.00 & 0.0464 & 0.0565 & 0.0382 & 0.1553 & 0.0668 & 0.2506 & 0.4409\\
92.00 & 0.0446 & 0.0543 & 0.0366 & 0.1512 & 0.0684 & 0.2427 & 0.4380\\
93.00 & 0.0428 & 0.0521 & 0.0352 & 0.1472 & 0.0713 & 0.2371 & 0.4301\\
94.00 & 0.0412 & 0.0501 & 0.0338 & 0.1434 & 0.0743 & 0.2338 & 0.4210\\
95.00 & 0.0396 & 0.0482 & 0.0325 & 0.1406 & 0.0719 & 0.2300 & 0.4113\\
96.00 & 0.0381 & 0.0463 & 0.0313 & 0.1378 & 0.0687 & 0.2257 & 0.4016\\
97.00 & 0.0366 & 0.0446 & 0.0301 & 0.1351 & 0.0656 & 0.2215 & 0.3925\\
98.00 & 0.0353 & 0.0429 & 0.0290 & 0.1357 & 0.0646 & 0.2171 & 0.3837\\
99.00 & 0.0339 & 0.0413 & 0.0279 & 0.1365 & 0.0637 & 0.2125 & 0.3753\\
100.00 & 0.0327 & 0.0398 & 0.0269 & 0.1355 & 0.0624 & 0.2081 & 0.3673\\
101.00 & 0.0315 & 0.0384 & 0.0259 & 0.1321 & 0.0605 & 0.2041 & 0.3596\\
102.00 & 0.0304 & 0.0370 & 0.0250 & 0.1287 & 0.0588 & 0.2003 & 0.3523\\
103.00 & 0.0293 & 0.0357 & 0.0241 & 0.1254 & 0.0572 & 0.1966 & 0.3453\\
104.00 & 0.0283 & 0.0344 & 0.0232 & 0.1223 & 0.0556 & 0.1943 & 0.3386\\
105.00 & 0.0273 & 0.0332 & 0.0224 & 0.1218 & 0.0542 & 0.1922 & 0.3321\\
106.00 & 0.0263 & 0.0320 & 0.0216 & 0.1258 & 0.0528 & 0.1901 & 0.3250\\
107.00 & 0.0254 & 0.0309 & 0.0209 & 0.1298 & 0.0514 & 0.1887 & 0.3164\\
108.00 & 0.0246 & 0.0299 & 0.0202 & 0.1292 & 0.0507 & 0.1873 & 0.3082\\
109.00 & 0.0237 & 0.0289 & 0.0195 & 0.1251 & 0.0501 & 0.1843 & 0.3003\\
110.00 & 0.0229 & 0.0279 & 0.0188 & 0.1211 & 0.0495 & 0.1821 & 0.2926\\
111.00 & 0.0222 & 0.0270 & 0.0182 & 0.1175 & 0.0491 & 0.1805 & 0.2852\\
112.00 & 0.0214 & 0.0261 & 0.0176 & 0.1141 & 0.0486 & 0.1779 & 0.2780\\
113.00 & 0.0208 & 0.0253 & 0.0170 & 0.1151 & 0.0482 & 0.1755 & 0.2710\\
114.00 & 0.0201 & 0.0244 & 0.0165 & 0.1139 & 0.0478 & 0.1739 & 0.2643\\
115.00 & 0.0194 & 0.0237 & 0.0160 & 0.1101 & 0.0474 & 0.1723 & 0.2578\\
116.00 & 0.0188 & 0.0229 & 0.0155 & 0.1065 & 0.0474 & 0.1707 & 0.2515\\
117.00 & 0.0182 & 0.0222 & 0.0150 & 0.1030 & 0.0471 & 0.1693 & 0.2460\\
118.00 & 0.0177 & 0.0215 & 0.0145 & 0.1037 & 0.0477 & 0.1678 & 0.2407\\
119.00 & 0.0171 & 0.0208 & 0.0141 & 0.1008 & 0.0482 & 0.1664 & 0.2356\\
120.00 & 0.0166 & 0.0202 & 0.0136 & 0.1029 & 0.0494 & 0.1642 & 0.2306\\
121.00 & 0.0161 & 0.0196 & 0.0132 & 0.1114 & 0.0527 & 0.1635 & 0.2258\\
122.00 & 0.0156 & 0.0190 & 0.0128 & 0.1093 & 0.0541 & 0.1614 & 0.2212\\
123.00 & 0.0151 & 0.0184 & 0.0124 & 0.1073 & 0.0540 & 0.1602 & 0.2167\\
124.00 & 0.0147 & 0.0179 & 0.0121 & 0.1053 & 0.0533 & 0.1602 & 0.2123\\
125.00 & 0.0143 & 0.0174 & 0.0117 & 0.1033 & 0.0526 & 0.1602 & 0.2080\\
126.00 & 0.0138 & 0.0168 & 0.0114 & 0.1014 & 0.0520 & 0.1592 & 0.2039\\
127.00 & 0.0134 & 0.0164 & 0.0110 & 0.0996 & 0.0514 & 0.1583 & 0.1999\\
128.00 & 0.0131 & 0.0159 & 0.0107 & 0.0978 & 0.0508 & 0.1573 & 0.1970\\
129.00 & 0.0127 & 0.0154 & 0.0104 & 0.0994 & 0.0503 & 0.1573 & 0.1949\\
130.00 & 0.0123 & 0.0150 & 0.0101 & 0.1012 & 0.0499 & 0.1575 & 0.1929\\
131.00 & 0.0120 & 0.0146 & 0.0098 & 0.1001 & 0.0498 & 0.1578 & 0.1910\\
132.00 & 0.0116 & 0.0141 & 0.0095 & 0.0973 & 0.0505 & 0.1569 & 0.1893\\
133.00 & 0.0113 & 0.0137 & 0.0093 & 0.0947 & 0.0500 & 0.1537 & 0.1877\\
134.00 & 0.0109 & 0.0133 & 0.0090 & 0.0921 & 0.0495 & 0.1536 & 0.1861\\
135.00 & 0.0106 & 0.0129 & 0.0087 & 0.0897 & 0.0493 & 0.1536 & 0.1846\\
136.00 & 0.0103 & 0.0126 & 0.0085 & 0.0893 & 0.0491 & 0.1531 & 0.1831\\
137.00 & 0.0100 & 0.0122 & 0.0082 & 0.0899 & 0.0475 & 0.1520 & 0.1817\\
138.00 & 0.0097 & 0.0119 & 0.0080 & 0.0965 & 0.0477 & 0.1519 & 0.1797\\
139.00 & 0.0095 & 0.0115 & 0.0078 & 0.0949 & 0.0472 & 0.1518 & 0.1771\\
140.00 & 0.0092 & 0.0112 & 0.0076 & 0.0932 & 0.0468 & 0.1523 & 0.1746\\
141.00 & 0.0090 & 0.0109 & 0.0074 & 0.0967 & 0.0508 & 0.1547 & 0.1722\\
142.00 & 0.0087 & 0.0106 & 0.0072 & 0.0937 & 0.0486 & 0.1557 & 0.1729\\
143.00 & 0.0085 & 0.0103 & 0.0070 & 0.0952 & 0.0466 & 0.1567 & 0.1688\\
144.00 & 0.0082 & 0.0100 & 0.0068 & 0.0933 & 0.0456 & 0.1547 & 0.1663\\
145.00 & 0.0080 & 0.0098 & 0.0066 & 0.0915 & 0.0446 & 0.1529 & 0.1661\\
146.00 & 0.0078 & 0.0095 & 0.0064 & 0.0908 & 0.0436 & 0.1530 & 0.1659\\
147.00 & 0.0076 & 0.0093 & 0.0062 & 0.0902 & 0.0427 & 0.1520 & 0.1657\\
148.00 & 0.0074 & 0.0090 & 0.0061 & 0.0873 & 0.0414 & 0.1446 & 0.1656\\
149.00 & 0.0072 & 0.0088 & 0.0059 & 0.0856 & 0.0408 & 0.1422 & 0.1635\\
150.00 & 0.0070 & 0.0086 & 0.0058 & 0.0839 & 0.0403 & 0.1399 & 0.1591\\
151.00 & 0.0068 & 0.0083 & 0.0056 & 0.0815 & 0.0377 & 0.1384 & 0.1826\\
152.00 & 0.0067 & 0.0081 & 0.0055 & 0.0800 & 0.0372 & 0.1362 & 0.1818\\
153.00 & 0.0065 & 0.0079 & 0.0053 & 0.0773 & 0.0346 & 0.1348 & 0.1975\\
154.00 & 0.0063 & 0.0077 & 0.0052 & 0.0761 & 0.0341 & 0.1331 & 0.1936\\
155.00 & 0.0062 & 0.0075 & 0.0051 & 0.0760 & 0.0384 & 0.1322 & 0.1898\\
156.00 & 0.0060 & 0.0073 & 0.0050 & 0.0752 & 0.0390 & 0.1307 & 0.1861\\
157.00 & 0.0059 & 0.0072 & 0.0048 & 0.0739 & 0.0387 & 0.1285 & 0.1825\\
158.00 & 0.0057 & 0.0070 & 0.0047 & 0.0727 & 0.0384 & 0.1264 & 0.1790\\
159.00 & 0.0056 & 0.0068 & 0.0046 & 0.0715 & 0.0379 & 0.1243 & 0.1756\\
160.00 & 0.0055 & 0.0066 & 0.0045 & 0.0703 & 0.0370 & 0.1222 & 0.1723\\

\hline
\end{tabular}}
\caption{Table \ref{tab:SEDs} continued. The full table (with data up to $\lambda=250\,\mu \rm m$) will be provided in electronic form.}
\end{sidewaystable}

    \bsp    
    \label{lastpage}
    \end{document}